\pdfoutput=1

\documentclass[a4paper,11pt]{article}
\usepackage{graphicx}
\usepackage{amsmath}
\usepackage{epsf}
\usepackage{epsfig}
\usepackage{amssymb}
\usepackage{epstopdf}
\usepackage{lmodern}
\usepackage{latexsym}
\usepackage{subfigure}
\usepackage{subfigure}
\usepackage{multirow}
\usepackage{jheppub}

\preprint{MITP/14-040}

\title{Higgs Production and Decay in Models of a Warped Extra Dimension with a Bulk Higgs}

\author[a]{Paul R.\ Archer,}
\author[b,c]{Marcela Carena,}
\author[d]{Adrian Carmona,}
\author[a,e]{and Matthias Neubert}

\affiliation[a]{PRISMA Cluster of Excellence \& Mainz Institute for Theoretical Physics,\\ 
Johannes Gutenberg University, 55099 Mainz, Germany}
\affiliation[b]{Theoretical Physics Department, Fermilab, Batavia, IL 60510, U.S.A.}
\affiliation[c]{Enrico Fermi Institute and KICP, University of Chicago, Chicago, IL 60637, U.S.A.}
\affiliation[d]{Institut f\"{u}r Theoretische Physik, ETH Z\"urich, 8093 Z\"urich, Switzerland}
\affiliation[e]{Department of Physics, LEPP, Cornell University, Ithaca, NY 14853, U.S.A.}

\emailAdd{archer@uni-mainz.de}
\emailAdd{carena@fnal.gov}
\emailAdd{carmona@itp.phys.ethz.ch}
\emailAdd{matthias.neubert@uni-mainz.de}

\abstract{Warped extra-dimension models in which the Higgs boson is allowed to propagate in the bulk of a compact AdS$_5$ space are conjectured to be dual to models featuring a partially composite Higgs boson. They offer a framework with which to investigate the implications of changing the scaling dimension of the Higgs operator, which can be used to reduce the constraints from electroweak precision data. In the context of such models, we calculate the cross section for Higgs production in gluon fusion and the $H\to\gamma\gamma$ decay rate and show that they are finite (at one-loop order) as a consequence of gauge invariance. The extended scalar sector comprising the Kaluza-Klein excitations of the Standard Model scalars is constructed in detail. The largest effects are due to virtual KK fermions, whose contributions to the cross section and decay rate introduce a quadratic sensitivity to the maximum allowed value $y_\ast$ of the random complex entries of the 5D anarchic Yukawa matrices. We find an enhancement of the gluon-fusion cross section and a reduction of the $H\to\gamma\gamma$ rate as well as of the tree-level Higgs couplings to fermions and electroweak gauge bosons. We perform a detailed study of the correlated signal strengths for different production mechanisms and decay channels as functions of $y_\ast$, the mass scale of Kaluza-Klein resonances and the scaling dimension of the composite Higgs operator.}

\begin{document}

\maketitle
\flushbottom

\section{Introduction}
\label{sec:intro}

The discovery of the Higgs boson marks the beginning of a new era in particle physics. It represents the completion of the standard model (SM), but it also means the gauge hierarchy problem ceases to be merely a theoretical puzzle. The measurements of the Higgs couplings introduce a new set of constraints on beyond-the-standard-model (BSM) scenarios,  which complement those coming from electroweak precision tests and flavour physics. The LHC, as well as other possible future colliders, will measure the Higgs couplings with ever increasing accuracy, thereby providing stringent constraints on models of electroweak symmetry breaking. Hence, if we are to have any hope of experimentally distinguishing between the multitude of BSM scenarios, it is important to have a clear understanding of how respective models modify the Higgs couplings.

This paper seeks to investigate how such Higgs couplings are modified in a class of models in which the Higgs propagates in a fifth dimension, with particular emphasis on a slice of five-dimensional (5D) anti-de Sitter space (AdS${}_5$). There has been considerable interest in the phenomenological implications of such a space following the proposal of the Randall-Sundrum (RS) model \cite{Randall:1999ee} as a non-supersymmetric resolution to the gauge hierarchy problem, as well as the associated description of flavour hierarchies \cite{Grossman:1999ra, Gherghetta:2000qt, Huber:2000ie}. In the most frequently studied version of the RS model, the Higgs is strictly localised to one of the branes on the boundary of the space. However, there is no compelling reason why the Higgs should have a special status as the only brane-localised particle, since a partially delocalized Higgs can also provide a resolution to the hierarchy problem.  In addition, allowing the Higgs to propagate in the bulk helps alleviating some of the tensions with electroweak precision tests \cite{Cabrer:2011fb, Cabrer:2011vu, Carmona:2011ib} and flavour physics \cite{Agashe:2008uz, Archer:2011bk, Cabrer:2011qb}, it also offers some explanation of the small scale of neutrino masses  \cite{Agashe:2008fe, Archer:2012qa, vonGersdorff:2012tt} and is consistent with extensions that include a dark matter candidate 
\cite{Agashe:2007jb,Panico:2008bx,Medina:2010mu}.

When considering the implications for Higgs physics, there are a couple of more fundamental motivations for considering a bulk Higgs. Firstly, under the AdS/CFT correspondence a bulk Higgs is conjectured to be dual to an elementary Higgs mixing with the bound states of a broken conformal field theory, i.e.\ a partially composite Higgs. By varying the 5D Higgs mass, one can vary the scaling dimension of the effective Higgs operator \cite{Witten:1998qj}. Hence, models with a bulk Higgs offer a relatively concrete framework for investigating the phenomenological implications of changing the scaling dimension of the Higgs operator, see also the discussion in \cite{Luty:2004ye}. Secondly, an important but subtle issue arises when one requires a four-dimensional (4D) low-energy effective chiral theory in scenarios with brane-localised Yukawa couplings. Fermions in 5D are vector-like objects, however a low-energy effective chiral theory can be obtained by imposing Dirichlet boundary conditions on one of the Weyl spinors \cite{Grossman:1999ra, Gherghetta:2000qt}. Such boundary conditions must be consistent with the variation of the action \cite{Csaki:2003sh}, and hence any brane-localised Yukawa couplings must be carefully regularised \cite{Azatov:2009na, Casagrande:2010si}. It is found that the calculation of the gluon-fusion cross section for Higgs production differs  depending on whether this regularisation is made before or after summing the Kaluza-Klein (KK) tower of fermions propagating in the loop \cite{Casagrande:2010si, Azatov:2010pf, Goertz:2011hj, Carena:2012fk, Frank:2013un}. In \cite{Malm:2013jia}, the physical origin of this effect was clarified, and detailed analytical predictions for the gluon-fusion cross section were derived in different RS scenarios, in which the scalar sector is either localised on or near the infrared (IR) brane. It was shown that the result found in \cite{Casagrande:2010si, Carena:2012fk} corresponds to the case of a strictly brane-localised Higgs field. The result obtained in \cite{Azatov:2010pf} instead corresponds to the case of a narrow bulk Higgs, whose profile along the extra dimension can however be resolved by high-mass KK states coupling to the Higgs boson. A scenario in which the Higgs sector lives in the bulk should connect to the narrow bulk-Higgs results obtained in \cite{Azatov:2010pf, Malm:2013jia} in the limit where the scaling dimension of the Higgs operator is taken to be very large.

With this in mind, this paper aims to conduct a comprehensive study of the modification of the Higgs couplings in scenarios where the Higgs propagates in a fifth dimension. Although our phenomenological discussions will focus on AdS${}_5$, in sections~\ref{sec:EWbreak} and \ref{sect:HiggsProd} we work with a generic 5D geometry, and hence this work can be applied to a wide range of scenarios. In section~\ref{sec:EWbreak} we examine the Higgs mechanism and derive the equations of motion of the model. In section~\ref{sect:HiggsProd} we derive the relevant Feynman rules and use them to compute the $gg\to H$ production cross section and the decay rates for $H\to WW^*$, $H\to ZZ^*$, and $H\to\gamma\gamma$. In section~\ref{sect:AdS5Geo} we specialise our general results to the case of an AdS$_5$ space and compute the size of the corrections to electroweak precision observables. We also estimate the size of 5D Yukawa couplings at which one looses perturbative control of the theory. In section~\ref{sect:phenom} we apply these results and calculate the modifications to the individual Higgs production rates and decay widths and to the relative signal strengths measurable at the LHC. Finally, we conclude in section~\ref{sect:Conclusions}.

\section{Electroweak Symmetry Breaking with a Bulk Higgs}
\label{sec:EWbreak}

The primary focus of this work is to study Higgs physics in the context of RS-type scenarios, i.e.\ spaces in which the dimensionful parameters exist at the Planck scale, but an effective 4D electroweak scale is generated via gravitational red-shifting (warping). In the following two sections we shall work with a generic 5D space. In particular, in this section we shall study a minimal 5D version of the SM, in which all fields propagate in the bulk and electroweak symmetry is broken by the vacuum expectation value (VEV) of the Higgs field.      

\subsection{Higgs Mechanism in a Generic 5D Space}
\label{sect:HiggsMech}

In the interest of generality, we shall consider 5D spaces described by the metric
\begin{equation}\label{genMetric}
   ds^2 = a^2(r)\,\eta^{\mu\nu} dx_\mu dx_\nu - b^2(r)\,dr^2 \,,
\end{equation}
which are cut off in the IR and ultraviolet (UV), i.e.\ $r\in [r_{\rm{UV}}, r_{\rm{IR}}]$. Note that without loss of generality $b(r)$ can be set to 1 with the coordinate transformation $r\to\tilde{r}=\int_c^r b(\hat{r})\,d\hat{r}$, but by not doing so the following expressions can be easily adapted to alternative geometries. Our 5D coordinates run over $X_M=\{x_\mu,r\}$ with $\mu=0,\dots,3$, and we adopt the Minkowski metric $\eta_{\mu\nu}=\mathrm{diag}(+1,-1,-1,-1)$. As already mentioned, we shall be primarily interested in an AdS$_5$ geometry, for which
\begin{equation}\label{ADS5def}
   a(r) = b(r) = \frac{R}{r} \,, \quad \mbox{with} \quad 
   r_{\rm{UV}} = R \,, \quad r_{\rm{IR}}=R^{\prime} \,.
\end{equation}
The space is assumed to have been stabilised such that one obtains a large warp factor, $\Omega\equiv R'/R\approx 10^{15}$ \cite{Goldberger:1999uk}. It is also useful to define the parameter $M_{\mathrm{KK}}\equiv 1/R'$, which sets the mass scale for low-lying KK excitations, as well as the AdS curvature $k=1/R$. In \cite{Casagrande:2010si,Goertz:2011hj,Carena:2012fk,Frank:2013un,Malm:2013jia}, the dimensionless variable $t=r/R'$ was used instead of $r$, and the warp factor was denoted by $\epsilon=R/R'=1/\Omega$. In other works \cite{Carena:2002dz,Carena:2003fx,Carena:2004zn,Carmona:2011rd,Carmona:2011ib}, a coordinate $y$ defined via $r=R\,e^{ky}$ is used, which takes values $y=0$ on the UV brane and $y=L/k$ on the IR brane.

We shall consider a minimal model with an $\mathrm{SU}_c(3)\times\mathrm{SU}_L(2)\times \mathrm{U}_Y(1)$ gauge symmetry in the bulk. The action is given by
\begin{equation}\label{FullLag}
   S = \int d^{5}x\,\sqrt{G} \left[ \mathcal{L}_{\rm{Higgs}} + \mathcal{L}_{\rm{gauge}}
      + \mathcal{L}_{\rm{ferm.}} + \frac{\delta(r-r_{\rm{IR}})}{b}\,\mathcal{L}_{\rm{IR}}
      + \frac{\delta(r-r_{\rm{UV}})}{b}\,\mathcal{L}_{\rm{UV}} \right] ,
\end{equation}
with 
\begin{align}
   \mathcal{L}_{\rm{Higgs}} &= g^{MN} (D_M\Phi^\dagger) (D_N\Phi) - V(\Phi) \,, \label{HiggLag} \\
   \mathcal{L}_{\rm{gauge}} &= - \frac{1}{4} G_{MN}^bG^{MN\,b} - \frac{1}{4} F_{MN}^aF^{MN\,a}
    - \frac{1}{4}B_{MN}B^{MN} \,, \label{GaugeLag} \\
   \mathcal{L}_{\rm{ferm.}} &= \sum_\Psi
    \bar\Psi \big( i\Gamma^M\Delta_M - M_{\Psi} \big) \Psi
    - \left(  \bar Q\,\mathbf{Y}_d^{5D} \Phi d + \bar Q\,\mathbf{Y}_u^{5D} \epsilon\Phi^\dag u 
    + \bar L\,\mathbf{Y}_e^{5D} \Phi e + \textrm{h.c.} \right) , \label{FermLag} \\
    \mathcal{L}_{\rm{IR},\rm{UV}} & =-V_{\rm{IR},\rm{UV}}(\Phi) \,, \label{BraneLag}
\end{align}
where $G$ is the determinant of the metric in (\ref{genMetric}), while $G_{MN}^b$, $F_{MN}^a$ and $B_{MN}$ are the field strength tensors for the $\mathrm{SU}_c(3)$, $\mathrm{SU}_L(2)$ and $\mathrm{U}_Y(1)$ gauge symmetries, and $\Phi$ is the 5D scalar doublet. In the first term of the fermion Lagrangian $\Psi=Q,L,u,d,e$ represents any of the 5D fields. The fermion content is the same as in the SM, with the $\mathrm{SU}_L(2)$ doublets denoted by $Q=(U,D)^T$ and $L=({\cal V},E)^T$, and the singlets denoted by $u$, $d$, and $e$. The 5D fermion fields are 4-component Dirac spinor fields, whose chiral zero modes correspond to the SM fermions. The covariant derivative includes a spin connection term as well as the usual covariant derivative, i.e.\ $i\Delta_M=iD_M+\omega_M$, where
\begin{equation}\label{covder}
   i D_M = i \partial_M + g_{s5}\,\mathcal{G}_M^b\frac{\lambda^b}{2} + g_5 A_M^a T^a
    + g_5^{\prime} Y B_M \,.
\end{equation}

Here we are focusing on arguably the most minimal scenario. In particular, we have not considered an extended custodial gauge symmetry. It is well known that models with a brane-localised Higgs and no custodial symmetry suffer from large constraints from electroweak precision tests, see for example \cite{Csaki:2002gy,Carena:2003fx}. However, these constraints are significantly reduced when the Higgs is free to propagate in the bulk \cite{Cabrer:2011fb, Carmona:2011ib,Cabrer:2011vu}. Also, in order to protect the $Z\to\bar{b}b$ vertex, such custodial models require a considerably extended quark sector, which would affect Higgs production in a much more pronounced way compared with the minimal model \cite{Casagrande:2010si, Goertz:2011hj, Malm:2013jia}. The principle focus of this work is to study the implications of a bulk Higgs field on Higgs-boson phenomenology. Hence, in order to be able to distinguish the effects coming from a bulk Higgs from those coming from an extended quark sector, it is useful to consider the simpler scenario first. We should also note that, for simplicity, the only brane-localised operators included in our analysis are those belonging to the Higgs potentials in (\ref{BraneLag}). In principle, all operators allowed by the symmetries of the model can be included on both branes and in the bulk \cite{Georgi:2000ks,Carena:2002dz,Carena:2004zn}. Indeed, such operators will in general be required as counterterms removing the UV singularities of divergent loop graphs with fields propagating in the bulk \cite{Georgi:2000ks}. Therefore, it is not inconsistent to work under the assumption that their coefficients are loop suppressed. In this way we avoid having to deal with a significantly enhanced parameter space, which would make phenomenological studies more challenging.

In the following we shall not consider the possibility that the bulk-Higgs scalar mixes with the radion field present in RS scenarios in which the distance between the two branes is stabilised dynamically by means of the Goldberger-Wise mechanism \cite{Goldberger:1999uk}. The phenomenology of Higgs-radion mixing has been studied by several authors \cite{Giudice:2000av,Csaki:2000zn,Dominici:2002jv,Gunion:2003px,Rizzo:2002pq}. A detailed analysis of Higgs phenomenology in the context of bulk-Higgs models with Higgs-radion mixing, in which the back reaction on the geometry is taken into account, has been performed in \cite{Cox:2013rva}. The most general set of kinetic mixing and mass mixing terms can be parameterised in terms of an effective Lagrangian involving three coefficient functions $c_i$, which might be suppressed in some scenarios but which in general could be of ${\cal O}(1)$. Irrespective of the values of these couplings, one finds that the effects of Higgs-radion mixing on the effective couplings of the physical Higgs boson to SM fermions and gauge bosons are suppressed by $v^2/\Lambda_{\rm TeV}^2$, where $\Lambda_{\rm TeV}\sim 10 M_{\mathrm{KK}}$ denotes the warped-down UV cutoff of the theory. These effects are parametrically smaller than the corrections of order $v^2/M_{\rm KK}^2$, which we compute in this work. 
 Note, in particular, that to an excellent approximation the Higgs couplings to $W$ and $Z$ bosons are reduced by a factor $\cos\theta_r$ when Higgs-radion mixing is taken into account, where $\theta_r={\cal O}(v/\Lambda_{\rm TeV})$ is the mixing angle \cite{Cox:2013rva}. The fact that the measured values of these couplings appear to be close to their SM values supports our assumption that mixing effects are numerically very small. We might add that, at order $v^2/\Lambda_{\rm TeV}^2$, one could in principle consider a larger set of higher-dimensional operators localised on the IR brane (see \cite{Malm:2013jia} for a discussion of such operators in the context of Higgs production in gluon fusion), which would give small corrections to basically every observable.

The Higgs potentials in (\ref{FullLag}) result in the field $\Phi$ gaining a non-zero VEV, $\langle\Phi\rangle^T=\frac{1}{\sqrt{2}}\,\Big( 0 ~~ v(r) \Big)$, such that the position-dependent value $v(r)$ satisfies \cite{Cacciapaglia:2006mz,Cabrer:2011fb,Archer:2012qa}
\begin{equation}\label{hvevEqn}
   \partial_r \left( a^4 b^{-1}\partial_r\,v \right)
    - a^4 b\,\frac{\delta V(\langle\Phi\rangle)}{\delta v} = 0 \,.
\end{equation}
The two consistent boundary conditions are either $v(r_{\rm{IR}})=v(r_{\rm{UV}})=0$, or
\begin{equation}\label{hvevBCs}
   \left[ b^{-1}\partial_r\,v + \frac{\delta V_{\rm{IR}}(\langle\Phi\rangle)}{\delta v}
    \right]_{r=r_{\rm{IR}}} = 0 \,, \qquad
   \left[ b^{-1}\partial_r\,v - \frac{\delta V_{\rm{UV}}(\langle\Phi\rangle)}{\delta v}
    \right]_{r=r_{\rm{UV}}} = 0 \,.
\end{equation}
While there is some model dependence in what Higgs potentials are considered, clearly, in order to actually break electroweak symmetry, the solution of (\ref{hvevEqn}) and (\ref{hvevBCs}) must be $v(r)\neq 0$. It is also worth pointing out that, in order to achieve a Higgs VEV that is constant with respect to $r$, one would need to require a fine-tuning between the three potentials $V$, $V_{\rm IR}$ and $V_{\rm UV}$. As shall be demonstrated in section~\ref{sect:HiggsVEV}, in AdS$_5$ or asymptotically AdS spaces, one finds that, without fine-tuning, the Higgs VEV is heavily peaked towards the IR side of the space \cite{Archer:2012qa, Luty:2004ye}. Hence, bulk Higgs scenarios still offer a potential resolution to the gauge hierarchy problem. It is convenient to rewrite the position-dependent VEV as $v(r)\equiv\tilde v\,h(r)$, where the profile $h(r)$ satisfies the normalization condition
\begin{equation}\label{eq42}
   \int dr\,a^2 b\,h^2 = 1 \,.
\end{equation}  
Here and below, all integrals over $r$ run from $r_{\rm UV}$ to $r_{\rm IR}$. With this definition, $\tilde v\approx v=246.2$\,GeV coincides with the SM Higgs VEV $v$ up to higher-order corrections in an expansion in powers of $\tilde v^2/M_{\rm KK}^2$ (see section~\ref{sec:EWpars} for more details).

We can expand the complex doublet $\Phi$ around the Higgs VEV by writing
\begin{equation}\label{PHIDEF}
   \Phi(x,r) = \left(\begin{array}{c} -i\pi^+(x,r) \\ 
    \frac{1}{\sqrt{2}} \left[ v(r)+H(x,r)+i\pi_3(x,r) \right] \end{array} \right) .
\end{equation}
From here on, we shall refer to $H(x,r)$ as the 5D Higgs field and make the orthonormal KK expansion $H(x,r)=\sum_n f_n^{(H)}(r)\,H^{(n)}(x)$, such that the KK Higgs particles are canonically normalised, i.e.\
\begin{equation}\label{HiggsOrthog}
   \int dr\,a^2 b\,f^{(H)}_n f^{(H)}_m = \delta_{nm} \,.
\end{equation}  
The profile functions satisfy the equations of motion \cite{Cacciapaglia:2006mz,Cabrer:2011fb,Archer:2012qa}
\begin{equation}\label{HProfEOM}
   \partial_r \left( a^4 b^{-1}\partial_r f_n^{(H)} \right)
    - a^4 b\,\frac{\delta^2 V(\langle\Phi\rangle)}{\delta v^2}\,f_n^{(H)}    
    + a^2 b\,m_n^{(H)\,2} f_n^{(H)} = 0 \,,
\end{equation}
where $\partial_\mu\partial^\mu H^{(n)}(x)\equiv\Box_4 H^{(n)}(x)=-m_n^{(H)\,2}\,H^{(n)}(x)$. Note that the functional derivative in the second term isolates the term linear in $f_n^{(H)}$. Again one could in principle impose Dirichlet boundary conditions on the Higgs field, $f_n^{(H)}(r_{\mathrm{IR}})=f_n^{(H)}(r_{\mathrm{UV}})=0$, but this would not allow for a light zero mode, which can be identified with the SM Higgs boson. Hence, we impose the boundary conditions
\begin{equation}\label{HiggsBCs}
   \left[ b^{-1}\partial_r f_n^{(H)}
    + \frac{\delta^2 V_{\rm IR}(\langle\Phi\rangle)}{\delta v^2}\,f_n^{(H)}
    \right]_{r=r_{\rm{IR}}} = 0 \,, \qquad
   \left[ b^{-1}\partial_r f_n^{(H)}
    - \frac{\delta^2 V_{\rm UV}(\langle\Phi\rangle)}{\delta v^2}\,f_n^{(H)}
    \right]_{r=r_{\rm{UV}}} = 0 \,.
\end{equation}  
In the limit where $m_0^{(H)}\ll M_{\rm KK}$, with $m_H\equiv m_0^{(H)}\approx 125.5$\,GeV being the Higgs-boson mass, one finds that $f_0^{(H)}(r)\approx h(r)$ up to small higher-order corrections of order $m_H^2/M_{\rm KK}^2$, see relation (\ref{VevProfApp}) in section~\ref{sect:HiggsVEV}.

As $\Phi$ acquires a non-zero VEV, the $W$ and $Z$ fields gain masses via the Higgs kinetic term (\ref{HiggLag}), and hence we make the usual field redefinitions 
\begin{equation}
   W_M^\pm = \frac{1}{\sqrt{2}} \left( A_M^1\mp iA_M^2 \right) , \quad
   A_M = s_w A_M^3 + c_w B_M \,, \quad 
   Z_M = c_w A_M^3 - s_w B_M \,,
\end{equation}     
where
\begin{equation}\label{swdef}
   s_w = \frac{g_5^{\prime}}{\sqrt{g_5^2+g_5^{\prime\,2}}} \,, \qquad
   c_w = \frac{g_5}{\sqrt{g_5^2+g_5^{\prime\,2}}} \,,
\end{equation} 
with $g_5$ and $g_5^{\prime}$ being the 5D gauge couplings associated with $\mathrm{SU}_L(2)$ and $\mathrm{U}_Y(1)$. It is also convenient to define the 5D gauge-boson masses
\begin{equation}\label{MWZdef}
   M_W \equiv \frac{g_5\tilde v}{2} \,, \qquad M_Z \equiv \frac{M_W}{c_w} \,,
\end{equation} 
which are the 5D analogues of the physical $W$ and $Z$ boson masses. Before we can proceed we must include gauge-fixing terms, chosen in order to cancel the terms in the Lagrangian which mix the 4D gauge fields with scalar fields. They are
\begin{eqnarray}\label{GFLag}
   S_{\rm GF} &=& \int d^5x\,\bigg\{ 
    - \frac{b}{2\xi} \left| \partial_\mu Z^\mu - \xi b^{-1} \left( \partial_r (a^2 b^{-1} Z_5)
    + a^2 b M_Z h \pi_3 \right) \right|^2 \nonumber\\
   &&\mbox{}- \frac{b}{2\xi} \left| \partial_\mu A^\mu - \xi b^{-1}\partial_r (a^2 b^{-1} A_5) 
     \right|^2 \nonumber\\
   &&\mbox{}- \frac{b}{\xi} \left| \partial_\mu W^{\mu\,+}
    - \xi b^{-1} \left( \partial_r (a^2 b^{-1} W_5^+) 
    + a^2bM_W h\pi^+ \right) \right|^2 \bigg\} \,,
\end{eqnarray}
where we use the short-hand notation $|\pi^+|^2=\pi^+\pi^-$ etc. Finally we make an orthonormal KK decomposition of the 4D gauge fields,
\begin{equation}\label{VectorKKdec}
   \mathbb{A}_\mu(x,r) = \sum_n f_n^{(\mathbb{A})}(r)\,\mathbb{A}_\mu^{(n)}(x) \,, 
    \quad \mbox{with} \quad
   \partial^\nu \mathbb{A}_{\nu\mu}^{(n)} + \frac{1}{\xi}\,\partial_\mu 
    (\partial^\nu \mathbb{A}_\nu^{(n)}) = - m_n^{(A)\,2} \mathbb{A}_\mu^{(n)} \,,
\end{equation}
where $\mathbb{A}\in[A,\, W^\pm,\, Z,\,\mathcal{G}]$, such that 
\begin{equation}\label{GaugeOrtho}
   \int dr\,b\,f_n^{(\mathbb{A})} f_m^{(\mathbb{A})} = \delta_{nm} \,.
\end{equation}
The profiles and masses can then be found by solving the equations of motion \cite{Cacciapaglia:2006mz,Falkowski:2008fz,Cabrer:2011fb,Archer:2012qa}
\begin{align}
   \partial_r \left( a^2 b^{-1} \partial_r f_n^{(W,Z)} \right) - a^2 b M_{W,Z}^2
    h^2 f_n^{W,Z}+bm_n^{(W,Z)\,2}f_n^{(W,Z)} &=0 \,, \label{fWZEOM} \\
   \partial_r (a^2 b^{-1} \partial_r f_n^{(A,\mathcal{G})}) + b m_n^{(A,\mathcal{G})\,2}
    f_n^{(A,\mathcal{G})} &= 0 \,. \label{fAEOM}  
\end{align}
Again, in order to ensure the existence of zero modes, we impose Neumann boundary conditions $\partial_r f_n^{(\mathbb{A})}|_{r=r_{\mathrm{IR}},r_{\mathrm{UV}}}=0$. Note that for the case of the photon and the gluon the zero mode is massless, $m_0^{(A,\mathcal{G})}=0$, and the corresponding profiles are constant along the extra dimension and given by $f_0^{(A,\mathcal{G})}=\left[\int dr\,b\,\right]^{-1/2}$.

\subsection{Extended Scalar Sector} 
\label{sec:ExtScalar}
  
An important feature of bulk Higgs models, which allows them to be distinguished from brane Higgs scenarios, is the existence of additional physical scalar fields. The existence of such scalars can be seen by simply counting the degrees of freedom in the model. It is well known that such scalars exist in models with universal extra dimensions \cite{Appelquist:2000nn}, and they have been previously studied in the context of warped extra dimensions in \cite{Falkowski:2008fz, Cabrer:2011fb, Archer:2012qa}. After choosing the gauge symmetries and imposing the relevant boundary conditions to ensure the existence of the $W$, $Z$ and photon zero modes, the scalar sector is completely fixed. In particular, in the electroweak sector there are four KK towers of unphysical Goldstone bosons in addition to the KK tower of Higgs particles (denoted by $H^{(n)}$). Three with zero modes, which are then eaten by the longitudinal degrees of freedom of the $W$ and $Z$ bosons and their KK excitations, and one without a zero mode corresponding to the longitudinal degrees of freedom of the KK photons. In addition to this, there are three KK towers of physical scalars which do not have zero modes.

We identify the Goldstone bosons as the linear combinations of the scalar fields $\mathbb{A}_5$ and $\pi$, which multiply the gauge parameter $\xi$ in (\ref{GFLag}). This implies 
\begin{align}\label{GFdef}
   G^A &= b^{-1}\,\partial_r (a^2 b^{-1} A_5) \,, \\
   G^Z &= b^{-1} \left( \partial_r (a^2 b^{-1} Z_5) + a^2 b M_Z h \pi_3 \right) , 
    \label{GZdef} \\
   G^\pm &= b^{-1} \left( \partial_r (a^2 b^{-1} W^\pm_5)
    + a^2 b M_W h \pi^\pm \right) \label{Gpmdef},
\end{align}  
where the first equation holds for both the photon and the gluon. The physical scalars are found by taking the linear combinations of $\mathbb{A}_5$ and $\pi$ that are gauge independent. The equations of motion for $W_5^{\pm}$ and $\pi^\pm$ are
\begin{align}
   \Box_4 W_5^\pm + a^2 M_W^2 h^2 W_5^\pm
    + a^2 M_W h^2 \partial_r (h^{-1} \pi^\pm) 
    - \xi \partial_r G^{\pm} &=0 \,, \label{W5eom} \\  
   \Box_4 \pi^\pm - a^{-2} b^{-1} \partial_r (a^4 b^{-1} \partial_r \pi^\pm)
    - a^{-2} b^{-1} M_W h^{-1} 
     \partial_r (a^4 b^{-1} h^2 W_5^\pm) \hspace{1cm} & \nonumber\\
   + \frac{a^2}{v}\,\frac{\delta V(\langle\Phi\rangle)}{\delta v}\,\pi^\pm
    + \xi M_W h G^{\pm} &= 0 \,, \label{pieom}
\end{align}
and corresponding equations hold for the neutral scalars. These results imply that the physical scalar fields are given by
\begin{equation}\label{phiW}
   \phi^\pm = W_5^\pm + M_W^{-1}\,\partial_r (h^{-1} \pi^\pm) \,, \qquad
   \phi^Z = Z_5 + M_Z^{-1} \,\partial_r (h^{-1} \pi_3) \,.
\end{equation}

A non-trivial cross check of these definitions can be made by comparing the equations of motion (\ref{fWZEOM}) of the $W$ and $Z$ bosons with those of the Goldstone bosons, which are obtained by adding $\partial_r(a^2b^{-1}(\ref{W5eom}))$ to $a^2b M_W h\,(\ref{pieom})$, and then using that $\partial_r(a^4b^{-1} h^2 \partial_r(h^{-1} \pi^\pm))-h\,\partial_r(a^4b^{-1}\partial_r\pi^\pm)=-\big(\partial_r(a^4b^{-1}\partial_r h)\big) \pi^\pm$. Employing then relation (\ref{hvevEqn}) to cancel the gauge-independent terms, we obtain
\begin{equation}\label{GpmEOM}
   b\,\Box_4 G^\pm - \xi\partial_r (a^2 b^{-1} \partial_r G^\pm)
    + \xi a^2 b M_W^2 h^2 G^\pm = 0 \,,
\end{equation} 
and a similar equation holds for $G^Z$. Comparing this equation with the equation of motion for the profiles of the gauge bosons in (\ref{fWZEOM}), where the mass term for the KK modes corresponds to minus the box operator in the equation above, we conclude that the Goldstone bosons and their KK excitations have the same profiles as the corresponding gauge bosons (apart from the normalization, which below we will choose slightly differently), and their masses are related by 
\begin{equation}\label{eq:massrel}
   m_n^{(G^\pm)\,2} = \xi m_n^{(W)\,2} \,, \qquad
   m_n^{(G^Z)\,2} = \xi m_n^{(Z)\,2} \,.
\end{equation}
This is a consequence of the gauge-Goldstone equivalence theorem. Likewise, the equations of motion for the charged physical scalars can be found by adding $M_W^{-1}\,\partial_r(h^{-1} (\ref{pieom}))$ to (\ref{W5eom}), and using similar relations to the ones employed above we obtain \cite{Falkowski:2008fz,Cabrer:2011fb,Archer:2012qa}
\begin{equation}\label{PhiEOM}
   \Box_4 \phi^\pm - \partial_r \left( a^{-2} b^{-1} h^{-2} 
    \partial_r (a^4 b^{-1} h^2 \phi^\pm) \right) + a^2 M_W^2 h^2 \phi^\pm = 0 \,,
\end{equation}
and a similar equation for $\phi^Z$. Note that the scalar potential does not enter in this result.
 
Before we can compute the Feynman rules for such scalars, we must invert the expressions in (\ref{GZdef}), (\ref{Gpmdef}) and (\ref{phiW}). Up to now we have worked with the 4D effective KK expansion rather the 5D position-momentum propagators often favoured in analogous extra-dimensional loop calculations \cite{Csaki:2010aj,Blanke:2012tv,Beneke:2012ie,Malm:2013jia,Hahn:2013nza}. One of the reasons for doing so is that the inversions are significantly simpler when using the KK expansion, since one can then use the equations of motion. In particular, we make the KK expansions 
\begin{align}
   \phi^{\pm,Z}(x,r) &= a^{-4} b\,M_{W,Z}^{-2}\,h^{-2}
    \sum_n m_n^{(\phi^{\pm,Z})\,2} f_n^{(\phi^{\pm,Z})}(r)\,\phi_{\pm,Z}^{(n)}(x) \,, \\
   G^{\pm,Z}(x,r) &= \sum_n m_n^{(W,Z)\,2} f_n^{(G^{\pm,Z})}(r)\,G_{\pm,Z}^{(n)}(x) \,,
\end{align}    
where $\Box_4\phi_{\pm,Z}^{(n)}=-m_n^{(\phi^{\pm,Z})\,2}\,\phi_{\pm,Z}^{(n)}$ and $\Box_4 G_{\pm, Z}^{(n)}=-\xi m_n^{(W,Z)\,2}\,G_{\pm, Z}^{(n)}$ from (\ref{eq:massrel}). The profiles $f_n^{(G^{\pm,Z})}$ of the Goldstone bosons obey the same equation as $f_n^{(W,Z)}$ in (\ref{fWZEOM}), while from (\ref{PhiEOM}) we find that the profiles of the physical scalar satisfy
\begin{equation}
	\partial_r \left( a^{-2} b^{-1} M_{W,Z}^{-2} h^{-2} 
	 \partial_r f_n^{(\phi^{\pm,Z})} \right)
	 - a^{-2} b f_n^{(\phi^{\pm,Z})} + a^{-4} b M_{W,Z}^{-2} h^{-2}
	 m_n^{(\phi^{\pm,Z})\,2} f_n^{(\phi^{\pm,Z})} =0 \,, \label{PhiWEOM}
\end{equation} 
along with the boundary conditions $f_n^{(\phi^{\pm, Z})}|_{r=r_{\mathrm{IR}},r_{\mathrm{UV}}}=0$. This allows us to invert (\ref{GZdef}), (\ref{Gpmdef}) and (\ref{phiW}) to get
\begin{align}
   \pi^\pm &=\sum_n \left[ - a^{-2} b^{-1} M_W^{-1} h^{-1}
    \big( \partial_r f_n^{(\phi^\pm)} \big)\,\phi_\pm^{(n)} 
    + M_W h f_n^{(G^\pm)}\,G_\pm^{(n)} \right] , \label{pipm}\\
   W_5^\pm &= \sum_n \left[ a^{-2} b f_n^{(\phi^\pm)}\,\phi_\pm^{(n)}
    - \big( \partial_r f_n^{(G^\pm)} \big)\,G_\pm^{(n)} \right] , \label{W5} \\
   \pi_3 &= \sum_n \left[ - a^{-2} b^{-1} M_Z^{-1} h^{-1}
    \big( \partial_r f_n^{(\phi^Z)} \big)\,\phi_Z^{(n)}
    + M_Z h f_n^{(G^Z)}\,G_Z^{(n)} \right] , \label{pi3} \\
   Z_5 &= \sum_n \left[ a^{-2} b f_n^{(\phi^Z)}\,\phi_Z^{(n)}
    - \big( \partial_r f_n^{(G^Z)} \big)\,G_Z^{(n)} \right] . \label{Z5}
\end{align}
The orthonormality conditions follow from requiring the fields to be canonically normalised. To be explicit, substituting (\ref{pipm})--(\ref{Z5}) into the kinetic terms
\begin{equation}
   S\supset\int d^5x \left( \frac12\,a^2 b\,\big| \partial_\mu\pi_3 \big|^2
    + a^2 b\,\big| \partial_\mu\pi^+ \big|^2
    + \frac12\,a^2 b^{-1} \big| \partial_\mu Z_5 \big|^2
    + a^2 b^{-1} \big| \partial_\mu W_5^+ \big|^2 \right)
\end{equation}
implies, after a partial integration, the orthogonality relations
\begin{align}
	\int dr\,a^{-4} b M_{W,Z}^{-2} h^{-2} m_n^{(\phi^{\pm,Z})\,2}
	 f_n^{(\phi^{\pm, Z})} f_m^{(\phi^{\pm, Z})} &= \delta_{nm} \,, \label{phiOrthog} \\
    \int dr\,b\,m_n^{(W,Z)\,2} f_n^{(G^{\pm, Z})} f_m^{(G^{\pm, Z})}
    &= \delta_{nm} \,. \label{Gorthog}
\end{align}
Comparison with (\ref{GaugeOrtho}) shows that
\begin{equation}\label{eq:243}
   f_n^{(W,Z)} = m_n^{(W,Z)} f_n^{(G^{\pm, Z})} \,,
\end{equation} 
and it would be possible to absorb the factor $m_n^{(W,Z)}$ into the normalisation of the profile functions. However, the resulting Feynman rules are slightly simplified by using the normalization in (\ref{Gorthog}). 

\subsection{Ghost Sector}

The final aspect of electroweak symmetry breaking concerns the definition of the ghost fields associated with the gauge-fixing terms. Focusing just on the electroweak gauge invariance, the 5D bulk Lagrangian (\ref{FullLag}) is invariant under the 5D gauge transformations \cite{Bohm:2001yx}
\begin{eqnarray}
	A_M &\to& A_M + \partial_M\delta\theta^A
	 - ig_5 s_w \left( W_M^+\delta\theta^- - W_M^-\delta\theta^+ \right) , \\
	Z_M &\to& Z_M + \partial_M\delta\theta^Z 
	- ig_5 c_w \left( W_M^+\delta\theta^- - W_M^-\delta\theta^+ \right) , \\
	W_M^\pm &\to& W_M^\pm + \partial_M\delta\theta^\pm \pm ig_5 \left[ 
	 W_M^\pm (s_w\delta\theta^A + c_w\delta\theta^Z) - (s_w A_M + c_w Z_M) \delta\theta^\pm \right] ,
	 \quad
\end{eqnarray}
and 
\begin{eqnarray}
   H &\to& H + \frac{g_5}{2c_w} \pi_3\delta\theta^Z
    + \frac{g_5}{2} \left( \pi^+\delta\theta^- + \pi^-\delta\theta^+ \right) , \\
   \pi_3 &\to& \pi_3 - \frac{g_5}{2c_w} (\tilde vh+H) \delta\theta^Z
	- \frac{ig_5}{2} \left( \pi^+\delta\theta^- - \pi^-\delta\theta^+ \right) , \\
   \pi^\pm &\to& \pi^\pm \pm ig_5 \pi^\pm \left( s_w\delta\theta^A
    - \frac{s_w^2-c_w^2}{2c_w}\delta\theta^Z \right)
    - \frac{g_5}{2} \left( \tilde vh + H\pm i\pi_3 \right) \delta\theta^\pm \,.
\end{eqnarray}
After introducing the gauge-fixing term (\ref{GFLag}), in order to complete the Lagrangian we must consider the variation of the gauge-fixing condition, i.e.\ $\delta F^a/\delta\theta^b$ with
\begin{equation}
	F^a = \frac{1}{\sqrt{\xi}} \left( \partial^\mu \mathbb{A}^a_\mu - \xi G^a \right) ; \qquad 
	a,b = +,-,Z,A.
\end{equation}
Of relevance to the calculation of the $H\to\gamma\gamma$ decay rate will be the charged ghosts obtained from (\ref{GFLag}) and (\ref{Gpmdef}), for which
\begin{eqnarray}\label{GpmVar}
   \frac{\delta F^\pm}{\delta\theta^\pm}
   &=& \frac{1}{\sqrt{\xi}}\,\bigg\{ \Box_4 
    \mp i g_5\partial_\mu (s_w A^\mu + c_w Z^\mu) \\
   &&\hspace{-2mm}\mbox{}- \xi b^{-1} \bigg[ \partial_r (a^2 b^{-1} \partial_r)
    - a^2 b M_W h \frac{g_5}{2} (\tilde vh+H\pm i\pi_3)
    \mp ig_5 \partial_r\!\left(\! a^2 b^{-1} (s_w A_5 + c_w Z_5) \!\right) \!\bigg] \bigg\} \,. \nonumber
\end{eqnarray}
This gives rise to the effective ghost action 
\begin{eqnarray}\label{GhostLag}
   S_{\rm ghost} &\supset& \int d^5x\,b\,\bar{u}_\pm \bigg\{ 
    - \Box_4 \pm ig_5 \partial_\mu (s_w A^\mu + c_w Z^\mu) \nonumber\\
   &&\mbox{}+\xi b^{-1}\bigg[ \partial_r (a^2 b^{-1} \partial_r)
    - a^2 b M_W^2 h^2 \left( 1 + \frac{H}{\tilde vh}
    \pm i\frac{\pi_3}{\tilde vh} \right) \nonumber\\
   &&\mbox{}\mp ig_5\partial_r \left( a^2 b^{-1} (s_w A_5 + c_w Z_5) \right) \bigg] 
    \bigg\}\,u_\pm \,.
\end{eqnarray}
Once again, we can make the KK decomposition $u_\pm(x,r)=\sum_n f_n^{(u^\pm)}(r)\,u _\pm^{(n)}(x)$, such that $\int dr\,b f_n^{(u^\pm)} f_m^{(u^\pm)}=\delta_{nm}$, and hence the ghost profiles and mass eigenvalues can be found from solving
\begin{equation}
   \partial_r \left( a^2 b^{-1} \partial_r f_n^{(u^\pm)} \right)
    - a^2 b M_W^2 h^2 f_n^{(u^\pm)} + b m_n^{(u^\pm)\,2} f_n^{(u^\pm)} = 0 \,,
\end{equation} 
with $\partial_r f_n^{(u^{\pm})}|_{r=r_{\mathrm{IR}},r_{\mathrm{UV}}}=0$. Comparison with (\ref{fWZEOM}) shows that, not surprisingly, the charged ghosts have the same profiles and masses as the $W$ bosons and their KK excitations, 
\begin{equation}
\label{eqn:ghostmass}
   m_n^{(u^\pm)} = m_n^{(W)} \,, \qquad
   f_n^{(u^\pm)} = f_n^{(W)} \,.
\end{equation}
This can be considered as a consistency check of our definition of the Goldstone fields.    

\subsection{Fermion Masses and Yukawa Couplings}
\label{sect:FermMasses}

Allowing for three generations, the fermion sector described by (\ref{FermLag}) is made more complicated due to the effects of flavour mixing. As usual we shall make the KK decomposition in the flavour basis, in which the bulk mass parameters $M_\Psi$ are flavour diagonal. In fact, when considering an AdS$_5$ geometry, we shall use the usual parametrisation $M_\Psi^{ij}=(c_\Psi^i/R)\,\delta^{ij}$, with $i,j$ running over the flavour indices,  and such that left-handed fermions are localised towards the UV (IR)  brane for $c_\Psi>\frac12$ ($c_\Psi<\frac12$), while right-handed fermions are localised towards the UV (IR) brane for $c_\Psi<-\frac12$ ($c_\Psi>-\frac12$). Even working in this basis, one still has 15 bulk mass parameters $c_\Psi^i$ and three complex $3\times 3$ Yukawa matrices $\mathbf{Y}_u^{5D}$, $\mathbf{Y}_d^{5D}$, $\mathbf{Y}_e^{5D}$ to fit to the masses and mixing angles. It is important to note that this basis is not equal to the physical mass basis. Upon electroweak symmetry breaking, the Yukawa couplings will give rise to mass terms, which will not generically be aligned with the bulk mass parameters. 

Unlike for the gauge fields, in the fermion sector we carry out the KK decomposition before electroweak symmetry breaking, i.e., we treat the Yukawa couplings of the fermions as a perturbation. To be explicit, we start by splitting the 5D vector-like fermions $\Psi=Q,L,u,d,e$ into the chiral components $\Psi=\Psi_L+\Psi_R$, such that $i\Gamma^5\Psi_{L,R}=b^{-1}\gamma_5 \Psi_{L,R}=\mp b^{-1} \Psi_{L,R}$. We then make the KK decomposition $\Psi_{L,R}(x,r)=\sum_n a^{-2} f_n^{(\Psi_{L,R})}\!(r)\,\psi_{L,R}^{(n)}(x)$, subject to the orthonormality conditions
\begin{equation}
   \int dr\,\frac{b}{a}\,f_n^{(\Psi_L)*} f_m^{(\Psi_L)}
   = \int dr\,\frac{b}{a}\,f_n^{(\Psi_R)*} f_m^{(\Psi_R)}
   = \delta_{nm} \,. 
\end{equation}
The fermion profiles can then be found by solving the coupled system of equations of motion \cite{Grossman:1999ra,Gherghetta:2000qt}\footnote{Note that we use a different sign convention for $\Gamma^5$ compared with \cite{Grossman:1999ra}.}
\begin{equation}\label{FermProfEom}
   \pm\partial_r f_n^{(\Psi_{L,R})} + b M_\Psi f_n^{(\Psi_{L,R})}
   = \frac{b}{a}\,m_n^{(\Psi)} f_n^{(\Psi_{R,L})} \,,
\end{equation}
where $\big(i\gamma^\mu\partial_\mu-m_n^{(\Psi)}\big) \big(\psi_L^{(n)}+\psi_R^{(n)}\big)=0$. A low-energy 4D effective chiral theory can be obtained by imposing boundary conditions such that only the left-handed doublets and right-handed singlets have zero modes. This is realised by the boundary conditions
\begin{equation}
   f_n^{(Q_R)}|_{r=r_{\mathrm{IR}},r_{\mathrm{UV}}} = 0 \,, \qquad
   f_n^{(u_L)}|_{r=r_{\mathrm{IR}},r_{\mathrm{UV}}} = 0 \,, \qquad
   f_n^{(d_L)}|_{r=r_{\mathrm{IR}},r_{\mathrm{UV}}} = 0 \,,
\end{equation}
and similarly for leptons. This is most naturally implemented by compactifying the space over a $S^1/\mathbb{Z}_2$ orbifold but could also be imposed on an interval. The boundary conditions for the remaining profile functions follow from the field equations (\ref{FermProfEom}) and are of mixed type.

We can now define the up-type quark Yukawa couplings  
\begin{equation}\label{YukawaDef}
   \tilde{Y}^{(n,m)}_{Q^i_{L,R} u^j_{R,L}}
   \equiv \left( Y_u^{5D}\right)_{ij} \int dr\,b\,f_0^{(H)} f_n^{(Q^i_{L,R})\ast} 
    f_m^{(u^j_{R,L})} \,, 
\end{equation}
and analogously for the Yukawa couplings of the down-type quarks and charged leptons. Similarly, we define the corresponding Yukawa mass terms
\begin{equation}\label{YukawaMassDef}
   Y^{(n,m)}_{Q^i_{L,R} u^j_{R,L}}
   \equiv \left( Y_u^{5D}\right)_{ij} \int dr\,b\,h\,f_n^{(Q^i_{L,R})\ast} f_m^{(u^j_{R,L})} \,.
\end{equation}
Note that these effective 4D Yukawa matrices are dimensionless, while the original 5D Yukawa couplings $\big(Y_u^{5D}\big)_{ij}$ have mass dimension $-1/2$. 

If we now take as an example the up-type quarks, the Yukawa couplings to the Higgs zero mode will be given by 
\begin{equation}\label{UpYukawa}
   \frac{1}{\sqrt{2}} 
    \left( \bar{U}_L^{(0)\,i},\, \bar{U}_L^{(1)\,i},\, \bar{u}_L^{(1)\,i}, \dots\right)
    \left(\renewcommand{\arraystretch}{1.5} \begin{array}{cccc} 
    \tilde{Y}^{(0,0)}_{{Q^i_{L}u^j_{R}}} & 0 & \tilde{Y}^{(0,1)}_{Q^i_{L}u^j_{R}} & \dots \\
    \tilde{Y}^{(1,0)}_{Q^i_{L}u^j_{R}} & 0 & \tilde{Y}^{(1,1)}_{Q^i_{L}u^j_{R}} & \\
    0 & \tilde{Y}^{(1,1)\ast}_{Q^j_{R}u^i_{L}} & 0 & \\
    \vdots & & & \ddots \end{array} \right)
    \left(\renewcommand{\arraystretch}{1.5}\begin{array}{c}
     u_R^{(0)\,j} \\ U_R^{(1)\,j} \\ u_R^{(1)\,j} \\ \vdots
     \end{array}\right)
    \equiv \bar{\mathbf U}_L\, \frac{\mathbf{\tilde Y}_u}{\sqrt 2}\,\mathbf{U}_R \,,
\end{equation}
and the mass matrix will be given by
\begin{equation}\label{UpMass}
   \left( \bar{U}_L^{(0)\,i},\, \bar{U}_L^{(1)\,i},\, \bar{u}_L^{(1)\,i}, \dots\right)
    \left(\renewcommand{\arraystretch}{1.5}\begin{array}{cccc} 
    \frac{\tilde v}{\sqrt{2}}{Y}^{(0,0)}_{{Q^i_{L}u^j_{R}}} & 0 & 
     \frac{\tilde v}{\sqrt{2}} {Y}^{(0,1)}_{Q^i_{L}u^j_{R}} & \dots \\ 
    \frac{\tilde v}{\sqrt{2}} {Y}^{(1,0)}_{Q^i_{L}u^j_{R}} & m_1^{(Q^i)}\delta_{ij} &
     \frac{\tilde v}{\sqrt{2}} {Y}^{(1,1)}_{Q^i_{L}u^j_{R}} & \\
    0 & \frac{\tilde v}{\sqrt{2}} Y^{(1,1)\ast}_{Q^j_{R}u^i_L} & 
     m_1^{(u^i)}\delta_{ij} & \\
    \vdots & & & \ddots \end{array}\right)
    \left(\renewcommand{\arraystretch}{1.5}\begin{array}{c} u_R^{(0)\,j} \\ U_R^{(1)\,j} \\
     u_R^{(1)\,j} \\ \vdots \end{array}\right)
    \equiv \bar{\mathbf U}_L\,\mathbf{M}_u\mathbf{U}_R \,,
\end{equation}
In the next step we diagonalise the mass matrix $\mathbf{M}_u$ by means of a bi-unitary transformation, such that 
\begin{equation}
   \mathbf{V}_u\,\mathbf{M}_u\mathbf{W}_u^\dag
   = \mbox{diag}\left( m_1^u, m_2^u, m_3^u, m_4^u, \dots, m_9^u, \dots \right) 
\end{equation}
is a diagonal matrix containing the physical masses of the up-type quarks and their KK excitations, $m_N^u$ with $N\ge 1$. They are the positive, real square roots of the eigenvalues of the squared mass matrices $\mathbf{M}_u\,\mathbf{M}_u^\dag$ (diagonalised by $\mathbf{V}_u$) and $\mathbf{M}_u^\dag\,\mathbf{M}_u$ (diagonalised by $\mathbf{W}_u$). We recall that for each SM fermion there are two KK excitations at each KK level \cite{Grossman:1999ra,Gherghetta:2000qt}. Our notation is such that the first three entries refer to the SM quarks, i.e.\ $m_1^u\equiv m_u$, $m_2^u\equiv m_c$, $m_3^u\equiv m_t$. The next six entries $\{ m_4^u, \dots, m_9^u \}$ refer to the states making up the first level of KK excitations, and so on. Note that in general the states in each KK level can have large flavour mixings \cite{Casagrande:2008hr}, so in the mass basis it is not meaningful to distinguish between the KK excitations of, e.g., the top, charm and up quarks. Rather, in each KK level there are six charge-$\frac23$ Dirac states of heavy quarks. For example, the masses $\{ m_4^u, \dots, m_9^u \}$ correspond to perturbations around the masses $\{ m_1^{Q_1}, m_1^{Q_2}, m_1^{Q_3}, m_1^{u_1}, m_1^{u_2}, m_1^{u_3} \}$ of first-level KK fermions in the basis of (\ref{UpMass}), which are induced by the mixings of the various fields induced by the Yukawa interactions. In our discussion below we will use a capital index $N$ to count KK states in the mass basis to indicate that this index is different from the KK level $n$. The state vectors $\mathbf{V}_u\mathbf{U}_L=(u_1^L, u_2^L, u_3^L, \dots)$ and $\mathbf{W}_u\mathbf{U}_R=(u_1^R, u_2^R, u_3^R, \dots)$ contain the left- and right-handed components of the 4D Dirac spinors corresponding to the physical mass eigenstates. Again, the first three entries refer to the up, charm and top quarks of the SM, while the remaining entries belong to the KK excitations. The ``physical'' Yukawa matrices in the mass basis are then given by 
\begin{equation}\label{eqn:YmassDef}
   \mathbf{\tilde Y}_u^{\mathrm{mass}} \equiv \mathbf{V}_u\mathbf{\tilde Y}_u\mathbf{W}_u^\dag \,.
\end{equation}
Analogous expressions exist for the down-type quarks and charged leptons. We shall return to these physical Yukawa couplings in section~\ref{sect:YukawaCouplings} in the context of Higgs physics in an AdS$_5$ warped extra-dimension model. For the moment it is productive to continue looking at other couplings and Higgs phenomenology while still working with a generic geometry.

\section{Higgs Production and Decay}
\label{sect:HiggsProd}

Having defined the initial KK decompositions, we can now move on to explore the implications on Higgs physics. An important point that we wish to emphasise here is the role that gauge invariance plays in making processes, in which the leading-order contribution is a loop amplitude such as  gluon fusion and $H\to \gamma\gamma$, perturbatively calculable. This is not a particularly original point, as it is well known that the Ward identities can reduce the superficial degree of divergence of a process, see for example \cite{Csaki:2010aj}. Nonetheless, we shall try to demonstrate from a more practical perspective how this is realised in extra-dimensional models.   

Before we can proceed with our study of Higgs production processes and decay rates, we must compute the Feynman rules of relevance to such processes and  specify the Higgs potentials introduced in (\ref{FullLag}). Here we shall consider the potentials 
\begin{equation}\label{HiggsPoten}
   V(\Phi)=M_\Phi^2|\Phi|^2 \,, \qquad 
   V_{\rm{IR}}(\Phi)=-M_{\rm{IR}}|\Phi|^2+\lambda_{\rm{IR}}|\Phi |^4 \,, \qquad 
   V_{\rm{UV}}(\Phi)=M_{\rm{UV}}|\Phi|^2.
\end{equation}   
The motivation for such a choice is related to the fact that we are primarily considering RS-type scenarios. In particular, if one requires that in the 5D theory all dimensionful parameters are at the Planck scale, then one would anticipate that a possible UV $|\Phi|^4$ term would have $\lambda_{\rm{UV}} \sim \mathcal{O}(M_{\rm{Pl}}^{-2})$ and hence can be neglected. In the case of $\lambda_{\rm{IR}}$, although being also $\sim \mathcal{O}(M_{\rm Pl}^{-2})$, the warping factor will enhance it such that it effectively behaves like $\sim \mathcal{O}(M_{\rm{KK}}^{-2})$. A bulk $|\Phi|^4$ term would be suppressed by an intermediate scale, although the inclusion of such a term would result in the Higgs VEV being the solution of a non-linear differential equation and hence result in a significantly more involved scenario.

The following calculations have been made in the 4D effective theory, valid at momenta lower than the warped-down UV cutoff $\Lambda_{\rm TeV}\sim 10 M_{\mathrm{KK}}$ \cite{Csaki:2008zd}. Hence, the derivation of the relevant 4D effective Feynman rules involves substituting the KK expansions obtained in the previous section into the original action and integrating out the 5th dimension. 

\subsection{Derivation of the Feynman Rules}
\label{sect:FeynRules}

Given the Lagrangian of the 5D warped extra-dimension model discussed in section~\ref{sec:EWbreak}, it is a straightforward exercise to derive the Feynman rules. Those of relevance to our discussion are collected in table~\ref{HGGfeynRules}. In the following we will comment on details of particular importance on the computation of the $H\to \gamma\gamma$ amplitude, which involves the interactions of the charged vectors and scalars with the photon and the Higgs boson. They originate from four terms: the Higgs kinetic term, the triple gauge couplings, the quartic gauge couplings, and the $|\Phi|^4$ term in the Higgs potential on the IR brane. Here we shall give a few key examples demonstrating the importance of gauge invariance. 

Let us consider, as a first example, the charged scalar--$W$--photon vertex originating from the Higgs kinetic term and the triple gauge vertex. One finds
\begin{align}
   \phi_\pm^{(l)}\,W_\mu^{\mp(m)} A_\nu^{(n)}\!\!: 
   &\quad \pm 2 g_5 s_w \eta_{\mu\nu} \int dr\,f_l^{(\phi^\pm)} f_m^{(W)} 
    \partial_r f_n^{(A)} \,, \label{Feyn:PhiWA}\\
   G_\pm^{(l)}\,W_\mu^{\mp(m)} A_\nu^{(n)}\!\!: 
   &\quad \pm g_5 s_w \eta_{\mu\nu} \int dr \left[
    m_l^{(W)\,2} b f_l^{(G^\pm)} f_m^{(W)} f_n^{(A)}
    - 2 a^2 b^{-1}\,\big( \partial_r f_l^{(G^\pm)} \big) f_m^{(W)} \partial_r f_n^{(A)} 
    \right]\label{Feyn:GWA} .
\end{align}
Hence the $l^{\rm th}$ KK Goldstone boson can couple with the $m^{\rm th}$ KK $W$ boson and the $n^{\rm th}$ KK photon even when $l\neq m\neq n$, i.e., KK number is not conserved. However, when we consider the massless photon zero mode, then the profile $f_0^{(A)}$ is constant with respect to $r$, and using (\ref{GaugeOrtho}) and (\ref{eq:243}) we can reduce the above rules to
\begin{align}
   \phi_\pm^{(l)}\,W_\mu^{\mp(m)} A_\nu^{(0)}\!\!: &\quad \,\, 0 \,,  \label{PhiWA0}\\
   G_\pm^{(l)}\,W_\mu^{\mp(m)} A_\nu^{(0)}\!\!: 
   &\quad \mp e \eta_{\mu\nu}\,\delta_{lm} m_m^{(W)} \,, \label{GWA0}
\end{align}    
where $e\equiv -g_5 s_w f_0^{(A)}$ is the 4D electromagnetic coupling constant. In other words, KK number is now conserved. Conservation of KK number in the 4D effective theory is completely equivalent to conservation of five momentum in the 5D theory. Since the 5D Lorentz symmetry is broken by the IR and UV branes, one would not generically expect KK number to be conserved. However, in all cases including both 3- and 4-particle vertices, we find that the vertices that conserve KK number are those between the photon or gluon zero mode (with an unbroken 4D gauge symmetry) and particles with the same profiles. The conservation of KK number has direct consequences on the convergence or divergence of a given diagram, since the delta functions act in the numerator of a given loop integral. Practically this amounts to determining the number of KK sums and physically this can be traced to the Ward identities reducing the superficial degree of divergence. This has been discussed in some detail in \cite{Csaki:2010aj}.  

A similar behaviour can be observed considering the coupling of charged scalars to KK photons. These interactions in general do not conserve KK number. However, in the case of the photon zero mode, the photon profile can be pulled outside the integrals and one finds that KK number is conserved,
\begin{align}
   \phi_-^{(l)}(p_{\phi_-})\,\phi_+^{(m)}(p_{\phi_+})\,A_\mu^{(0)}\!\!: 
   &\quad -ie\delta_{lm} \big(p_{\phi_+}-p_{\phi_-}\big)_\mu \,, \label{PhPhA0} \\
   \phi_\pm^{(l)}(p_{\phi_\pm})\,G_\mp^{(m)}(p_{G_\mp})\,A_\mu^{(0)}\!\!:
   &\quad \,\, 0 \,, \label{PhiGA} \\
   G_-^{(l)}(p_{G_-})\,G_+^{(m)}(p_{G_+})\,A_\mu^{(0)}\!\!:
   &\quad - ie\delta_{lm} \big(p_{G_+}-p_{G_-}\big)_\mu \,. \label{GGA0}
\end{align} 
On the other hand, it is worth noting from table~\ref{HGGfeynRules} that vertices involving the Higgs boson are typically more involved than others and do not conserve KK number.   

\begin{table}[t]
\begin{tabular}{|c|c|c|c|}
\hline
Vertex & Feynman rule & $g^{(n,m)}$ in bulk Higgs model & $g^{(0,0)}$ in SM \\
\hline
$H^{(0)} W_\mu^{(n)+} W_\nu^{(m)-}$ & $ig_{HWW}^{(n,m)}\,\eta_{\mu\nu}$
 & $g_5\int dr\,a^2 b M_W h f_0^{(H)} f_n^{(W)} f_m^{(W)}$ & $gm_W$ \\
$G^{(n)\pm} W_\mu^{(m)\mp} A^{(0)}_\nu$ & $\pm g_{GWA}^{(n,m)}\,\eta_{\mu\nu}$
 & $-e m_n^{(W)} \delta_{nm}$ & $-em_W$ \\
$\phi^{(n)\pm} W_\mu^{(m)\mp} A^{(0)}_\nu$ & $\pm g_{\phi WA}^{(n,m)}\,\eta_{\mu\nu}$
 & 0 & $-$ \\
\hline
$A_\mu^{(0)} G^{(n)+} G^{(m)-}$ & $-ig_{AGG}^{(n,m)}\,(p_{G^+}-p_{G^-}\!)_\mu$
 & $e\delta_{nm}$ & $e$ \\
$H^{(0)} G^{(n)\pm} W_\mu^{(m)\mp}$ & $g_{HGW}^{(n,m)}\,(p_{G^\pm}-p_H)_\mu$ 
 & $\frac{1}{2m_n^{(W)}}\,g_{HWW}^{(n,m)}$ & $\frac{g}{2}$ \\
$H^{(0)} \phi^{(n)\pm} W_\mu^{(m)\mp}$ & $g_{H\phi W}^{(n,m)}\,(p_{H}-p_{\phi^\pm}\!)_\mu$
 & $\int dr\,h^{-1} f_m^{(W)} f_0^{(H)} \partial_r f_n^{(\phi)}$ & $-$ \\
$A_\mu^{(0)} \phi^{(n)+} \phi^{(m)-}$ & $-ig_{A\phi\phi}^{(n,m)}\,
 (p_{\phi^+}-p_{\phi^-}\!)_\mu$ & $e\delta_{nm}$ & $-$ \\
$A_\mu^{(0)} \bar{u}^{(n)\pm} u^{(m)\pm}$ & $\pm ig_{Auu}^{(n,m)}\,(p_{\bar{u}^\pm}\!)_\mu$
 & $e\delta_{nm}$ & $e$ \\
\hline
$H^{(0)} G^{(n)+} G^{(m)-}$ & $-ig_{HGG}^{(n,m)}$
 & $\frac{m_H^2}{2m_n^{(W)} m_m^{(W)}}\,g_{HWW}^{(n,m)}$ & $\frac{gm_H^2}{2m_W}$ \\
$H^{(0)} \bar{u}^{(n)\pm} u^{(m)\pm}$ & $-ig_{Huu}^{(n,m)}\xi$
 & $\frac12 g_{HWW}^{(n,m)}$ & $\frac{gm_W}{2}$ \\
$H^{(0)} \phi^{(n)+} \phi^{(m)-}$ & $-ig_{H\phi\phi}^{(n,m)}$
 & $\frac{1}{\tilde v} \int dr\,\bigg[ 
    \frac{b}{a^2} \Big( m_m^{(\phi)\,2} + m_n^{(\phi)\,2} \Big) 
    \frac{f_{0}^{(H)}}{h} f_m^{(\phi)} f_n^{(\phi)}$ & $-$ \\
 & & $+ \frac{1}{b} \Big( \partial_r\frac{f_0^{(H)}}{h} \Big)
    \partial_r \Big( f_m^{(\phi)} f_n^{(\phi)} \Big) \bigg]$ & \\
 & & $+ \frac{2\lambda_{\rm{IR}} \tilde v}{M_W^2 b^2}\,\frac{f_0^{(H)}}{h}
    \left( \partial_r f_m^{(\phi)} \right) \left( \partial_r f_n^{(\phi)} \right) 
    \big|_{r=r_{\rm IR}}$ & \\ 
$H^{(0)} \phi^{(n)\pm} G^{(m)\mp}$ & $ig_{H\phi G}^{(n,m)}$
 & $\frac{g_5}{2} \int dr\,\bigg[ a^2 M_W^2 h^2 f_n^{(\phi)} 
 \partial_r \bigg( \frac{f_m^{(G)} f_0^{(H)}}{M_W h} \bigg)$ & $-$ \\
 & & $- \frac{a^4}{b} f_0^{(H)\,2} \left( \partial_r f_m^{(G)} \right)
 \partial_r \bigg( \frac{\partial_r f_n^{(\phi)}}{a^2 b M_W h f_0^{(H)}} \bigg) \bigg]$ & \\
 & & $+\frac{2\lambda_{\rm{IR}}a^2}{b}\,\tilde v h f_0^{(H)} 
 \left( \partial_r f_n^{(\phi)} \right) f_m^{(G)} \big|_{r=r_{\rm IR}}$ & \\
\hline
$H^{(0)} G^{(n)\pm} W_\mu^{(m)\mp} A_\nu^{(0)}$ & $\pm g_{HGWA}^{(n,m)}\,\eta_{\mu\nu}$
 & $-\frac{e}{2m_n^{(W)}}\,g_{HWW}^{(n,m)}$ & $-\frac{eg}{2}$ \\
$G^{(n)+} G^{(m)-} A^{(0)}_\mu A^{(0)}_\nu$ & $ig_{GGAA}^{(n,m)}\,\eta_{\mu\nu}$
 & $2e^2\delta_{nm}$ & $2e^2$ \\
$H^{(0)} \phi^{(n)\pm} W_\mu^{(m)\mp} A_\nu^{(0)}$ & $\pm g_{H\phi WA}^{(n,m)}\,\eta_{\mu\nu}$
 & $\frac{eg_5}{2M_W}\,g_{H\phi W}^{(n,m)}$ & $-$ \\
$\phi^{(n)+} \phi^{(m)-} A^{(0)}_\mu A^{(0)}_\nu$ & $ig_{\phi\phi AA}^{(n,m)}\,\eta_{\mu\nu}$
 & $2e^2\delta_{nm}$ & $-$ \\
\hline
$W_\nu^{(n)+} A_\mu^{(0)} W_\lambda^{(m)-}$ & $ig_{AWW}^{(n,m)}\,S_{\nu\mu\lambda}^{W^+\!AW^-}$
 & $e\delta_{nm}$ & $e$ \\
$W_\lambda^{(n)+} W_\rho^{(m)-} A_\mu^{(0)} A_\nu^{(0)}$ & $-ig_{WWAA}^{(n,m)}\, 
 S_{\mu\nu,\lambda\rho}$ & $e^2\delta_{nm}$ & $e^2$ \\
\hline
$H^{(0)}\,\bar q_N^L\,q_M^R$ & $\frac{-i}{\sqrt2} \left( \tilde Y_q^{\rm mass} \right)_{NM}$ & see (\ref{eqn:YmassDef}) & \\
$A_\mu^{(0)} \bar{\psi}_N \psi_M$ & $ig_{A\bar{\psi}\psi}^{(N,M)} \gamma_\mu$
 & $-eQ_\psi\delta_{NM}$ & $-eQ_\psi$ \\
$\mathcal{G}_\mu^{(0)} \bar{q}_N q_M$ & $ig_{\mathcal{G}\bar{\psi}\psi}^{(N,M)} 
 \gamma_\mu\frac{\lambda^b}{2}$ & $g_s\delta_{NM}$ & $g_s$ \\
\hline
\end{tabular}
\centering 
\caption{Relevant Feynman rules needed for the calculations in this paper. We have defined $e\equiv -g_5 s_w f_0^{(A)}$ and $g_s=g_{s5} f_0^{(\mathcal{G})}$, as well as $S_{\mu\nu,\lambda\rho}\equiv 2\eta_{\mu\nu}\eta_{\lambda\rho}-\eta_{\mu\lambda}\eta_{\nu\rho}-\eta_{\mu\rho}\eta_{\nu\lambda}$ and $S_{\nu\mu\lambda}^{W^+\!AW^-}\equiv(p_A-p_{W^+})_\lambda\eta_{\mu\nu}+(p_{W^+}-p_{W^-})_\mu\eta_{\nu\lambda}+(p_{W^-}-p_{A})_\nu\eta_{\lambda\mu}$, where all momenta are assumed to be flowing into the vertex. In the last three Feynman rules $q=u,d$ represents a generic quark field in the mass basis, while $\psi_N$ is a generic fermion field.}
\label{HGGfeynRules}
\end{table}

\subsection{\boldmath $H\to WW^*$ and $H\to ZZ^*$ Decay Rates}
\label{sect:HtoWW}

Having found the relevant Feynman rules, we can now proceed and compute Higgs processes of interest to us. We start by considering two simple processes, in which the leading-order contribution occurs at tree level, namely $H\to WW^*$ and $H\to ZZ^*$, where the second gauge boson is produced off shell. A detailed analysis of these decay processes in the context of RS models is presented in \cite{Malm:2014gha}, where it is shown that the decay rate can be written in the form
\begin{equation}\label{HWW_rate}
   \Gamma_{H\to W W^*} 
   = \frac{m_H^3}{16\pi v^2}\,\frac{\kappa_{\Gamma_W}\Gamma_W}{\pi m_W} 
    \left\{ \bigg( \frac{v g_{HWW}^{(0,0)}}{2m_W^2} \bigg)^2\,
    g\bigg( \frac{m_W^2}{m_H^2} \bigg) + \mbox{KK contributions} \right\} , 
\end{equation}
where $m_W=m_0^{(W)}$ and $\Gamma_W$ are the physical mass and total decay width of the $W$ boson, $m_H=m_0^{(H)}\approx 125.5$\,GeV is the mass of the Higgs boson, and the quantity $\kappa_{\Gamma_W}\approx 1$ accounts for a small correction to the total decay width of the $W$ boson. An analogous formula, with $m_W$, $\Gamma_W$ and $g_{HWW}^{(0,0)}$ replaced by $m_Z$, $\Gamma_Z$ and $g_{HZZ}^{(0,0)}$, and with an overall symmetry factor of 1/2, holds for the decay $H\to ZZ^*$. Explicit expressions for $\kappa_{\Gamma_{W,Z}}$ and the KK-tower contributions can be found in \cite{Malm:2014gha} and are included in our numerical analysis. Numerically, it is a good approximation to only consider the first term inside the parenthesis in (\ref{HWW_rate}), which accounts for the contribution of the $W$-boson zero mode. The relevant phase-space function reads \cite{Keung:1984hn}
\begin{equation}
\begin{aligned}
   g(x) &= \frac{6x(1-8x+20x^2)}{\sqrt{4x-1}}
    \arccos\bigg( \frac{3x-1}{2x^{3/2}} \bigg) \\
   &\quad\mbox{}- 3x (1-6x+4x^2) \ln x - (1-x)(2-13x+47x^2) \,.
\end{aligned}
\end{equation}
Numerically, one finds $g(m_W^2/m_H^2)\approx 0.157$ and $g(m_Z^2/m_H^2)\approx 0.033$. To excellent approximation we now obtain from (\ref{HWW_rate}) the simple expressions
\begin{equation}    
\label{HtoWWRw}
   \frac{\Gamma_{H\to WW^*}}{\Gamma_{H\to WW^*}^{\rm (SM)}} 
    \approx \left( \frac{v g_{HWW}^{(0,0)}}{2m_W^2} \right)^2 , \qquad   
   \frac{\Gamma_{H\to ZZ^*}}{\Gamma_{H\to ZZ^*}^{\rm (SM)}} 
    \approx \left( \frac{v g_{HZZ}^{(0,0)}}{2m_Z^2} \right)^2 .
\end{equation}   

In table~\ref{HGGfeynRules}, the relevant coupling $g_{HWW}^{(0,0)}$ is expressed in terms of an integral over a product of profile functions for the Higgs VEV, the Higgs boson, and the $W$ boson. The overlap integral can be simplified by treating the ratio $\tilde v^2/M_{\rm KK}^2$ as a small parameter (recall that $\tilde v\approx v=246.2$\,GeV). To this end, we split up the Higgs-boson profile as $f_0^{(H)}=h+\delta f_0^{(H)}$, where the deviation $\delta f_0^{(H)}$ from the VEV profile is an effect of order $\tilde v^2/M_{\rm KK}^2$, see relation (\ref{VevProfApp}) in section~\ref{sect:HiggsVEV}. Inserting this ansatz into the overlap integral and using the equation of motion (\ref{fWZEOM}) for the gauge-boson profile, we obtain after an integration by parts
\begin{equation}
   g_{HWW}^{(0,0)} 
   = \frac{g_5}{M_W} \left[ m_W^2\!\int dr\,b f_0^{(W)\,2}
    - \int dr\,a^2 b^{-1} (\partial_r f_0^{(W)})^2
    + M_W^2\!\int dr\,a^2 b\,h\,\delta f_0^{(H)} f_0^{(W)\,2} \right] .
\end{equation}
The first integral on the right-hand side equals~1 due to the normalization condition (\ref{GaugeOrtho}). The second term is explicitly negative. Next, up to higher-order corrections we can set $f_0^{(W)}=\mbox{const.}$ in the last term, which is then proportional to the integral $\int dr\,a^2 b\,h\,\delta f_0^{(H)}$, which vanishes up to higher-order terms. This follows from the normalization conditions $\int dr\,a^2 b\,h^2=1$ in (\ref{eq42}) and $\int dr\,a^2 b\,f_0^{(H)\,2}=\int dr\,a^2 b\,(h^2+2h\,\delta f_0^{(H)}+\dots)=1$ in (\ref{HiggsOrthog}). Using finally that $M_W=g_5\tilde v/2$, we obtain the relation
\begin{equation}
\label{HtoWWPertApprox}
   \frac{v g_{HWW}^{(0,0)}}{2m_W^2} = \frac{v}{\tilde v} \left[ 1
    - \frac{1}{m_W^2} \int dr\,a^2 b^{-1} (\partial_r f_0^{(W)})^2
    + {\cal O}\bigg(\frac{\tilde v^4}{M_{\rm KK}^4} \bigg) \right] ,
\end{equation}
which is valid for any choice of metric. An analogous expression holds for $g_{HZZ}^{(0,0)}$. Notice that the quantity inside the bracket is necessarily smaller than~1. In RS models based on AdS$_5$ geometry, the modifications of the Higgs VEV are such that the ratio $v/\tilde v$ is also smaller than~1, and hence the $H\to WW^*$ and $H\to ZZ^*$ decay rates are reduced compared with their SM values. Further still, analogous studies of bulk Higgs scenarios have found that, when one enhances the gauge symmetry to an $\mathrm{SU}_L(2)\times\mathrm{SU}_R(2)$ custodial symmetry, this suppression is typically significantly enhanced \cite{Cacciapaglia:2006mz} (see also \cite{Hahn:2013nza}). This is an important result, which implies that any possible enhancement in the $H\to WW^*, ZZ^*$ signal strengths should be a consequence either of an enhanced Higgs production rate or a reduced Higgs total width. This finding also has implications for the $H\to\gamma\gamma$ decay rate, which in the SM receives its dominant contribution from vector-boson loops. These effects are  proportional to the $HWW$ coupling and are reduced in extra-dimensional models. As we shall see, in bulk-Higgs models new physics effects in the fermion sector also act to decrease the decay amplitude, so that a reduction of the $H\to\gamma\gamma$ decay rate is a generic feature of these scenarios.

\subsection{Higgs Production via Gluon Fusion}
\label{sect:GGFusion}

\begin{figure}
\begin{center}
\includegraphics[width=0.28\textwidth]{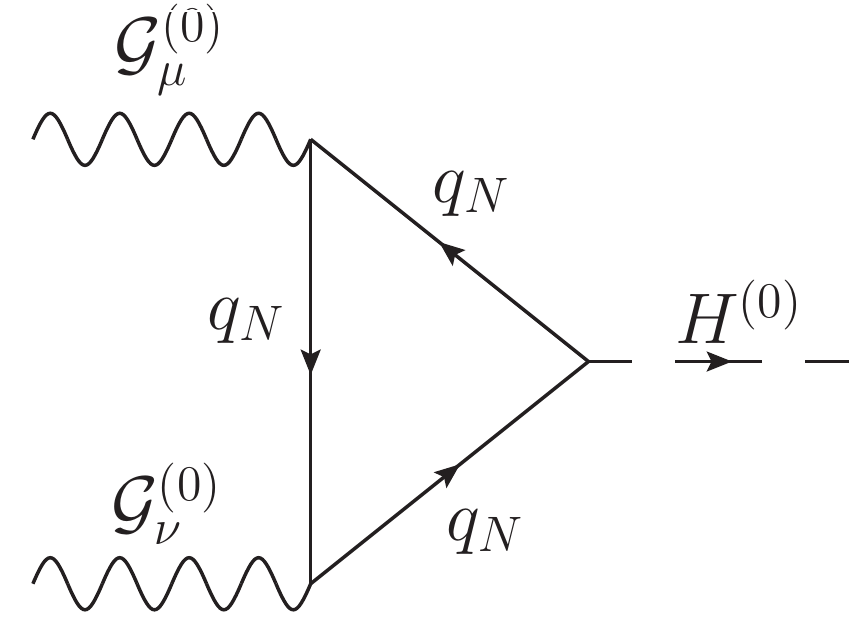}
\caption{\label{fig:GluonFusion} 
Feynman diagram contributing to Higgs production via gluon fusion.}
\end{center}
\end{figure}

It is well known that the dominant method of Higgs production at the LHC is via gluon fusion, pictured in figure~\ref{fig:GluonFusion}. We parametrise the corresponding scattering amplitude in the form 
\begin{equation}
   - {\cal A}_{gg\to H} 
   = C_q\,\frac{\alpha_s}{16\pi\tilde v}\,
    \langle\,0\,|G_{\mu\nu}^a G^{\mu\nu,a}|gg\rangle
    + C_q'\,\frac{\alpha_s}{16\pi\tilde v}\,
    \langle\,0\,|G_{\mu\nu}^a \widetilde G^{\mu\nu,a}|gg\rangle \,,
\end{equation}
where $\widetilde G_{\mu\nu}^a$ is the dual field-strength tensor, and $\tilde v$ denotes the Higgs VEV in the RS model, which differs from the parameter $v$ of the SM by  a small amount (see section~\ref{sec:EWpars}). Evaluating the diagram in figure~\ref{fig:GluonFusion}, we find
\begin{align}
   C_q &= \sum_{q=u,d} \sum_N \frac{\tilde v}{\sqrt2}\,
    \frac{\mbox{Re}\,(\tilde Y_q^{\rm mass})_{NN}}{m_N^q}\,
    \bigg(\! - \frac43 \bigg)\,A_{\frac12}(\tau_N^q) \,, \\
   C_q' &= \sum_{q=u,d} \sum_N \frac{\tilde v}{\sqrt2}\,
    \frac{\mbox{Im}\,(\tilde Y_q^{\rm mass})_{NN}}{m_N^q}\,\,2 B_{\frac12}(\tau_N^q) \,,
\end{align}
where $m_N^q$ are the mass eigenvalues of the physical up- and down-type quarks and their KK excitations, $\tau_N^q=4(m_N^q)^2/m_H^2$, and the loop functions are \cite{Gunion:1989we}
\begin{equation}
   A_{\frac12}(\tau) = \frac{3\tau}{2} \left[ 1 + (1-\tau)\,f(\tau) \right] , \qquad
   B_{\frac12}(\tau) = \tau\,f(\tau) \,,
\end{equation}
with
\begin{equation}\label{ftau}
   f(\tau) = \begin{cases}
    & \left( \arcsin\frac{1}{\sqrt\tau} \right)^2 ; \quad \text{for } \tau\ge 1 \,, \\[3mm]
    &-\frac14 \left( \ln\left (\frac{1+\sqrt{1-\tau}}{1-\sqrt{1-\tau}}\right )
     - i\pi \right)^2 ; \quad \text{for } \tau<1 \,.
    \end{cases}
\end{equation}
Recall that we label the physical quark mass eigenstates by $N=1,2,3$ for the SM quark (the zero modes), $N=4,\dots,9$ for the six states filling the first KK level, etc. Note also that the gluon vertices conserve KK number, so there is only a single sum over fermion states. Using the fact that $A_{\frac12}(\tau)\to 1$ and $B_{\frac12}(\tau)\to 1$ as $\tau\to\infty$, while both functions are of ${\cal O}(\tau)$ for $\tau\ll 1$, one finds that the light first- and second-generation SM quarks make negligible contributions to the sum, while the heavy KK quarks contribute with an approximately universal form factor. Hence we can approximate
\begin{align}
	\sum_{q=u,d}\,\frac{\tilde v}{\sqrt2}\,\sum_N
    \frac{\mbox{Re}\,(\tilde Y_q^{\rm mass})_{NN}}{m_N^q}\,A_{\frac12}(\tau_N^q) 
	&\approx \mbox{Re}(\kappa_{t,b})\,A_{\frac12}(\tau_{t,b}) 
	+ \sum_{q=u,d} \mbox{Re}(\kappa_{\rm KK}^q) \,, \\
	\sum_{q=u,d}\,\frac{\tilde v}{\sqrt2}\,\sum_N
    \frac{\mbox{Im}\,(\tilde Y_q^{\rm mass})_{NN}}{m_N^q}\,B_{\frac12}(\tau_N^q) 
	&\approx \mbox{Im}(\kappa_{t,b})\,B_{\frac12}(\tau_{t,b})
	+ \sum_{q=u,d} \mbox{Im}(\kappa_{\rm KK}^q) \,,
\end{align}
where $\kappa_t\equiv\kappa_3^u=(\tilde v/\sqrt 2)\,(\tilde Y_u^{\rm mass})_{33}/m_t$ and $\tau_t=4m_t^2/m_H^2$, and similarly for $\kappa_b$ and $\tau_b$. The gluon-fusion cross section, relative to its value in the SM, is then given by
\begin{equation}
   \frac{\sigma_{gg\to H}}{\sigma_{gg\to H}^{\rm (SM)}}
   = \frac{|\kappa_g|^2+|\kappa_{g5}|^2}{\kappa_v^2} \,,
\end{equation}
where 
\begin{equation}
   \kappa_g = - \frac34\,\frac{C_q}{A_{\frac12}(\tau_t)+A_{\frac12}(\tau_b)} \,, \qquad
   \kappa_{g5} = \frac34\,\frac{C_q'}{A_{\frac12}(\tau_t)+A_{\frac12}(\tau_b)} \,, 
\end{equation}
and $\kappa_v=\tilde v/v$ accounts for the corrections to the Higgs VEV arising in the RS model.

The infinite sum
\begin{equation}
	\kappa_{\rm KK}^q = \sum_{N\ge 4} \frac{\tilde v}{\sqrt2}\,
    \frac{(\tilde Y_q^{\rm mass})_{NN}}{m_N^q}
\end{equation}
contains the contributions from virtual KK quarks propagating in the fermion loop in figure~\ref{fig:GluonFusion}. In order to simplify the result, we add and subtract zero-mode terms in the following way:
\begin{equation}\label{KKferms}
	\kappa_{\rm KK}^q
   = \sum_{N\ge 1} \frac{\tilde v}{\sqrt2}\,\frac{(\tilde Y_q^{\rm mass})_{NN}}{m_N^q}
    - \sum_{N=1,2,3} \frac{\tilde v}{\sqrt2}\,\frac{(\tilde Y_q^{\rm mass})_{NN}}{m_N^q}
   = \frac{\tilde v}{\sqrt2}\,\mbox{Tr} \left( \mathbf{\tilde Y}_q\,\mathbf{M}_q^{-1} \right)
    - \sum_{N=1,2,3} \kappa_N^q \,,
\end{equation}
where $\kappa_N^q=(\tilde v/\sqrt2)\,(\tilde Y_q^{\rm mass})_{NN}/m_N^q$, and we have exploited the fact that the trace of the product $\mathbf{\tilde Y}_q\,\mathbf{M}_q^{-1}$ of the Yukawa and mass matrices defined in (\ref{UpYukawa}) and (\ref{UpMass}) is basis independent. In the limit where $m_H\ll M_{\mathrm{KK}}$, in which $h(r)\approx f_0^{(H)}(r)$ and hence $Y_{Q_{L,R}^i u_{R,L}^j}^{(n,m)}\approx\tilde Y_{Q_{L,R}^i u_{R,L}^j}^{(n,m)}$ up to corrections of $\mathcal{O}(m_H^2/M_{\mathrm{KK}}^2)$, one can simplify the result further by noting that $\frac{1}{\sqrt2}\,\mathbf{\tilde Y}_q\approx\partial\mathbf{M}_q/\partial\tilde v$. One thus obtains the relation \cite{Ellis:1975ap,Falkowski:2007hz,Azatov:2010pf,Carena:2012xa} 
\begin{equation}\label{TrMrelation}
   \frac{\tilde v}{\sqrt2}\,\mbox{Tr} \left( \mathbf{\tilde Y}_q\,\mathbf{M}_q^{-1} \right)
   = \tilde v\,\frac{\partial}{\partial\tilde v} \ln\det\mathbf{M}_q
    + {\cal O}\bigg( \frac{m_H^2}{M_{\mathrm{KK}}^2} \bigg) \,.
\end{equation}
Such an approximation, which is equivalent to assuming that the Higgs VEV is carried by just the Higgs zero mode, can be safely made when considering the contribution from the KK fermions in (\ref{KKferms}), since it turns out that in the difference of the two terms one is neglecting an $\mathcal{O}(\tilde v^2 m_H^2/M_{\mathrm{KK}}^4)$ correction to the overall result, which is indeed very small.\footnote{On the contrary, when evaluating the corrections $\kappa_{t,b}$ for the third-generation SM quarks, it is important not to neglect the misalignment between the couplings $Y_{Q_{L,R}^i u_{R,L}^j}^{(n,m)}$ and $\tilde Y_{Q_{L,R}^i u_{R,L}^j}^{(n,m)}$.} 
Working at first non-trivial order in the ratio $\tilde v^2/M_{\mathrm{KK}}^2$, it is possible to derive and approximate expression for $\kappa_{\rm KK}^q$ in terms of infinite sums involving the masses $m_n^{(\Psi)}$ from (\ref{FermProfEom}) and the corresponding $3\times 3$ Yukawa matrices  defined in (\ref{YukawaMassDef}). One obtains (with $q=u,d$)~\cite{Azatov:2010pf}
\begin{equation}
   \kappa_{\rm KK}^q
   = \frac{\tilde v^2}{2} \sum_{i,j} \! \sum_{n=1}^\infty \Bigg[
    \frac{Y_{Q_L^i q_R^j}^{(n,0)\ast}\,Y_{Q_L^i q_R^j}^{(n,0)}}{m_n^{(Q^i)2}} 
    + \frac{Y_{Q_L^i q_R^j}^{(0,n)}\,Y_{Q_L^i q_R^j}^{(0,n)\ast}}{m_n^{(q^j)2}} \Bigg]
    - \tilde v^2 \sum_{i,j} \sum_{n,m=1}^\infty\!\!
    \frac{Y_{Q_L^i q_R^j}^{(n,m)}\,Y_{Q_R^i q_L^j}^{(n,m)\ast}}{m_n^{(Q^i)} m_m^{(q^j)}}  
    + \mathcal{O}\bigg( \frac{\tilde v^4}{M_{\mathrm{KK}}^4} \bigg) \,. 
\end{equation}  
One can then go further and use completeness relations \cite{Hirn:2007bb} to evaluate the infinite sums in closed form \cite{Azatov:2010pf,Frank:2013un}. This is described in more detail in appendix~\ref{sec:Completness}, where we extend the work by these authors to the general case of three generations.

Previous work on the contributions of fermionic KK modes to the $gg\to H$ and $H\to\gamma\gamma$ amplitudes in RS models with a brane-localised scalar sector have shown that there is an ${\cal O}(1)$ sensitivity to the precise way in which the Higgs is localised on or near the IR brane \cite{Casagrande:2010si,Azatov:2010pf,Goertz:2011hj,Carena:2012fk,Frank:2013un,Malm:2013jia}. The results depend on whether the effective UV cutoff $\Lambda_{\rm TeV}$ near the IR brane is larger or smaller than the inverse width $\Delta_H\sim\beta\tilde v$ of the Higgs profile \cite{Malm:2013jia,Hahn:2013nza}. For $\Lambda_{\rm TeV}\gg\Delta_H$ high-momentum virtual particles can resolve the ``bulky nature'' of the Higgs profile, and this gives rise to an unsuppressed contribution. In \cite{Malm:2013jia}, these models are therefore referred to as ``narrow bulk-Higgs scenarios''. For $\Lambda_{\rm TeV}\ll\Delta_H$, on the other hand, the Higgs looks like a field that is strictly localised on the IR orbifold fixed point, and it is thus referred to as a ``brane Higgs''. 

\begin{figure}
\begin{center}
	\includegraphics[width=0.475\textwidth]{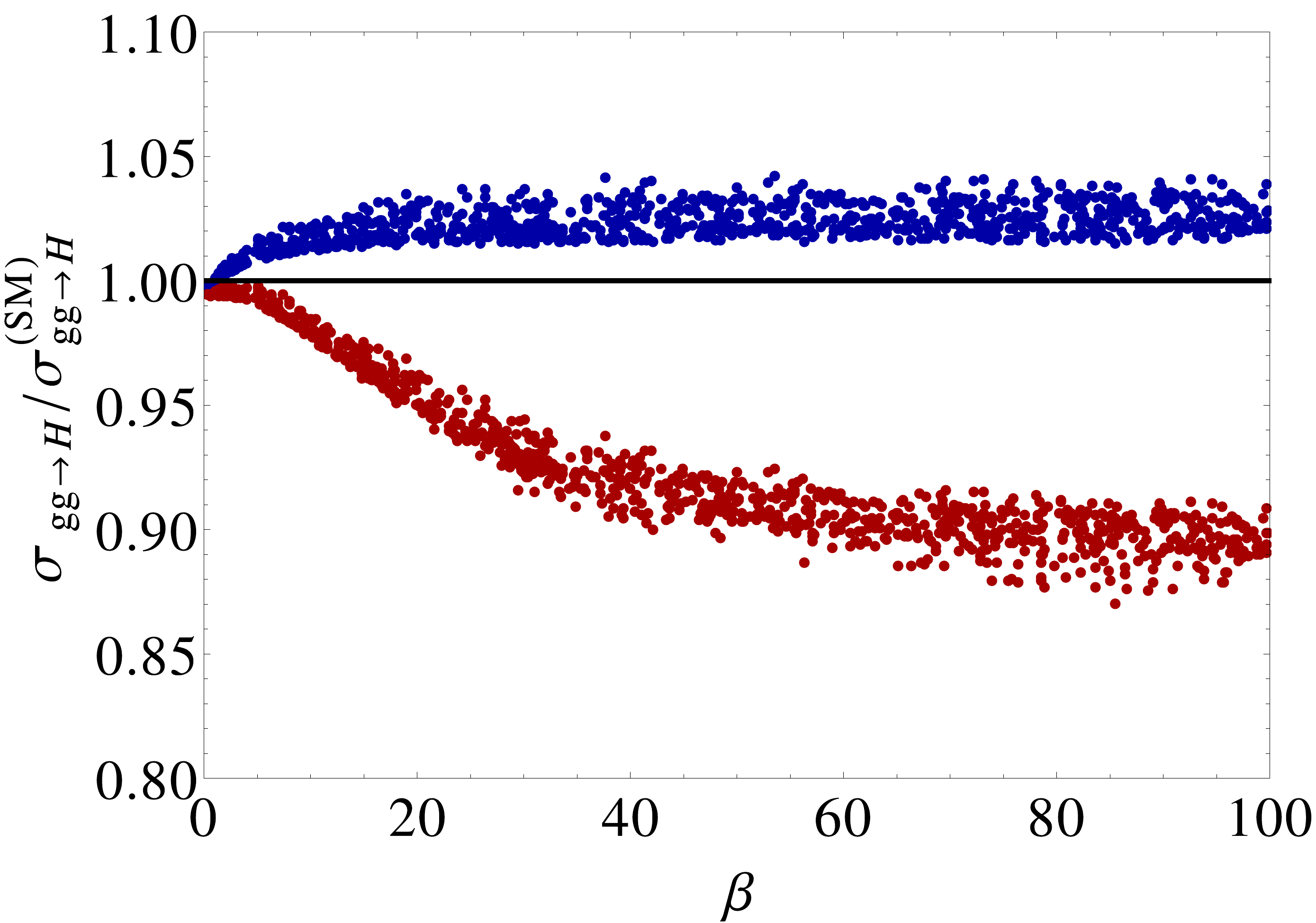}
	\includegraphics[width=0.515\textwidth]{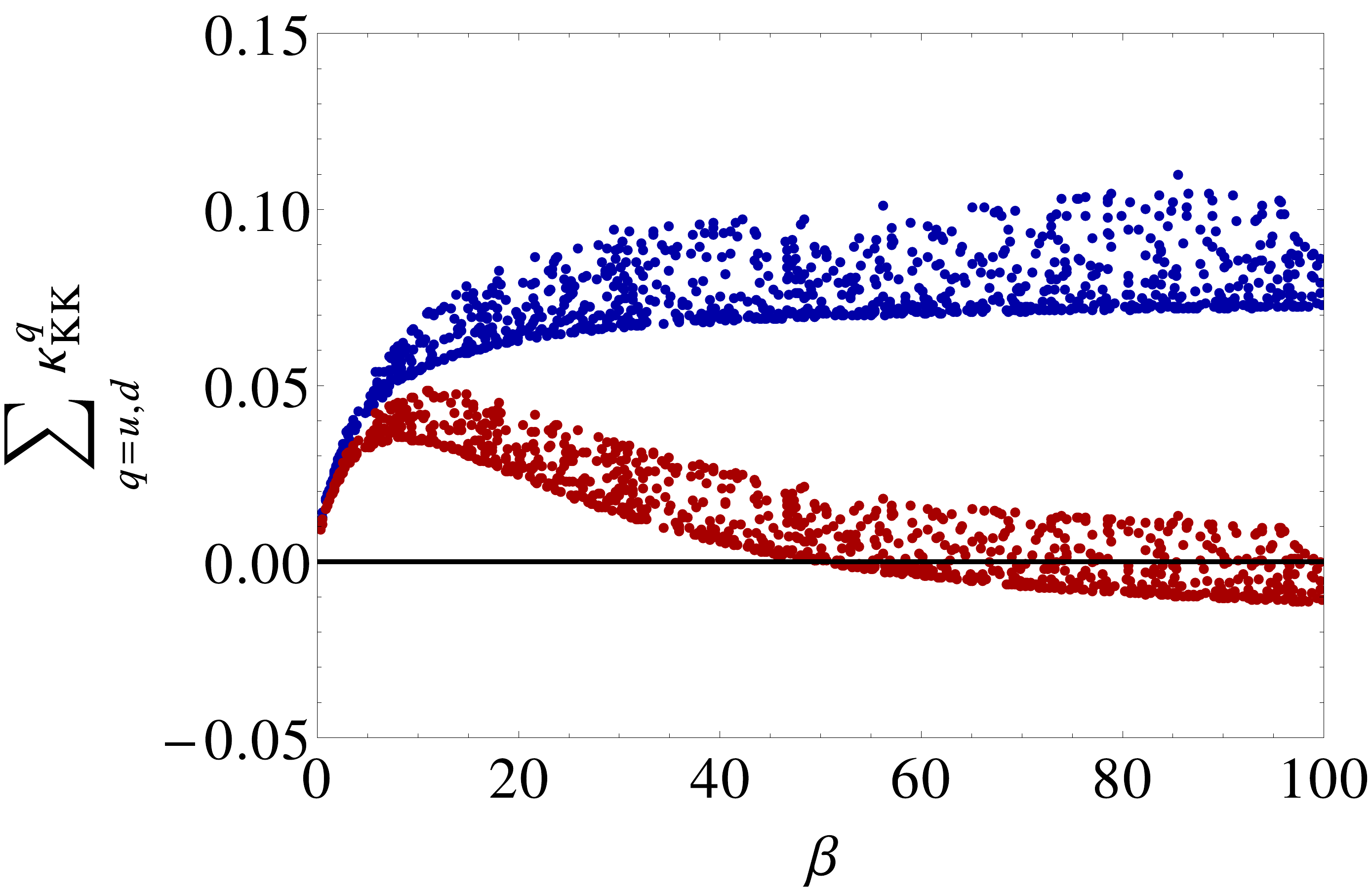}
\caption{\label{fig:conv}
Gluon-fusion cross section $\sigma_{gg\to H}/\sigma_{gg\to H}^{(\rm SM)}$ (left) and KK contribution $\sum_{q=u,d}\kappa_{\rm KK}^q$ (right) as a function of $\beta$ for the AdS$_5$ case with one fermion generation. In blue we show the effect of including all KK resonances, while red points correspond to summing over just the first three levels of physical KK modes. We use $c_Q^3\in[0,0.47], M_{\rm KK}=1.5$ TeV and $Y_u^{5D}=Y_d^{5D}=\sqrt{R(1+\beta)}$.}
\end{center}
\end{figure}

If we naively take the limit $\beta\to\infty$ in our analysis of the gluon fusion amplitude (and likewise for the fermionic contribution to the $H\to\gamma\gamma$ decay amplitude to be discussed in the following section), then our results converge toward the ``narrow bulk-Higgs scenario'' studied in \cite{Azatov:2010pf,Frank:2013un,Malm:2013jia,Hahn:2013nza}. The numerical results in section~\ref{sect:phenom} indicate that the asymptotic regime is reached for values $\beta\gtrsim 10$. If on the other hand we implement a fixed UV cutoff, e.g.\ by only summing over a finite number of KK levels, then for $\beta\to\infty$ our results converge toward the ``brane Higgs scenario'' studied in \cite{Casagrande:2010si,Goertz:2011hj,Carena:2012fk,Malm:2013jia,Hahn:2013nza}. This can be seen from figure~\ref{fig:conv}, where we show the gluon-fusion cross section $\sigma_{gg\to H}/\sigma_{gg\to H}^{(\rm SM)}$ and the KK contribution $\sum_{q=u,d}\kappa_{\rm KK}^q$ to the $gg\to H$ amplitude from (\ref{KKferms}) as a function of $\beta$, for the case of an AdS$_5$ geometry and one fermion generation (for simplicity). We use fixed values $M_{\rm KK}=1.5$\,TeV and $Y_u^{5D}=Y_d^{5D}=\sqrt{R(1+\beta)}$ and vary the bulk mass parameter $c_Q^3\in[0,0.47]$. The other two relevant parameters $c_u^3$ and $c_d^3$ are then fixed by the requirement that we reproduce the correct values for the masses of the top and bottom quarks. The blue scatter points are obtained when one sums over the infinite tower of KK resonances, while the red points refer to the case where the sum is truncated after the first three levels of physical KK states. Using the analytic formulas valid for $\beta\gg 1$ derived in \cite{Carena:2012fk,Malm:2013jia}, one finds $\sum_{q=u,d}\kappa_{\rm KK}^q\in[0.075,0.11]$ (narrow bulk Higgs) and $[-0.03,0.01]$ (brane Higgs) for the two cases, in excellent agreement with our numerical results. 

In the remainder of this paper we will refrain from taking the limit of very large $\beta$ values, since our main interest is in the phenomenology of RS models featuring a bulk Higgs. It is then most natural to consider values $\beta={\cal O}(1)$.

\subsection{\boldmath $H\to\gamma\gamma$ Decay Rate}
\label{sect:HiggstoGammaGamma}

From a technical point of view, the calculation of the $H\to\gamma\gamma$ decay amplitude is the most challenging aspect of this work. Our strategy is to expand the 4D effective action in order to obtain all possible vertices and then use \texttt{FeynArts} \cite{Hahn:2000kx} to generate the Feynman diagrams contributing to the decay $H\to \gamma\gamma$ at one-loop order and in the Feynman gauge. These diagrams are then evaluated for generic couplings (e.g.\ the second column of table~\ref{HGGfeynRules}) using \texttt{FormCalc} \cite{Hahn:1998yk}. Finally, the couplings of relevance are computed explicitly (e.g.\ the third column of table~\ref{HGGfeynRules}) in order to obtain the final expression for the decay amplitude. 

\begin{figure}
\begin{center}
\subfigure[]{%
           \label{fig:floop}
           \includegraphics[width=0.28\textwidth]{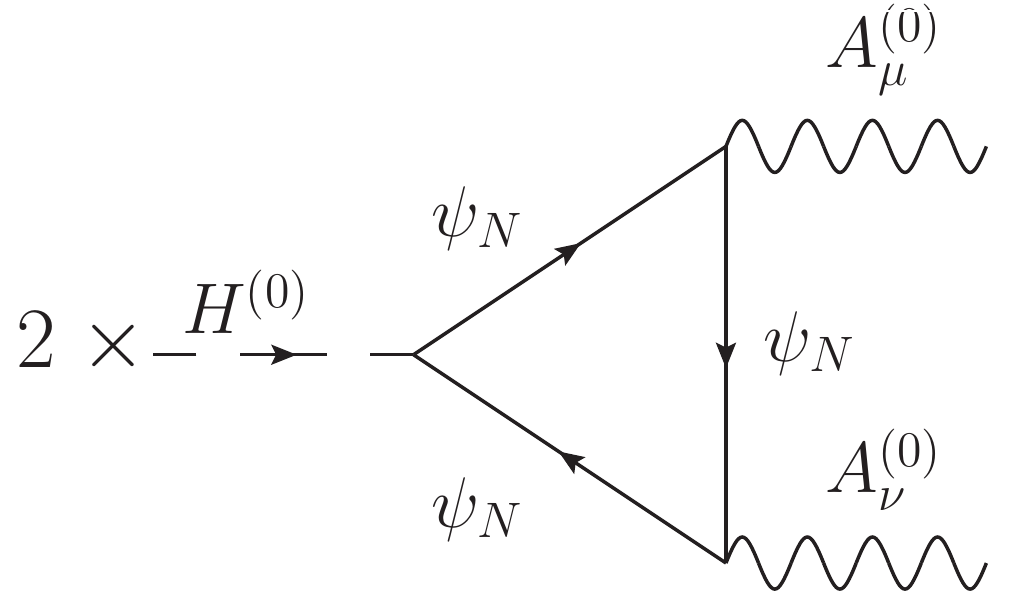}
        }
        \subfigure[]{%
           \label{fig:Wloop}
           \includegraphics[width=0.28\textwidth]{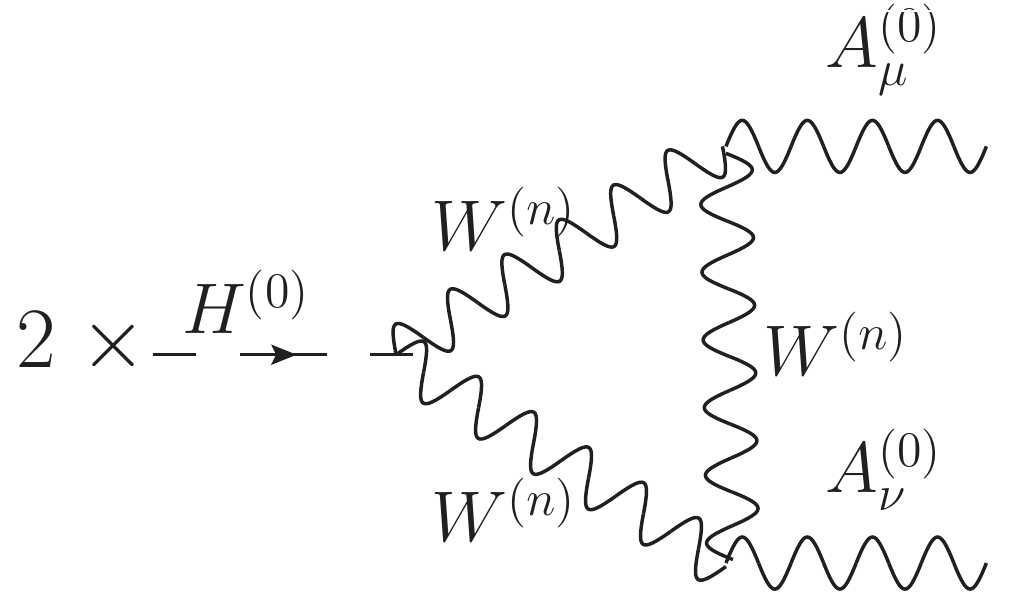}
        }
        \subfigure[]{%
           \label{fig:WWloop}
           \includegraphics[width=0.28\textwidth]{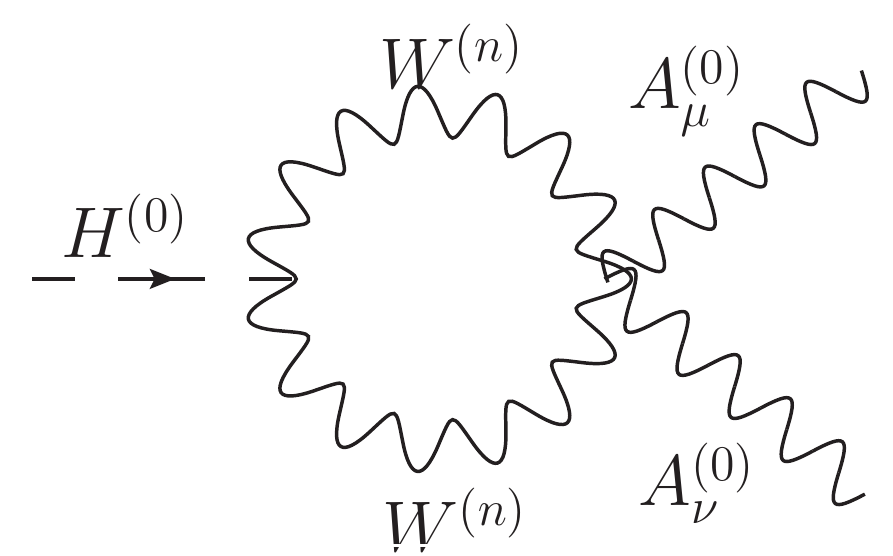}
        }
        \\
          \subfigure[]{%
           \label{fig:WGGloop}
           \includegraphics[width=0.28\textwidth]{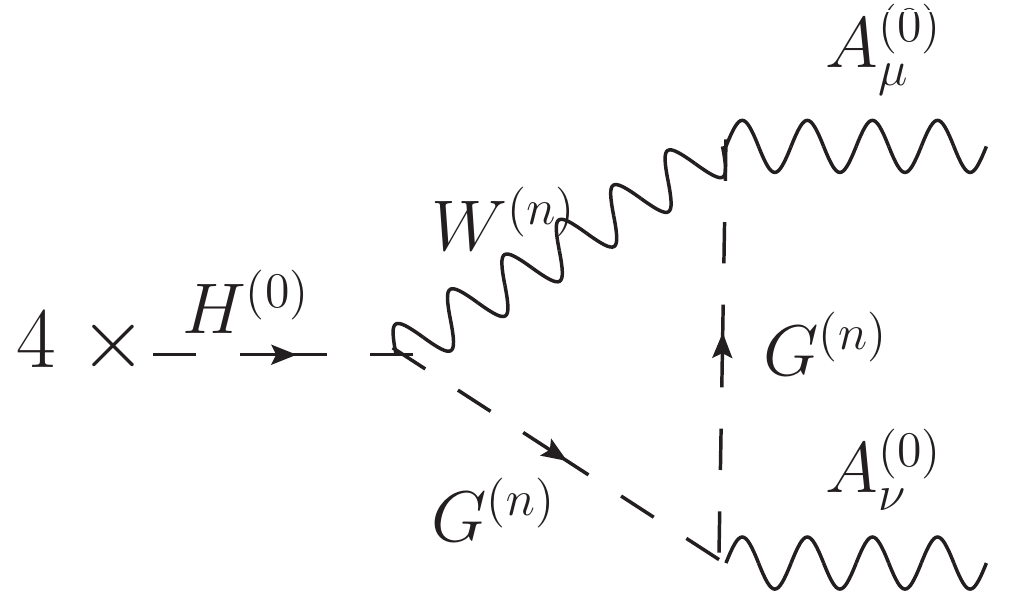}
        } 
        \subfigure[]{%
           \label{fig:WGWloop}
           \includegraphics[width=0.28\textwidth]{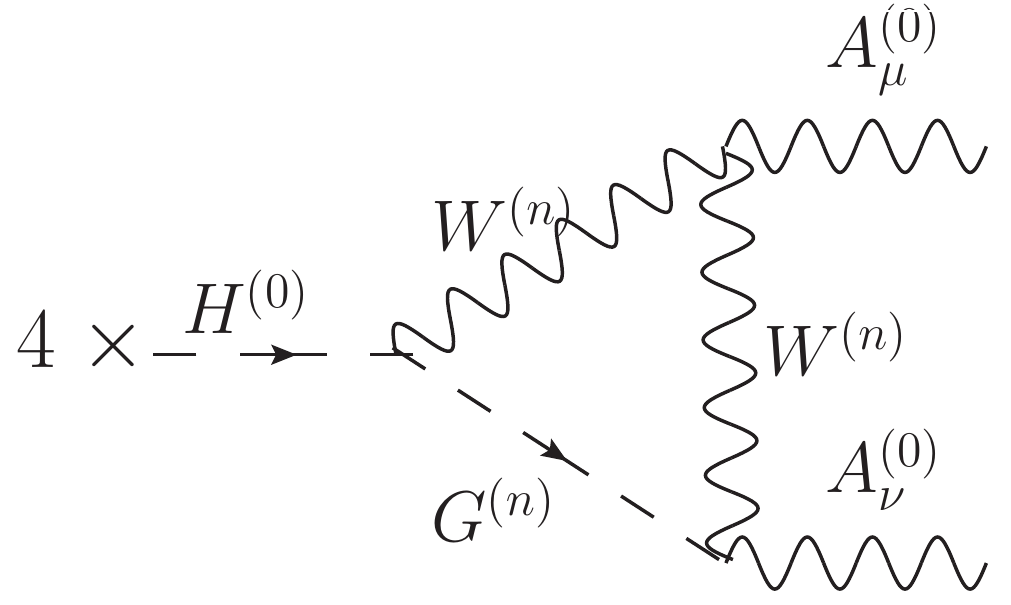}
        }
        \subfigure[]{%
           \label{fig:WWGloop}
           \includegraphics[width=0.28\textwidth]{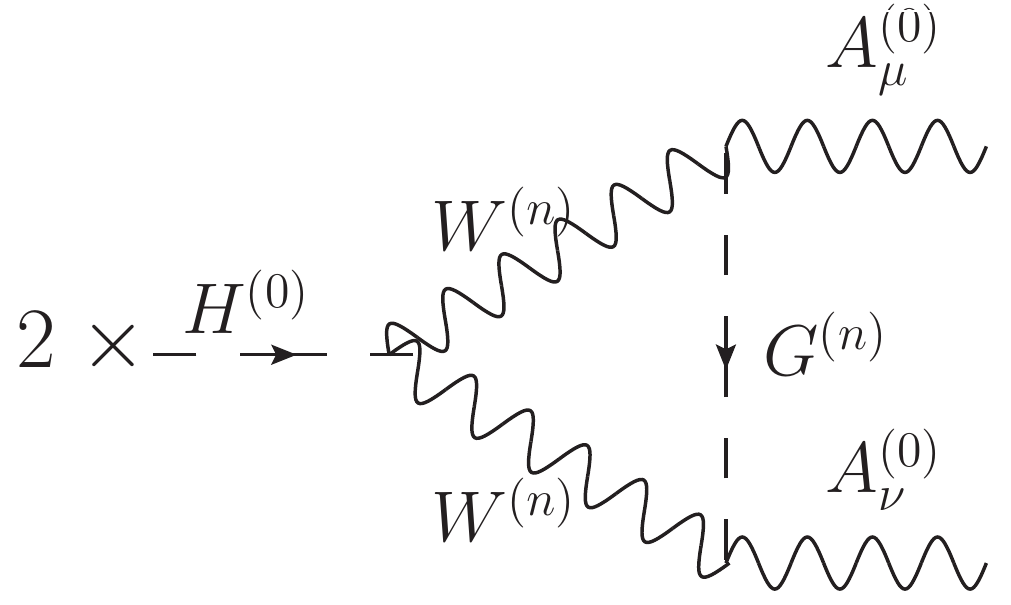}
        }
        \\
        \subfigure[]{%
           \label{fig:GGWloop}
           \includegraphics[width=0.28\textwidth]{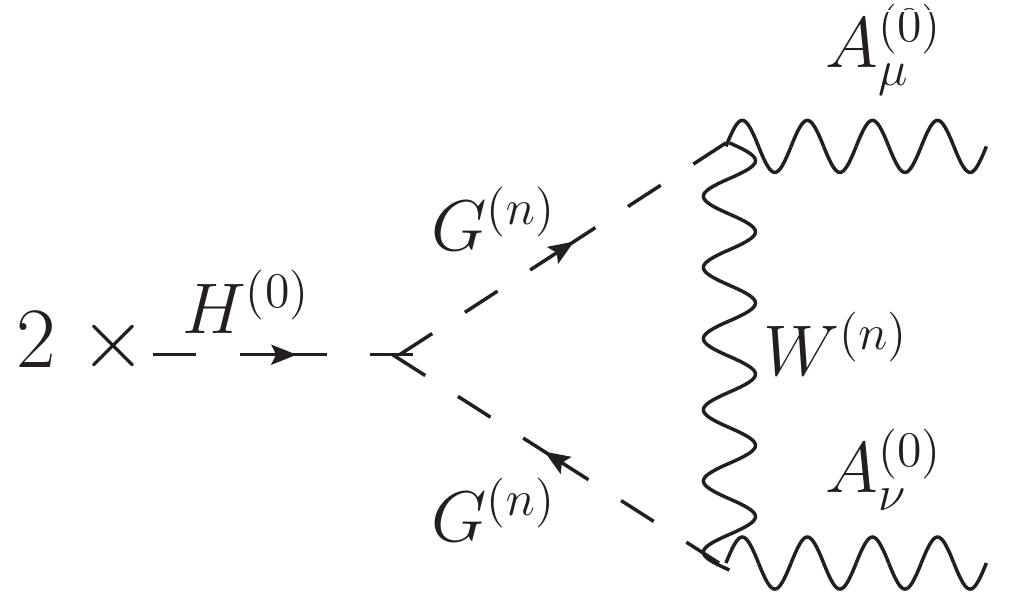}
        }
        \subfigure[]{%
           \label{fig:Gloop}
           \includegraphics[width=0.28\textwidth]{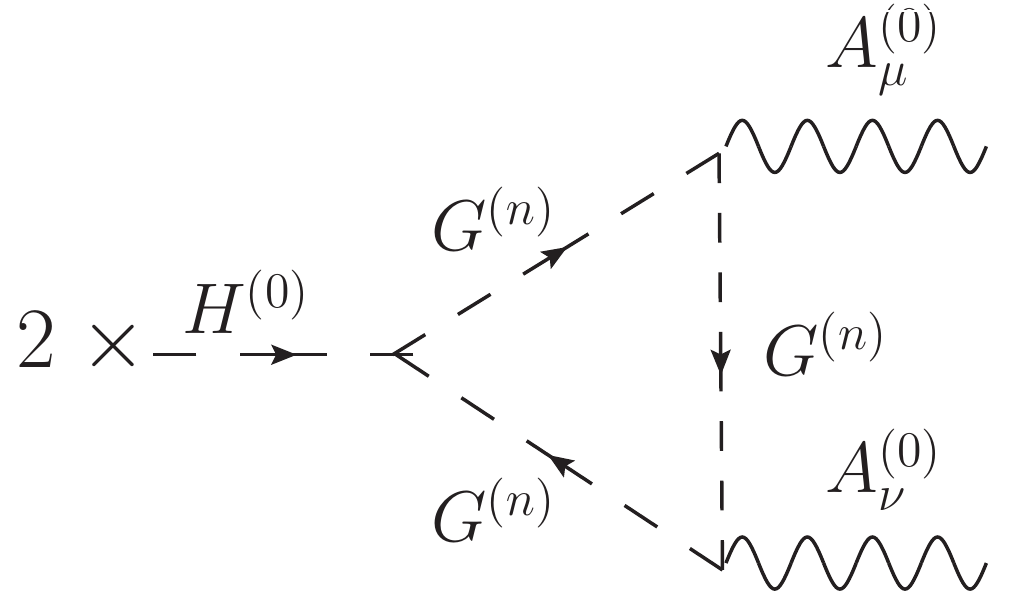}
        }
        \subfigure[]{%
           \label{fig:uloop}
           \includegraphics[width=0.28\textwidth]{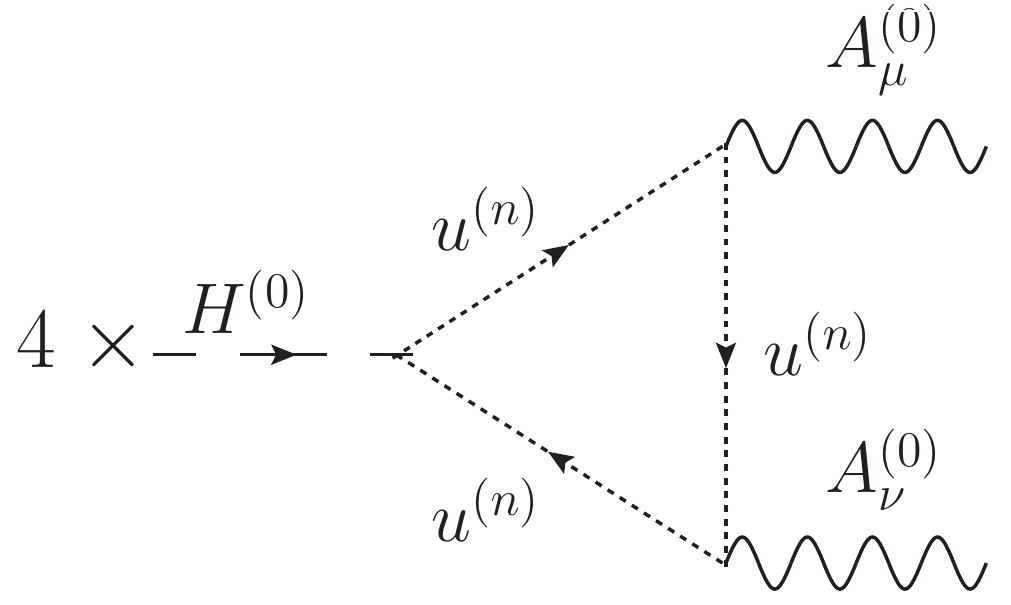}
        }
        \\       
        \subfigure[]{%
           \label{fig:GGloop}
           \includegraphics[width=0.28\textwidth]{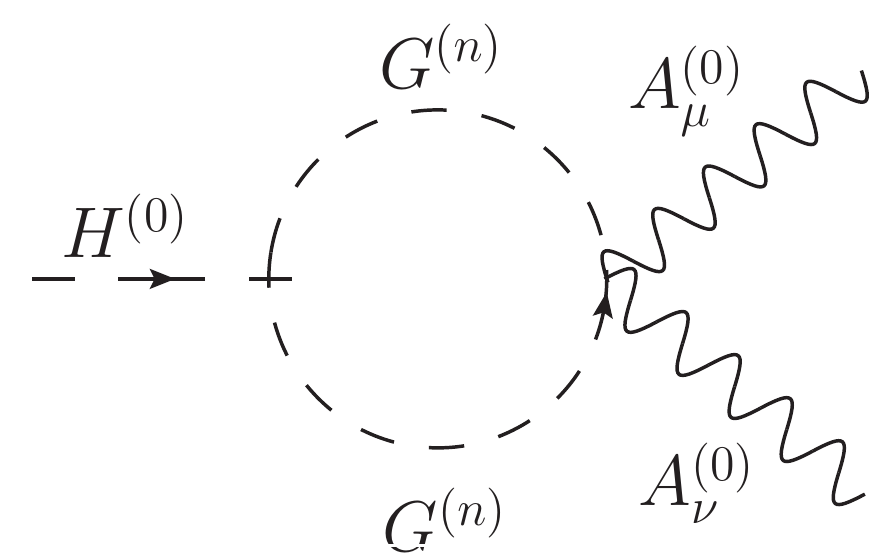}
        }
        \subfigure[]{%
           \label{fig:WGloop}
           \includegraphics[width=0.28\textwidth]{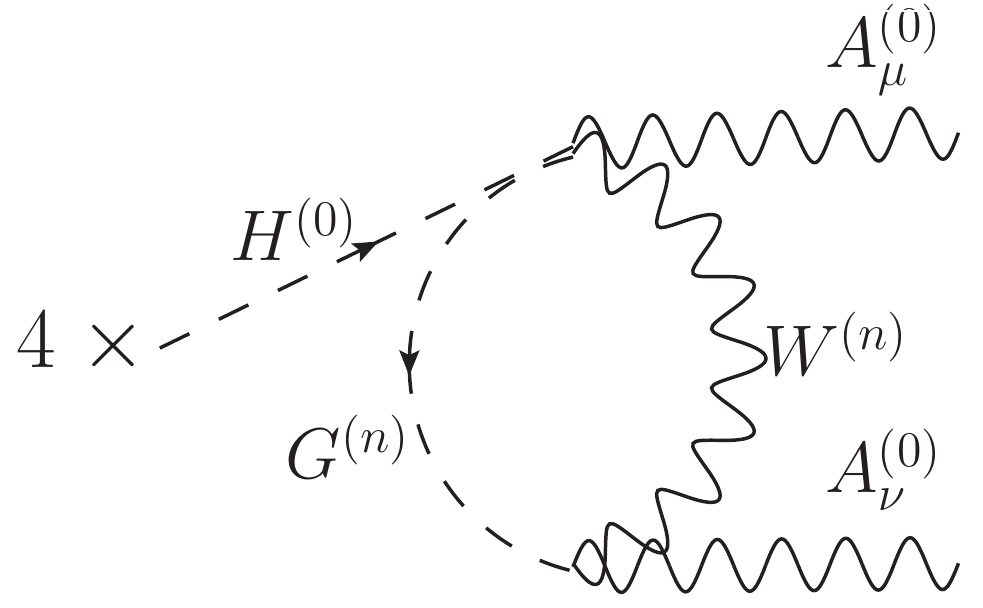}
          }
           \subfigure[]{%
           \label{fig:Ploop}
           \includegraphics[width=0.28\textwidth]{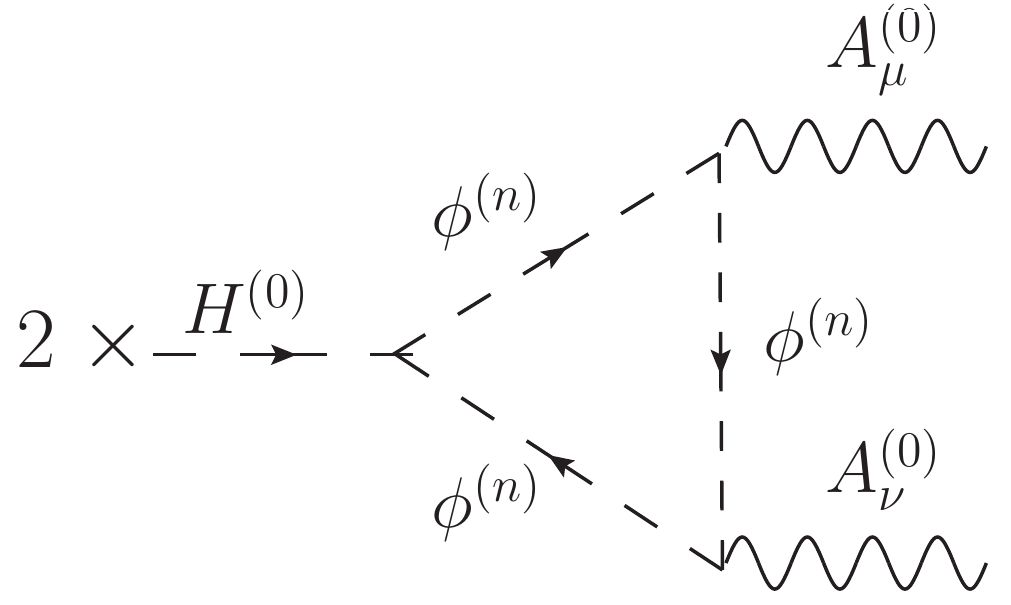}
        }\\
         \subfigure[]{%
           \label{fig:PPloop}
           \includegraphics[width=0.28\textwidth]{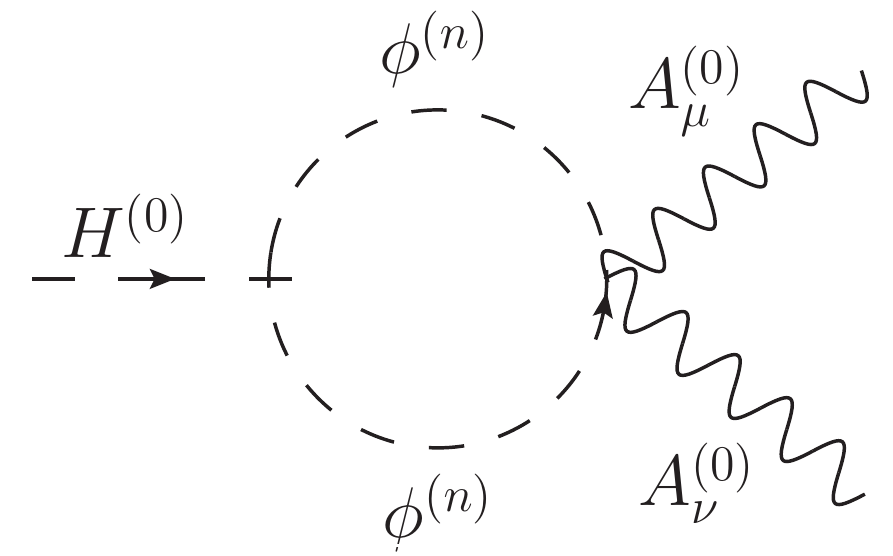}
        }
\caption{Feynman diagrams contributing to the $H\to\gamma\gamma$ decay amplitude in the Feynman gauge.}
\label{fig:GaugeContrib}
\end{center}
\end{figure}

The leading-order contribution to the Higgs to diphoton decay rate again occurs via a loop process, including all charged scalars, fermions and vectors that couple to the Higgs and the photon. For a model with a bulk Higgs, the relevant diagrams are shown in figure~\ref{fig:GaugeContrib}. With the Feynman rules listed in table~\ref{HGGfeynRules}, the partial decay width is calculated to be
\begin{equation}\label{GammaHgaga}
   \Gamma_{H\to\gamma\gamma}
   = \left( \big| C_V + C_S + C_f \big|^2 + \big| C_f' \big|^2 \right) 
    \frac{\alpha^2 m_H^3}{256\pi^3\tilde v^2} \,,
\end{equation}
where the vector, scalar and fermion contributions are found to be 
\begin{align}
   C_V &= \sum_n\,\frac{\tilde v g_{HWW}^{(n,n)}}{2m_n^{(W)\,2}}\,\,
    7 A_1(\tau_n^W) \,, \label{CVres} \\
   C_S &= \sum_n\,\frac{\tilde v g_{H\phi\phi}^{(n,n)}}{2m_n^{(\phi)\,2}}\,
    \left( -\frac13 \right) A_0(\tau_n^\phi) \,, \\
   C_f &= \!\sum_{f=u,d,e}\!N_{c,f}\,Q_f^2\,\sum_N\,\frac{\tilde v}{\sqrt2}\,
    \frac{\mbox{Re}\,(\tilde Y_f^{\rm mass})_{NN}}{m_N^f} 
    \left( -\frac43 \right) A_{\frac12}(\tau_N^f) \,, \\ 
   C_f' &= \!\sum_{f=u,d,e}\!N_{c,f}\,Q_f^2\,\sum_N\,\frac{\tilde v}{\sqrt2}\,
    \frac{\mbox{Im}\,(\tilde Y_f^{\rm mass})_{NN}}{m_N^f}\,\,2B_{\frac12}(\tau_N^f) \,.
    \label{Cfprimeres} 
\end{align}
The loop functions for the vector and scalar contributions are given by \cite{Gunion:1989we}
\begin{align}
   A_1(\tau) &= \frac17\,\Big[ 2 + 3\tau + 3\tau (2-\tau) f(\tau) \Big] \,, \\
   A_0(\tau) &= -3\tau \left[ 1 - \tau f(\tau) \right] ,
\end{align}
and are normalised such that $A_{1,0}(\tau)\to 1$ for $\tau\to\infty$. Relative to its value in the SM, it follows from (\ref{GammaHgaga}) that the $H\to\gamma\gamma$ decay rate is given by
\begin{equation}
   \frac{\Gamma_{H\to\gamma\gamma}}{\Gamma_{H\to\gamma\gamma}^{\rm (SM)}}
   = \frac{|\kappa_\gamma|^2+|\kappa_{\gamma 5}|^2}{\kappa_v^2} \,,
\end{equation}
where 
\begin{equation}
   \kappa_\gamma 
   = \frac{C_V+C_S+C_f}%
          {7A_1(\tau_W)-\frac{16}{9} A_{\frac12}(\tau_t)-\frac49 A_{\frac12}(\tau_b)} \,, \qquad
   \kappa_{\gamma 5} 
   = \frac{-C_f'}{7A_1(\tau_W)-\frac{16}{9} A_{\frac12}(\tau_t)-\frac49 A_{\frac12}(\tau_b)} \,.
\end{equation}

As is well known, the decay width in the SM is dominated by the vector contribution. On the other hand, we will find that in our model the most significant deviation from the SM result arises from the effect of KK quarks and charged leptons in the fermion contribution $C_f$, unless the 5D Yukawa couplings are taken to be very small. The reason is twofold. Firstly, for large Yukawa couplings the fermionic corrections are parametrically dominant over the bosonic corrections. Secondly, in the asymptotic limit $\beta\gg 1$ there is a partial cancellation between the effects of the KK tower of $W$-boson resonances and the correction to the coupling of the $W$-boson zero mode \cite{Hahn:2013nza}. The two effects would cancel exactly in the limit where $A_1(\tau_0^W)\to 1$, while in practice $A_1(\tau_0^W)\approx 1.19$. Further still, the KK fermion contribution is strongly correlated with the corresponding effect in the gluon-fusion amplitude (even though it now includes the additional lepton contributions), but it interferes destructively with the dominant vector amplitude. Hence an enhancement (suppression) in the gluon-fusion cross section would typically result in a suppression (enhancement) in the $H\to\gamma\gamma$ decay rate.

The relative simplicity of the results (\ref{CVres}) to (\ref{Cfprimeres}) reflects the significant cancellations that occur as a result of gauge invariance. While here we shall not give the details of such cancellations, it is nonetheless worth commenting on them. In particular one finds:
\begin{itemize}
\item
The $H\to\gamma\gamma$ amplitude includes contributions from naively logarithmically divergent bubble diagrams (figures~\ref{fig:WWloop}, \ref{fig:GGloop}, \ref{fig:WGloop} and \ref{fig:PPloop}). However, one finds that all the divergent scalar integrals cancel. This cancellation occurs at the level of working with completely generic couplings (i.e.\ working with the second column of table~\ref{HGGfeynRules}). 
\item
One finds further that a gauge-dependent tensor structure cancels as a result of the Goldstone boson equivalence theorem. In particular, this requires that the Higgs to Goldstone coupling follows the relation: $g_{HGG}^{(n,n)}=g_{HWW}^{(n,n)}\,(m_H^2/2m_n^{(W)2})$. This occurs for generic 5D geometries.
\item
The flatness of the profile of the photon zero mode implies that the three-point vertices including one photon and the four-point vertices including two photons conserve KK number. Hence, the $H\to\gamma\gamma$ amplitude contains only single sums over KK states. This is important in ensuring that the sums converge. If there were double sums over KK modes, then a convergent result could only arise after non-trivial cancellations.   
\item
There are no loop contributions analogous to figures~\ref{fig:WGGloop}, \ref{fig:WGWloop}, \ref{fig:WWGloop}, \ref{fig:GGWloop} and \ref{fig:WGloop}, but with the Goldstone bosons replaced by charged physical scalars $\phi^\pm$. This is due to the photon--scalar--$W$ boson vertex being forbidden, see (\ref{PhiWA0}). This has a more intuitive explanation, and one would anticipate this to be true in a wide range, if not all, BSM scenarios. Consider a generic scenario containing many different mass scales $v_i$ (for example VEVs of multiple Higgs fields), as well as (before gauge fixing) a number of scalar fields $\phi_j$. By definition, the gauge-fixing terms will be constructed in order to cancel the terms in the 4D effective Lagrangian that mix one $W$ boson with one scalar, i.e., $\mathcal{L}\supset\sum_{i,j} \alpha_{ij} v_i\partial^\mu W_\mu^\mp\phi_j^\pm$, where $\alpha_{ij}$ are dimensionless couplings. Hence the Goldstone bosons will be proportional to $G^\pm\sim\sum_{i,j}\alpha_{ij}v_i\phi_j^\pm$.  Likewise, the scalar--photon--$W$ vertex will arise from the terms $\mathcal{L}\supset\sum_{i,j} \beta_{ij} v_i A^\mu W_\mu^\mp\phi_j^\pm$, again with dimensionless couplings $\beta_{ij}$. However, when the gauge symmetry is unbroken, the Lagrangian must be invariant under the transformation $A_\mu\to A_\mu+\partial_\mu\theta^A$, which implies that $\alpha_{ij}$ and $\beta_{ij}$ should be aligned, and hence generically the three-point interactions with scalars, a photon and a $W$ boson should just include Goldstone bosons.      
\end{itemize}
It is also worth commenting that, for $H\to Z\gamma $, the last three points would not naively apply. The vertex between the charged scalars, the $W$ boson and the $Z$ boson is both allowed and does not conserve KK number. Hence the calculation of the $H\to Z\gamma$ amplitude would be significantly more involved and lies beyond the scope of this paper.

\section{\boldmath Application to AdS$_5$ Space}
\label{sect:AdS5Geo}

Having derived the key expressions for a generic geometry, we shall now focus our attention on the particular case of an AdS${}_5$ geometry. This will allow us to solve for the relevant profiles and hence study the phenomenology in more detail. By considering a Higgs propagating in an AdS${}_5$ space, it is conjectured that we are studying a 4D theory of a fundamental Higgs mixing with a conformal composite dual theory, i.e.\ a partially composite Higgs \cite{Luty:2004ye}. Accepting this, then the model offers a compelling framework for investigating how the Higgs phenomenology would change as the scaling dimension of the dual Higgs operator (here related to a parameter $\beta$) is varied. Therefore, in the following two sections we will examine the $\beta$ dependence of the resulting physics.  

\subsection{Scalar Profiles, Vector Profiles, and Higgs VEV}
\label{sect:HiggsVEV}

Firstly, the Higgs VEV is found by solving (\ref{hvevEqn}) with (\ref{ADS5def}), (\ref{hvevBCs}) and (\ref{HiggsPoten}) (see for example \cite{Cacciapaglia:2006mz,Archer:2012qa,Malm:2013jia}). We obtain 
\begin{equation}\label{HiggsVeV}
   v(r) = N_v \left( \frac{r}{R'} \right)^{2+\beta}
   \left[ 1 + B_h \left( \frac{r}{R'} \right)^{-2\beta} \right] ,
\end{equation}
where 
\begin{align}
	B_h &= - \frac{2+\beta-R M_{\rm UV}}{2-\beta-R M_{\rm UV}}\,
	 \Omega^{-2\beta} \,, \\
	N_v^2 &= \frac{1}{R\lambda_{\rm IR}}\,
	\frac{(R M_{\rm IR}-2-\beta) + B_h (R M_{\rm IR}-2+\beta)}{(1+B_h)^3} \,. 
\end{align}
Above we have introduced the important parameter
\begin{equation}\label{betaDef}
   \beta = \sqrt{4+R^2M_\Phi^2} \,.
\end{equation}
In the conjectured dual theory, the corresponding 4D operator $\mathcal{O}_H$ of a partially composite Higgs would have a scaling dimension of $(2+\beta)$. It was argued in \cite{Luty:2004ye} that, in order to resolve the gauge hierarchy problem, one requires that the Higgs mass term $\mathcal{O}_H^\dag\mathcal{O}_H$ is not relevant and hence $\beta\ge 0$. It is quite interesting to note that this coincides with the Breitenlohner-Freedman bound, obtained independently by requiring that the flux of the energy-momentum tensor vanishes at the AdS boundary \cite{Breitenlohner:1982jf}. Further still, when this bound is saturated (i.e.\ $\beta\approx 0$), the allowed range of the fermion zero-mode Dirac mass term approximately corresponds to the observed fermion mass hierarchy \cite{Archer:2012qa,Agashe:2008fe,vonGersdorff:2012tt}.   

Rewriting the VEV profile in the form $v(r)=\tilde v\,h(r)$ and imposing the normalization condition (\ref{eq42}), we obtain
\begin{equation}
   \frac{\tilde v^2}{M_{\rm KK}^2} 
   = \frac{R^2}{2\lambda_{\rm IR}}
    \left( \frac{1}{1+\beta} + 2B_h + \frac{B_h^2}{1-\beta} \right)
    \frac{(R M_{\rm IR}-2-\beta) + B_h (R M_{\rm IR}-2+\beta)}{(1+B_h)^3} \,,
	\label{ratio}
\end{equation}
where we have introduced the KK scale $M_{\rm KK}=1/R'$, which sets the mass scale of the low-lying KK modes. Here and below we drop terms of ${\cal O}(\Omega^{-2})$ and higher, recalling that $\Omega^{-1}=R/R'\approx 10^{-15}$ is an exceedingly small parameter. Note that the natural scale for the 5D parameters $M_{\rm IR}$ and $\lambda_{\rm IR}$ is the Planck scale, and hence the dimensionless quantities $R M_{\rm IR}$ and $\lambda_{\rm IR}/R^2$ are expected to be of ${\cal O}(1)$. The natural scale for the Higgs VEV $\tilde v$ is set by the KK scale $M_{\rm KK}=1/R'$. We now note that, unless $\beta$ is very small or $R M_{\rm UV}$ is extremely fine-tuned to the value $(2-\beta)$, it is a safe approximation to set $B_h\propto\Omega^{-2\beta}\to 0$. It then follows from equation (\ref{HiggsVeV}) that the Higgs VEV will be peaked towards the IR brane. This is an important point, which allows the model to offer a potential resolution to the gauge hierarchy problem. Moreover, in this case relation (\ref{ratio}) simplifies to
\begin{equation}\label{rela1}
   \frac{\tilde v^2}{M_{\rm KK}^2}
   = \frac{R^2}{\lambda_{\rm IR}}\,\frac{R M_{\rm IR}-2-\beta}{2(1+\beta)} \,.
\end{equation}
The fact that the left-hand side must be positive implies the bound $R M_{\rm IR}\ge 2+\beta$, and in order to achieve that $\tilde v^2/M_{\rm KK}^2\ll 1$ the right-hand side must be fine-tuned to some extent. In this case one obtains the simple result \cite{Cacciapaglia:2006mz,Azatov:2009na,Malm:2013jia} 
\begin{equation}\label{hprofile}
   h(r) = \frac{R'}{R^{3/2}}\,\sqrt{\frac{2(1+\beta)}{1-\Omega^{-2-2\beta}}}
    \left( \frac{r}{R'} \right)^{2+\beta} .
\end{equation}
The term $\Omega^{-2-2\beta}$ can be dropped for all practical purposes, since $\beta\ge 0$.

The profile for the Higgs particle itself is found by solving (\ref{HProfEOM}) with (\ref{HiggsBCs}), which yields 
\begin{equation}\label{HiggsProfile}
   f_n^{(H)} = N_H r^2 \left[ J_\beta(m_n^{(H)}r) + B_H Y_\beta(m_n^{(H)}r) \right] ,
\end{equation} 
where $N_H$ is determined from (\ref{HiggsOrthog}), and
\begin{equation}
   B_H = -\frac{(2+\beta-R M_{\rm{UV}})\,J_\beta(m_n^{(H)} R)
                - m_n^{(H)} R\,J_{\beta+1}(m_n^{(H)} R)}%
               {(2+\beta-R M_{\rm{UV}})\,Y_\beta(m_n^{(H)} R)
                - m_n^{(H)} R\,Y_{\beta+1}(m_n^{(H)} R)} 
   \propto \Omega^{-2\beta} \,.
\end{equation}
Unless $\beta$ is very small, it is again a safe approximation to set $B_H\to 0$. The mass eigenvalues $m_n^{(H)}$ are determined by imposing the boundary condition on the IR brane shown in the first equation in (\ref{HiggsBCs}). For $B_H=0$, this yields the condition \cite{Malm:2013jia}
\begin{equation}\label{eigenvals}
   \frac{x_n J_{\beta+1}(x_n)}{J_\beta(x_n)} = 2 (R M_{\rm IR} - 2 - \beta) \,,
\end{equation}
where $x_n=m_n^{(H)}/M_{\rm KK}$. It follows from this equation that even the zero mode (the SM Higgs boson) would have a mass that is naturally of order the KK scale $M_{\rm KK}$, unless the right-hand side is tuned to be much smaller than ${\cal O}(1)$. The same tuning was require to obtain a value $\tilde v\ll M_{\rm KK}$, see (\ref{rela1}). 

It remains to relate the parameters $M_{\rm IR}$ and $\lambda_{\rm IR}$ to physical quantities. Equation~(\ref{rela1}) provides one useful relation to this end. To obtain a second relation, we evaluate the eigenvalue condition for the mass of the SM Higgs boson. Denoting $\delta=R M_{\rm IR}-2-\beta\ll 1$, we find from (\ref{eigenvals}) that \cite{Malm:2013jia}
\begin{equation}\label{rela2}
   \frac{m_H^2}{M_{\rm KK}^2} = 4(1+\beta)\delta \left[ 1 - \frac{\delta}{2+\beta}
   + {\cal O}(\delta^2) \right] .
\end{equation}
To leading order, we now obtain from (\ref{rela1}) and (\ref{rela2}) the solutions
\begin{equation}\label{lambdaIRapp}
   R M_{\rm IR} \approx 2+\beta + \frac{1}{4(1+\beta)}\,\frac{m_H^2}{M_{\rm KK}^2} \,, \qquad
   \frac{\lambda_{\rm IR}}{R^2} 
    \approx \frac{1}{8(1+\beta)^2}\,\frac{m_H^2}{\tilde v^2} 
    \approx \frac{G_F m_H^2}{4\sqrt2(1+\beta)^2} \,.
\end{equation}
Although we have presented, for illustrative purposes, the approximate relations, in the following analysis we use the exact numerical result for $\lambda_{\rm{IR}}$. In the limit that $m_H\ll M_{\rm{KK}}$ we can now make the expansion, alluded to a number of times already, of the ratio of the Higgs and VEV profiles. We find \cite{Azatov:2009na,Malm:2013jia}
\begin{equation}\label{VevProfApp}
   \frac{f_0^{(H)}(r)}{h(r)} 
   = 1 - \frac{m_H^2}{4M_{\rm KK}^2} \left[ \frac{(r/R')^2}{1+\beta}
    - \frac{1}{2+\beta} \right] + {\cal O}\bigg( \frac{m_H^4}{M_{\rm KK}^4} \bigg) \,.
\end{equation}   

Finally we must solve for the scalar and $W$ profiles. This is slightly more involved, since the equations (\ref{fWZEOM}) and (\ref{PhiWEOM}),
\begin{align}
   \partial_r^2 f_n^{(W)} - \frac{1}{r}\partial_r f_n^{(W)}
    - \frac{R^2 M_W^2 h^2}{r^2} f_n^{(W)} + m_n^{(W)2} f_n^{(W)} &=0 \,, \\
   \partial_r^2 f_n^{(\phi^{\pm})} + \left( \frac{3}{r} - \frac{2\partial_r h}{h} \right)
    \partial_r f_n^{(\phi^{\pm})} - \frac{R^2 M_W^2 h^2}{r^2} f_n^{(\phi^{\pm})}
    + m_n^{(\phi^\pm)2} f_n^{(\phi^{\pm})} &=0 \,,  
\end{align}   
do not have exact analytical solutions. However, when $m_n^{(W,\phi^\pm)}\gg\tilde v$ the fourth term dominates over the third term and one can use the approximate solutions
\begin{equation}\label{Wprofile}
   f_n^{(W)}\approx N_W r \left[ J_1(m_n^{(W)}r) + B_W Y_1(m_n^{(W)}r) \right] \quad
    \mbox{for}\quad n\ge 1 \,,
\end{equation}
where $B_W=-\frac{J_0(m_n^{(W)}R)}{Y_0(m_n^{(W)}R)}$. Likewise the scalar profiles can be approximated by
\begin{equation}\label{Psiprofile}
   f_n^{(\phi^\pm)}\approx N_\phi r^{1+\beta} \left[ J_{-1-\beta}(m_n^{(\phi^\pm)}r)
    + B_\phi Y_{-1-\beta}(m_n^{(\phi^\pm)}r) \right] ,
\end{equation}
where $B_\phi=-\frac{J_{-1-\beta}(m_n^{(\phi^\pm)}R)}{Y_{-1-\beta}(m_n^{(\phi^\pm)}R)}$. In both cases the mass eigenvalues are derived by imposing the IR boundary conditions $\partial_r f_n^{(W)}|_{r=r_{\rm IR}}=0$ and $f_n^{(\phi^\pm)}|_{r=r_{\rm IR}}=0$, respectively. We have tested these approximations against the numerical solutions and found them to be reasonable for $M_{\rm{KK}}\gtrsim 1$--2\,TeV. The KK masses are found by imposing the relevant boundary conditions and scale approximately as
\begin{equation}
\label{eqn:AppMasses}
   \frac{m_n^{(W)}}{M_{\rm KK}}\approx \left( n - \frac14 \right) \pi \,, \qquad
   \frac{m_n^{(\phi^\pm)}}{M_{\rm KK}}\approx \left( n + \frac14 + \frac{\beta}{2}
    \right) \pi \,.
\end{equation}   
These relations approximate reasonably well the numerical results found in \cite{Archer:2012qa}. Also note that as $\beta\to\infty$ the masses of the scalars go to infinity and hence in the brane-localised Higgs limit these scalars decouple from the theory.  

We emphasise that the approximation (\ref{Wprofile}) breaks down for the $W$ zero mode, in which case the equations of motion can be solved using a power series in $m_W^2/M_{\rm KK}^2$. We obtain
\begin{align}\label{mWmassrel}
   m_W^2 &= m_{W,0}^2 \left\{ 1 - \frac{m_{W,0}^2}{2M_{\rm KK}^2} \left[
    \frac{2L(1+\beta)^2}{(2+\beta)(3+2\beta)} - 1 + \frac{1}{(2+\beta)^2} + \frac{1}{2L}
    \right] + \dots \right\} \,, \\
   f_0^{(W)} &= \frac{1}{\sqrt{RL}}\,\Bigg\{ 1 + \frac{m_{W,0}^2}{2M_{\rm KK}^2}\,
    \Bigg[ \frac{L}{2+\beta} \left( \frac{r}{R'} \right)^{4+2\beta} \!\!
    - L \left( \frac{r}{R'} \right)^2 + \left( \frac{r}{R'} \right)^2
    \left( \frac12 - \ln\frac{r}{R'} \right) \nonumber\\
   &\hspace{3.8cm}\mbox{}+ \frac12 \left( 1 - \frac{1}{L} \right) - \frac{1}{2(2+\beta)^2} 
    \Bigg] + \dots \Bigg\} \,,
\end{align}
where $m_{W,0}^2=M_W^2/(RL)=g_5^2\tilde v^2/(4RL)$, and we have again used the parameter $L=\ln(R'/R)=\ln\Omega\approx 34.5$, which is a measure of the size of the warped extra dimension. Analogous formulas, with $m_{W,0}^2$ replaced by $m_{Z,0}^2=(g_5^2+g_5'^2)\tilde v^2/(4RL)$, hold for the $Z$ boson.

\subsection{\boldmath Electroweak Fit and Determination of 5D Parameters}
\label{sec:EWpars}

We now describe in detail how we determine the parameters in the 5D Lagrangian in terms of physical observables and then compute the electroweak precision observables $S$, $T$ and $U$ in our model. Due to the presence of heavy KK modes and the non-universality of the fermion profiles, new physics effects in RS models can in general not be uniquely described in terms of oblique corrections. However, by fitting to a selected subset of the most precisely measured observables one can still parametrise electroweak corrections in terms of $S$, $T$ and $U$. Specifically, in order to uniquely determine the parameter $\tilde v$ in (\ref{ratio}) and the 5D gauge coupling $g_5$ and $g_5'$ in (\ref{covder}) along with the electroweak parameters requires specifying six independent observables. We choose them to be $G_F$, $m_Z$, $m_W$ (or equivalently $s_W^2$), $\alpha$ (or equivalently $s_0^2$), $s_*^2$ and $\rho_*$, where the various definitions of the weak mixing angles and of $\rho_*$ will be specified below.

We begin by deriving explicit expressions for the observables $G_F$, $m_Z$, $m_W$ and $\alpha$ in our model. The first relation in (\ref{mWmassrel}) and the corresponding relation for the $Z$-boson mass determine $m_W$ and $m_Z$ in terms of the dimensionless ratios $g_5^2/(RL)$ and $g_5'^2/(RL)$ and the parameter $\tilde v$. In addition, using the fact that the photon has a flat profile along the extra dimension, one finds that the fine-structure constant is given in terms of the same dimensionless ratios as
\begin{equation}
   4\pi\alpha = \frac{1}{RL}\,\frac{g_5^2\,g_5^{\prime 2}}{g_5^2 + g_5^{\prime 2}} \,.
\end{equation}
The relation between the 5D parameter $\tilde v$ and the Fermi constant can be derived by constructing the effective four-fermion interaction mediating muon decay. Modifications to the muon decay amplitude arise because of the modification of the $W$-boson interaction with light fermions and due to the presence of the infinite tower of heavy KK resonances, which can be exchanged instead of the $W$-boson of the SM. At a technical level, both effects are encoded in the 5D gauge-boson propagator 
\begin{equation}
   D(r,r';p^2) = - \sum_n\,\frac{f_n^{(W)}(r)\,f_n^{(W)}(r')}{p^2-m_n^{(W)\;2}+i\epsilon}
\end{equation}
evaluated at zero momentum transfer. From (\ref{GaugeOrtho}) and (\ref{fWZEOM}), it follows that the 5D propagator in AdS$_5$ space obeys the differential equation
\begin{equation}
   \left( r\,\partial_r\,\frac{1}{r}\,\partial_r + p^2 - \frac{M_W^2 R^2}{r^2}\,h^2(r) \right)
   D(r,r';p^2) = - \frac{r}{R}\,\delta(r-r') \,,
\end{equation}
with the boundary conditions $\partial_r D(r,r';p^2)=0$ for $r=R,R'$. For $p^2=0$ and with the VEV profile $h(r)$ given in (\ref{hprofile}) this equation can be solved exactly in terms of Bessel functions. Expanding the solution to first non-trivial order in $m_W^2/M_{\rm KK}^2$, we find
\begin{equation}
   D(r,r';0) = \frac{1}{M_W^2} \left\{ 1 + \frac{L m_W^2}{2M_{\rm KK}^2} \left[
    \frac{2(1+\beta)^2}{(2+\beta)(3+2\beta)} - \frac{r_>^2}{R^{\prime\;2}} 
    + \frac{r^{4+2\beta}+r^{\prime\;4+2\beta}}{(2+\beta) R^{\prime\;4+2\beta}} \right]
    + \dots \right\} ,
\end{equation}
where $r_>=\mbox{max}(r,r')$. In the calculation of the muon decay amplitude this expression is convoluted with the profile functions of the light leptons of the SM. Then the $r$ and $r'$ dependent terms in the expression above give exponentially small contributions. We thus obtain
\begin{equation}
   \frac{G_F}{\sqrt2} \equiv \frac{1}{2v^2}
   = \frac{1}{2\tilde v^2} \left\{ 1 + \frac{m_W^2}{2M_{\rm KK}^2}\,
    \frac{2L(1+\beta)^2}{(2+\beta)(3+2\beta)} + \dots \right\} .
\end{equation}
This relation can be used to determine the Higgs VEV $\tilde v$ in our model in terms of the SM value $v=246.2$\,GeV. 

We now proceed to calculate a couple of relevant electroweak parameters. We consider three different definitions of the weak mixing angle, using the notation of Peskin and Takeuchi \cite{Peskin:1991sw}. The first definition employs the structure $(T_3-s_*^2 Q)$ in the weak neutral current, as measure from the $Z$-pole polarization asymmetries at LEP. The parameter $s_*^2$ coincides with our definition of $s_w^2$ in terms of 5D gauge couplings in (\ref{swdef}), namely
\begin{equation}
   s_*^2 = \frac{g_5^{\prime 2}}{g_5^2 + g_5^{\prime 2}} = s_w^2 \,.
\end{equation}
The second definition employs the ratio of the electroweak gauge-boson masses, for which we find
\begin{equation}
   s_W^2\equiv 1 - \frac{m_W^2}{m_Z^2}
   = \frac{g_5^{\prime 2}}{g_5^2 + g_5^{\prime 2}} \left\{ 1
    - \frac{m_W^2}{2M_{\rm KK}^2} \left[ \frac{2L(1+\beta)^2}{(2+\beta)(3+2\beta)}
    - 1 + \frac{1}{(2+\beta)^2} + \frac{1}{2L} \right] + \dots \right\} \,.
\end{equation}
Thirdly, one can relate the weak mixing angle to the precisely measured parameters $G_F$, $\alpha$ and $m_Z$, defining
\begin{equation}
   s_0^2\,c_0^2\equiv \frac{\pi\alpha}{\sqrt2 G_F m_Z^2} \,.
\end{equation}
This can be solved to give
\begin{equation}
   s_0^2 = \frac{g_5^{\prime 2}}{g_5^2 + g_5^{\prime 2}} \left\{ 1
    + \frac{m_W^2}{2M_{\rm KK}^2}\,\frac{1}{c_W^2-s_W^2} \left[ 
    s_W^2\,\frac{2L(1+\beta)^2}{(2+\beta)(3+2\beta)} 
    - 1 + \frac{1}{(2+\beta)^2} + \frac{1}{2L} \right] 
    + \dots \right\} \,.
\end{equation}
Note that either $m_W$ or $s_W^2$, and either $\alpha$ or $s_0^2$ can be considered independent observables in addition to $G_F$, $m_Z$ and $s_*^2$. As a final electroweak observable we consider the parameter $\rho_*$ defined via the structure of the low-energy effective frou-fermion Lagrangian ${\cal L}_{\rm eff}=-\frac{4G_F}{\sqrt2}\,\big[J_\mu^+ J^{-\mu}+\rho_* (J_3^\mu-s_*^2 J_Q^\mu)^2\big]$. This yields
\begin{equation}
   \rho_* = 1 + \frac{m_Z^2-m_W^2}{2M_{\rm KK}^2}\,
    \frac{2L(1+\beta)^2}{(2+\beta)(3+2\beta)} + \dots \,.
\end{equation}

We can now solve for $S$, $T$, $U$ using the first three relations in eq.~(3.13) in \cite{Peskin:1991sw}. In this way, we obtain
\begin{equation}
\label{SandTeqn}
   S = \frac{2\pi v^2}{M_{\rm KK}^2} \left[ 1 - \frac{1}{(2+\beta)^2} - \frac{1}{2L}
    \right] , \qquad
   T = \frac{\pi v^2}{2c_W^2 M_{\rm KK}^2}\,\frac{2L(1+\beta)^2}{(2+\beta)(3+2\beta)} ,
    \qquad
   U = 0 \,. ~~
\end{equation}
In the brane-Higgs limit $\beta\to\infty$, these results agree with the corresponding expressions derived in the literature \cite{Csaki:2002gy,Carena:2003fx,Agashe:2003zs,Carena:2004zn,Delgado:2007ne,Casagrande:2008hr,Fichet:2013ola} up to very small ${\cal O}(1/L)$ terms, which depend on how precisely one deals with non-oblique effects in the definition of $S,T,U$. Rather than using a definition in terms of self-energy functions, we have chosen to define these parameters in terms of a set of physical observables. Note the interesting fact that, compared with the brane-Higgs case, the $T$ parameter can be lowered by a factor 3 in the limit $\beta\to 0$, thereby softening the bound on $M_{\rm KK}$ implied by the electroweak precision data by about a factor $1/\sqrt{3}\approx 0.58$. This may be compared with the effect of implementing a custodial symmetry in the RS model by extending the gauge symmetry in the bulk \cite{Agashe:2003zs,Csaki:2003zu,Agashe:2006at}. This eliminates the large, $L$-enhanced corrections to the $T$ parameter, such that the dominant constraint from electroweak precision tests arises from the $S$ parameter. This mechanism reduces the bound on the KK mass scale by a factor of about 0.4 \cite{Malm:2013jia}, at the expense of significantly complicating the fermion and gauge sectors of the model. RS scenarios with a bulk Higgs thus provide a compelling alternative to RS models with a custodial protection mechanism \cite{Cabrer:2011vu,Cabrer:2011fb,Carmona:2011ib}.

\begin{figure}
\begin{center}
\includegraphics[width=0.6\textwidth]{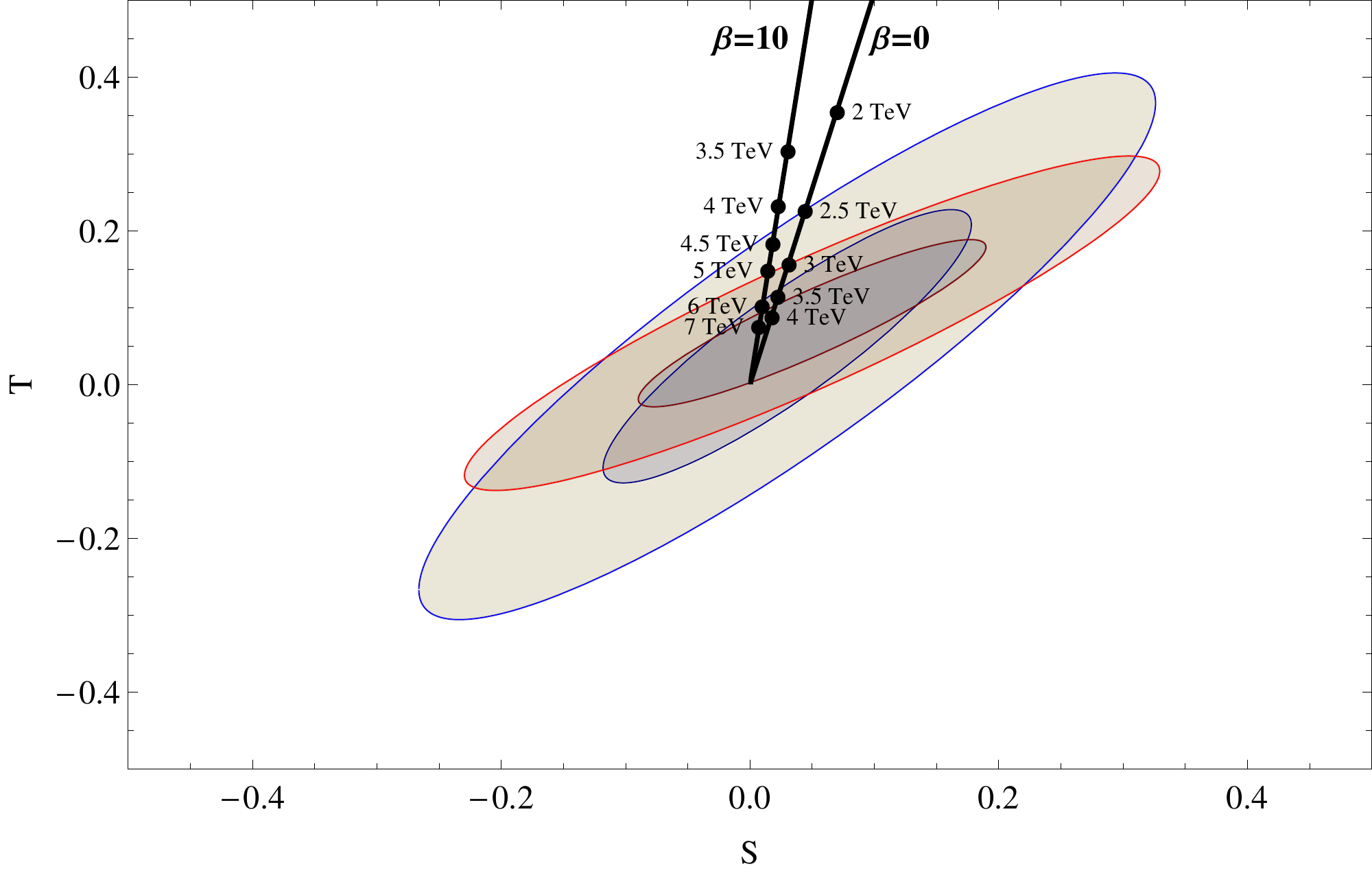}
\caption{\label{fig:STUellipses} 
68\% and 95\% confidence-level contours in the $S$--$T$ plane derived from a recent fit to electroweak precision data using $m_H=126$\,GeV and $m_t=173$\,GeV \cite{Baak:2012kk}. The blue ellipses show the result of a three-parameter fit with floating $U$, while the red ellipses are obtain with the constraint $U=0$. From the black lines one can extract the resulting lower bounds on the KK mass scale $M_{\rm KK}$ for different values of $\beta$.}
\end{center}
\end{figure}

In figure~\ref{fig:STUellipses} we illustrate the current constrains on the KK mass scale implied by the most recent analyses of electroweak precision data performed by the Gfitter group \cite{Baak:2012kk}. The lower bounds obtained depend on whether the fit is performed under the assumption that $U=0$ (red ellipses) or as a three-parameter fit (blue ellipses). Since at tree level in the RS model we find that $U=0$, it is likely that also in higher orders the parameter $U$ remains much smaller than $S$ and $T$, in which case performing a constrained fit might be more appropriate. We then obtain lower bounds varying between $M_{\rm KK}>3.0$\,TeV for $\beta=0$ to $M_{\rm KK}>5.1$\,TeV for $\beta=10$, both at 95\% confidence level. With an unconstrained fit, these bounds would be relaxed to $M_{\rm KK}>2.5$\,TeV and $M_{\rm KK}>4.3$\,TeV, respectively. For significantly larger values of $\beta$, the lower bounds rapidly tend towards the brane localised Higgs limits. For example, when $\beta=100$ these bounds increase to $M_{\rm KK}>5.5$\,TeV when $U$ is constrained to zero, and $M_{\rm KK}>4.6$\,TeV otherwise.

\subsection{Effective 4D Yukawa Couplings and Loss of Perturbative Control}
\label{sect:YukawaCouplings}

We are now in a position to evaluate the effective Yukawa couplings (\ref{YukawaMassDef}) for the fermion zero modes in an AdS${}_5$ space, which determine the masses of the light SM quarks. The equations of motion (\ref{FermProfEom}) imply for the properly normalised zero-mode profiles (with $q=u,d$)
\begin{equation}
   f_0^{(Q_L^i)}(r) = \sqrt{\frac{1-2c_Q^i}{R'\big(1-\Omega^{-1+2c_Q^i}\big)}}
    \left( \frac{r}{R'} \right)^{-c_Q^i} , \qquad
   f_0^{(q_R^i)}(r) = \sqrt{\frac{1+2c_q^i}{R'\big(1-\Omega^{-1-2c_q^i}\big)}}
    \left( \frac{r}{R'} \right)^{c_q^i} , 
\end{equation}
and when combined with the profile of the Higgs VEV from (\ref{hprofile}) one obtains
\begin{equation}\label{Y00LR}
   Y_{Q_L^i q_R^j}^{(0,0)} = \frac{\big(Y_q^{5D}\big)_{ij}}{\sqrt{R}} 
    \sqrt{\frac{2(1+\beta)(1-2c_Q^i)(1+2c_q^j)}{(\Omega^{1-2c_Q^i}-1)(\Omega^{1+2c_q^j}-1)}}\,
    \frac{\Omega^{1-c_Q^i+c_q^j}-\Omega^{-1-\beta}}{2+\beta-c_Q^i+c_q^j} \,.
\end{equation}
This expression has two interesting features. Firstly, when the fermion profiles are heavily peaked towards the UV brane (i.e.\ $1-c_Q^i+c_q^j\ll 0$), the fermion zero modes acquire a minimum Dirac mass of order $\tilde{v}\,\Omega^{-1-\beta}$. As has been discussed in \cite{Archer:2012qa, Agashe:2008fe, vonGersdorff:2012tt}, this might provide a natural framework for explaining the scale of neutrinos masses. More relevant for our discussion  is the situation where the bulk mass parameters are not too far away from the values $c_Q^i\approx\frac12$ and $c_q^j\approx-\frac12$ required to reproduce a realistic spectrum of quark masses. In this case we may approximate 
\begin{equation}\label{bulkYq}
   Y_{Q_L^i q_R^j}^{(0,0)} \approx \frac{\big(Y_q^{5D}\big)_{ij}}{\sqrt{R}}\,\sqrt{\frac{2}{1+\beta}}\,
    \sqrt{\frac{(1-2c_Q^i)(1+2c_q^j)}%
     {\big(1-\Omega^{-1+2c_Q^i}\big)\big(1-\Omega^{-1-2c_q^j}\big)}} \,.
\end{equation}
This result may be compared with the corresponding expression 
\begin{equation}\label{braneYq}
   Y_{Q_L^i q_R^j}^{(0,0)} \approx \big(Y_q\big)_{ij}\,
    \sqrt{\frac{(1-2c_Q^i)(1+2c_q^j)}%
     {\big(1-\Omega^{-1+2c_Q^i}\big)\big(1-\Omega^{-1-2c_q^j}\big)}}
\end{equation}
holding in RS models where the scalar sector is localised on the IR brane \cite{Grossman:1999ra,Gherghetta:2000qt,Huber:2003tu,Agashe:2004cp}, where the rescaled 5D Yukawa matrices
\begin{equation}\label{YqYstbrane}
   \mathbf{Y}_q\equiv \frac{\mathbf{Y}_{q,\rm brane}^{5D}}{\sqrt{2}R} 
\end{equation}
are dimensionless, and it is natural to expect that their entries should be of ${\cal O}(1)$. Matching expressions (\ref{bulkYq}) and (\ref{braneYq}), we now identify in the bulk Higgs model\footnote{Relations (\ref{YqYstbrane}) and (\ref{ydimlesb}) differ from corresponding expressions in (B.33) of \cite{Malm:2013jia} by a factor $\sqrt2$, which arises since in the current paper we consider an RS model on an interval rather than an $S^1/Z_2$ orbifold.}
\begin{equation}\label{ydimlesb}
	\mathbf{Y}_q^{5D} = \sqrt{\frac{R(1+\beta)}{2}}\,\mathbf{Y}_q
	= \sqrt{\frac{1+\beta}{4R}}\,\mathbf{Y}_{q,\rm brane}^{5D} \,.
\end{equation}
Hence, in the limit of large $\beta$ the original 5D Yukawa couplings $\mathbf{Y}_q^{5D}$ in a bulk Higgs model must be increased proportional to $\sqrt{1+\beta}$ so as to obtain the correct fermion mass spectrum \cite{Azatov:2009na,Malm:2013jia}. It can be checked that the effective 4D Yukawa couplings of the higher KK modes scale in a similar way as the zero-mode couplings $Y_{Q_L^i q_R^j}^{(0,0)}$ in (\ref{bulkYq}) and (\ref{braneYq}). 

Following much of the literature on RS models (see for example \cite{Huber:2003tu,Agashe:2004cp,Blanke:2008zb,Casagrande:2008hr}), we shall consider anarchic Yukawa couplings rather than considering flavour symmetries. The fermion hierarchies are then explained by the structures under the square root in (\ref{bulkYq}) and (\ref{braneYq}), which for ${\cal O}(1)$ differences of the bulk mass parameters $c_Q^i$ and $c_q^j$ exhibit exponential hierarchies. Concretely, we shall consider random 5D Yukawa couplings, such that 
\begin{equation}\label{eqn:YstarDef}
   \big|\big(Y_q\big)_{ij}\big| \le y_\ast
\end{equation}
for each entry of the complex, dimensionless Yukawa matrices $\mathbf{Y}_u$, $\mathbf{Y}_d$ and $\mathbf{Y}_e$ in (\ref{ydimlesb}). It has been shown in \cite{Azatov:2010pf,Goertz:2011hj,Carena:2012fk,Malm:2013jia} that the results for the $gg\to H$ and $H\to\gamma\gamma$ amplitudes are very sensitive to the value of the parameter $y_\ast$. Indeed, the dominant corrections from KK fermion contributions involve the traces
\begin{equation}
\label{eqn:TrYY}
   \frac{\tilde v^2}{2M_{\rm KK}^2}\,\mbox{Tr}\,\big(\mathbf{Y}_q \mathbf{Y}_q^\dagger\big)
   \approx \frac{\tilde v^2}{2M_{\rm KK}^2}\,\frac{N_g^2\,y_\ast^2}{2} \,,
\end{equation}
where $N_g=3$ is the number of fermion generations. The expression on the left-hand side is a sum over nine absolute squares of independent random complex numbers subject to the condition (\ref{eqn:YstarDef}), which by the central limit theorem obeys a gaussian distribution around the central value shown on the right.

The question how large $y_\ast$ can be is of particular relevance to the stringent constraints from flavour physics \cite{Csaki:2008zd, Blanke:2008zb, Bauer:2009cf}. In particular, it is well known that flavour-changing neutral currents are suppressed in RS-like scenarios. Such a suppression requires that the fermion zero modes are sufficiently peaked towards the UV brane. However, in order to obtain the correct quark masses, typically one finds that the top- and bottom-quark profiles need to be slightly peaked towards the IR brane. By increasing the size of $y_\ast$, it is possible to shift all of the fermion profiles slightly further towards the UV brane and thereby weaken the constraints derived from the flavour sector \cite{Agashe:2008uz,Archer:2011bk,Cabrer:2011qb}.

On the other hand, if $y_\ast$ is too large, we will loose perturbative control of the theory at energy scales below $M_{\mathrm{KK}}$. Following \cite{Csaki:2008zd}, one can use naive dimensional analysis to estimate when this occurs. To this end, one estimates the size of the one-loop correction to the Yukawa couplings. In RS models where the scalar sector is localised on (or very near) the IR brane, one finds that the relevant one-loop graphs diverge quadratically in the effective UV cutoff near the IR brane, $\Lambda_{\rm TeV}\equiv\Lambda_{\rm UV}(R')$. The masses of the KK modes are given by multiples of $M_{\rm KK}$; for example, the low-lying KK gluon states have masses $2.45\,M_{\rm KK}$, $5.57\,M_{\rm KK}$, $8.70\,M_{\rm KK}$, etc.\ \cite{Davoudiasl:1999tf,Gherghetta:2000qt}. We shall assume somewhat arbitrarily that at least three KK levels lie below the cutoff, such that $\Lambda_{\rm TeV}\gtrsim 8.7 M_{\rm KK}$. In RS models with a bulk Higgs, on the other hand, the divergence is only linear. Accounting carefully for phase-space factors and the number of fermion generations, one obtains the perturbativity bounds $y_\ast<y_{\rm max}$ with \cite{Malm:2013jia}
\begin{equation}\label{pertbound}
   y_{\rm max} \sim \left\{ 
    \begin{array}{cl}
     \displaystyle
     \frac{6\pi^2}{\sqrt 5}\,\frac{M_{\rm KK}}{\Lambda_{\rm TeV}}
      \sim 3.0 \,; & \mbox{brane Higgs,} \\[4mm]
     \displaystyle
     \sqrt{\frac{96\pi^3}{5}}\,\sqrt{\frac{M_{\rm KK}}{(1+\beta)\Lambda_{\rm TeV}}}
      \sim \frac{8.3}{\sqrt{1+\beta}} \,; \quad & \mbox{bulk Higgs.} 
    \end{array} \right.
\end{equation}
The transition between the two regimes occurs at $\beta\sim\Lambda_{\rm TeV}/M_{\rm KK}$. In RS models in which one considers Yukawa couplings in the bulk as well as one the branes, the perturbativity bound may fall in between the two values shown above.

In most of our phenomenological analysis we shall assume that $y_\ast$ in an ${\cal O}(1)$ parameter, which lies below the perturbativity bound and is independent of $\beta$, as suggested by the structure of (\ref{braneYq}). In particular, we will consider the three representative values $y_\ast=1$, 2 and 3. We emphasise that there is no firm theoretical reason why $y_\ast$ should be near the perturbativity bound in (\ref{pertbound}). However, for values of $y_\ast\lesssim 1$ it becomes increasingly difficult to fit to the top-quark mass.

\section{Phenomenology}
\label{sect:phenom}

We now bring together many of the results of the previous sections in order to numerically evaluate the size of any new physics effects in the Higgs sector arising in warped extra-dimension models based on AdS${}_5$. The main observables measured at the LHC are the Higgs signal strengths into various final states $X$, defined as
\begin{equation}\label{Hrates}
   \sigma_{pp\to H}\,\mbox{BR}_{H\to X}
   = \sigma_{pp\to H}\,\frac{\Gamma_{H\to X}}{\Gamma_{H,\,\rm tot}} \,.
\end{equation}
In the production process, one distinguishes between Higgs production in gluon fusion on the one hand (plus a tiny contribution from Higgs production in association with a $t\bar t$ pair, which we shall neglect), and Higgs production in weak vector-boson fusion or in associated production with a $W$ or $Z$ boson on the other. New physics effects can enter in any one of the three quantities on the right-hand side of (\ref{Hrates}). The resulting corrections to the $gg\to H$ production process will be discussed in detail in section~\ref{subsec:ggHgaga}. For the weak vector-boson fusion and associated production processes, we shall assume that the leading corrections to the cross section are those to the $HWW$ coupling, and hence (see \cite{Malm:2014gha} for a detailed discussion)
\begin{equation}\label{eq5.2}
   \frac{\sigma_{{\rm VBF}+V\!H}}{\sigma_{{\rm VBF}+V\!H}^{\rm(SM)}}
   \approx \left( \frac{v g_{HWW}^{(0,0)}}{2m_W^2} \right)^2
   \approx \frac{\Gamma_{H\to WW^*}}{\Gamma_{H\to WW^*}^{\rm(SM)}} \,.
\end{equation}

Next, one defines the normalised signal strengths as the Higgs signal strengths (\ref{Hrates}) normalised to their SM values. With the approximations just described, this yields
\begin{equation}\label{GGfusionSigApp}
   \mu_{\mathrm{ggF}+t\bar t H}^{H\to X}
   \approx \frac{\sigma_{gg\to H}}%
                {\sigma_{gg\to H}^{(\mathrm{SM})}}\,
     \frac{\Gamma_{H\to X}}{\Gamma_{H\to X}^{(\mathrm{SM})}}\,
     \frac{\Gamma_{H,\,\rm tot}^{(\mathrm{SM})}}{\Gamma_{H,\,\mathrm{tot}}} \,,
\end{equation}
and 
\begin{equation}\label{VBFSigApp}
   \mu_{\mathrm{VBF}+V\!H}^{H\to X}
   \approx \frac{\Gamma_{H\to WW^*}}{\Gamma_{H\to WW^*}^{(\mathrm{SM})}}\,
   \frac{\Gamma_{H\to X}}{\Gamma_{H\to X}^{(\mathrm{SM})}}\,
   \frac{\Gamma_{H,\,\rm tot}^{(\mathrm{SM})}}{\Gamma_{H,\,\mathrm{tot}}} \,. 
\end{equation}
For a Higgs boson with mass 125.5\,GeV, one has \cite{Denner:2011mq}
\begin{equation}\label{totrate}
   \frac{\Gamma_{H,\,\mathrm{tot}}}{\Gamma_{H,\,\mathrm{tot}}^{(\mathrm{SM})}}
   \approx 0.57\,\frac{\Gamma_{H\to b\bar b}}{\Gamma_{H\to b\bar b}^{(\mathrm{SM})}}
    + 0.22\,\frac{\Gamma_{H\to WW^*}}{\Gamma_{H\to WW^*}^{(\mathrm{SM})}}
    + 0.03\,\frac{\Gamma_{H\to ZZ^*}}{\Gamma_{H\to ZZ^*}^{(\mathrm{SM})}}
    + 0.09\,\frac{\Gamma_{H\to gg}}{\Gamma_{H\to gg}^{(\mathrm{SM})}}
    + 0.06\,\frac{\Gamma_{H\to\tau\tau}}{\Gamma_{H\to\tau\tau}^{(\mathrm{SM})}}
    + 0.03 \,,
\end{equation}
where we are neglecting possible new physics effects in the decays $H\to c\bar c,Z\gamma$ and even rarer modes. In the SM the combined contribution of these channels is about~0.03.

It is obvious from the structure of the above relations that it is instructive to consider the gluon-fusion cross section $\sigma_{gg\to H}$ and the individual Higgs decay rates $\Gamma_{H\to X}$, both normalised to their SM values, before making predictions for the various signal strength. We will do so in sections~\ref{subsec:WWZZ}--\ref{subsec:Hff}. However, we should first make some comments concerning the scan over the RS parameter space employed in our numerical analysis.

\subsection{Numerical Analysis}
\label{subsec:numanal}

This model, like most BSM scenarios, suffers from a large and under-constrained parameter space. In particular, the geometry is specified by two free parameters, the KK mass scale $M_{\mathrm{KK}}$ and the warp-factor $\Omega=10^{15}$. One could trade the scale $M_{\rm KK}$ for the mass of one of the KK resonances. For example, independently of the details of the localization of the scalar sector and the choice of the electroweak gauge group, the lightest KK gluon or photon states have mass $M_{g^{(1)}}=M_{\gamma^{(1)}}\approx 2.45 M_{\rm KK}$ \cite{Davoudiasl:1999tf}. The Higgs sector has four parameters, of which two are fixed by the Higgs mass and the electroweak scale $v$. A third parameter related to the potential on the UV brane has a negligible impact on weak-scale physics. This leaves one free parameter~$\beta$. For the fermions every $\mathrm{SU}_L(2)$ singlet and doublet has an associated bulk mass parameter, and every Yukawa coupling consists of a complex $3\times 3$ matrix. These must be constrained to yield acceptable results for the SM fermion masses and mixing matrices.\footnote{We do not consider neutrino masses or the PMNS matrix in our analysis, since this would require the specification of the neutrino sector, which is both model dependent and of little relevance to Higgs physics.}  
To be explicit, for a given value of $\beta$ and $M_{\mathrm{KK}}$, we perform a $\chi^2$ minimisation starting from a random point, with anarchic Yukawa matrix elements chosen as complex random numbers subject to the condition (\ref{eqn:YstarDef}), and bulk mass parameters in the intervals
\begin{align}
   c_Q^i &\in \left(\left[0.6, 0.66\right],\left[0.52, 0.62\right],
    \left[0, 0.66\right]\right) , \nonumber\\
   c_L^i &\in \left(\left[0.6, 0.76\right],\left[0.52, 0.72\right],
    \left[0.4, 0.64\right]\right) , \nonumber\\
   c_u^i &\in \left(\left[-0.72, -0.64\right],\left[-0.63, -0.39\right], 
	\left[-0.5, 1.5\right]\right) , \nonumber\\
   c_d^i &\in \left(\left[-0.68, -0.62\right],\left[-0.65, -0.59\right],
    \left[-0.61, -0.55\right]\right) , \nonumber\\
   c_e^i &\in \left(\left[-0.88, -0.62\right], \left[-0.73, -0.59\right],
    \left[-0.64, -0.55\right]\right) .
\end{align}
We reject all points deviating from the SM by more than 1$\sigma$ (i.e.\ for the quark sector, points with $\chi^2/\mathrm{d.o.f.}>11.5/10$). 

\begin{figure}
\begin{center} 
\subfigure[]{\label{fig:HGGvcQ3}
\includegraphics[width=0.47\textwidth]{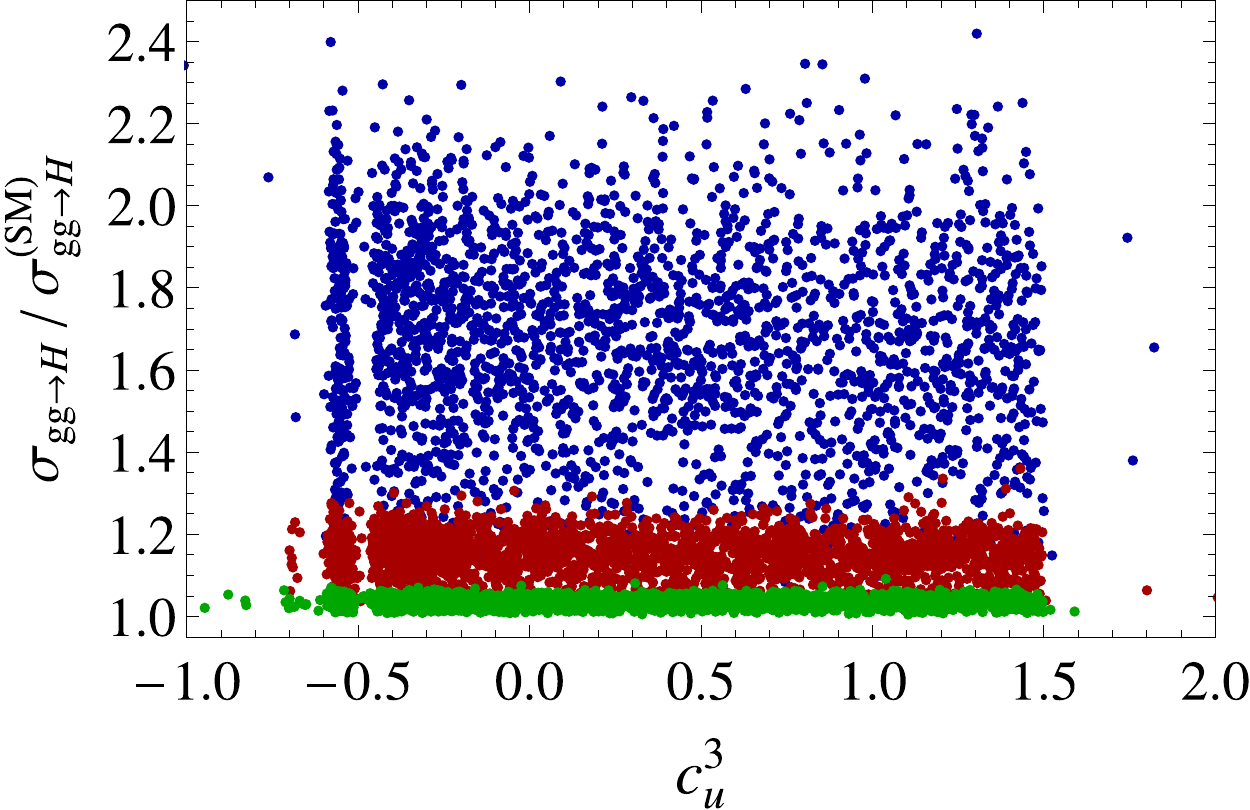}}
\subfigure[]{\label{fig:HYYvcQ3}
\includegraphics[width=0.47\textwidth]{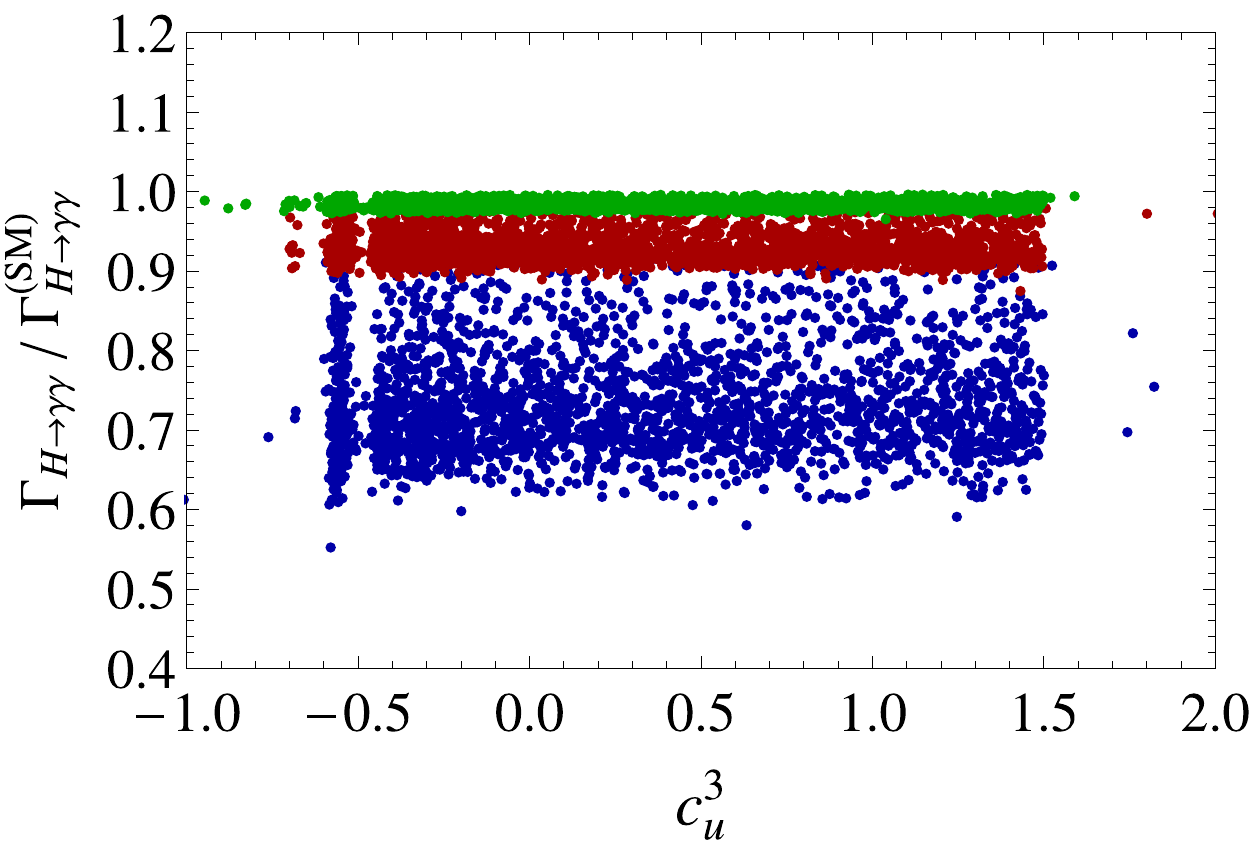}} \\
\subfigure[]{\label{fig:HGGvcbeta}
\includegraphics[width=0.47\textwidth]{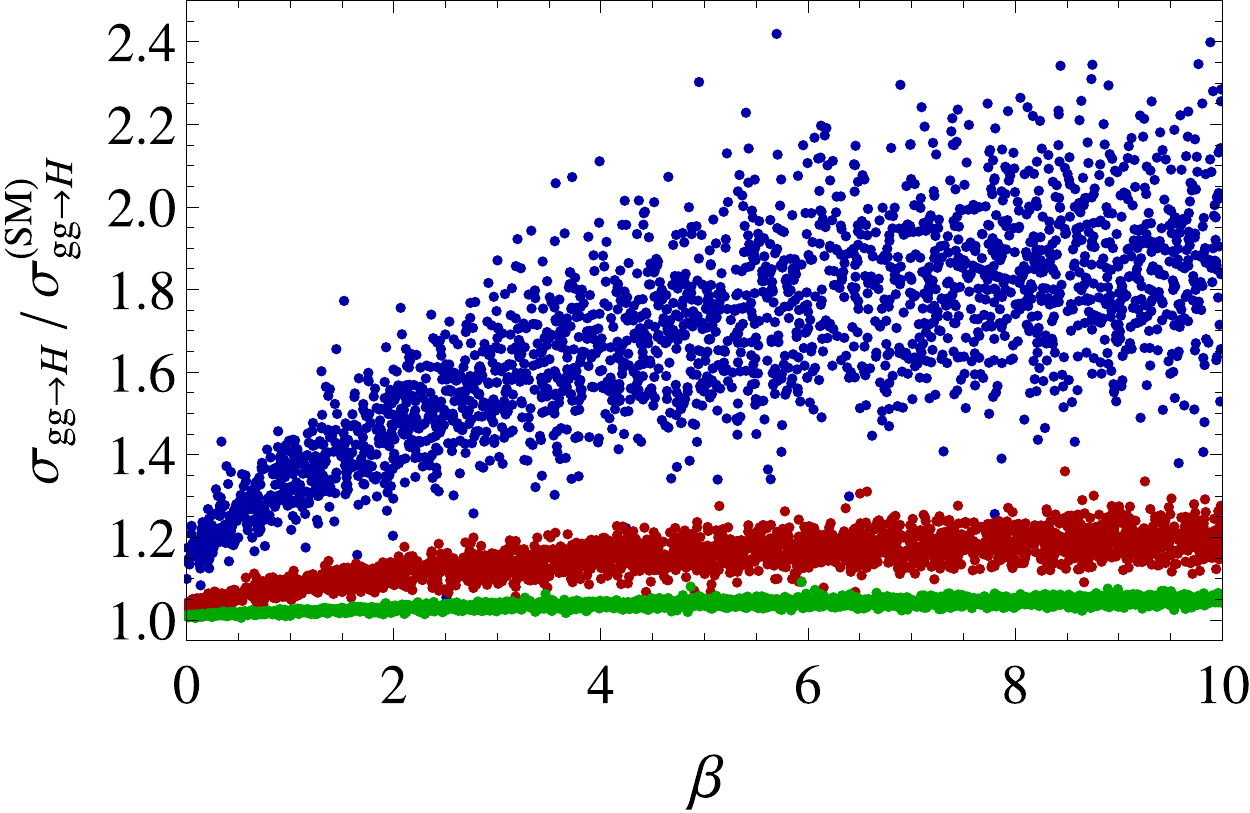}}
\subfigure[]{\label{fig:HYYvcbeta}
\includegraphics[width=0.47\textwidth]{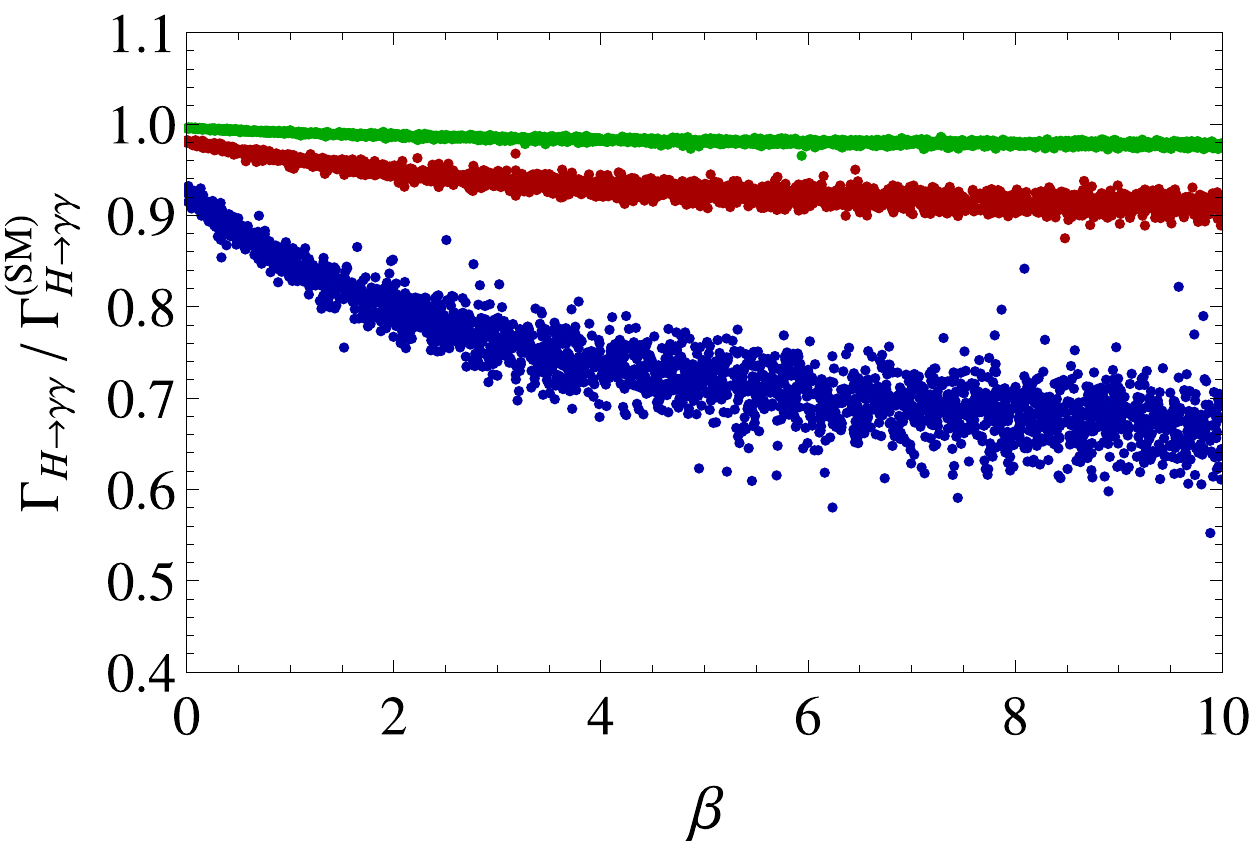}} 
\caption{\label{fig:IndependenceWRTbetaAndc} 
Dependence of the gluon-fusion cross section (left) and the $H\to\gamma\gamma$ decay rate (right) on one of the bulk mass parameters (top row) and on the parameter $\beta$ (bottom row). We use $y_\ast=3$, while $M_{\mathrm{KK}}=2$\,TeV (blue), 4\,TeV (red) and 8\,TeV (green). The regions with missing data points around $c_u^3=-0.5$ in the upper plots are artifacts of our scanning procedure.}
\end{center}
\end{figure} 

With such a large parameter space one may wonder if it is still possible to make meaningful predictions. As can be seen from (\ref{Y00LR}), the SM fermion masses and mixing angles are exponentially sensitive to the bulk mass parameters, but only linearly sensitive to the 5D Yukawa couplings. However, the dependence on the bulk mass parameters largely drops out of the expressions of relevance to Higgs physics \cite{Casagrande:2010si,Azatov:2010pf,Goertz:2011hj,Carena:2012fk,Frank:2013un,Malm:2013jia}. This fact is illustrated in figures~\ref{fig:HGGvcQ3} and \ref{fig:HYYvcQ3}, where we show the dependence of the $gg\to H$ cross section and the $H\to\gamma\gamma$ decay rate on the third-generation bulk mass parameter $c_u^3$ as an example. This is an important result, indicating that the modifications to Higgs physics are largely independent of the degree of compositeness of the fermions. The physical reason is that the main effects are due to KK resonances, whose profiles are rather insensitive to the bulk mass parameters. Further still, particularly in the large $\beta$ limit, the loop-induced fermion contributions to Higgs physics are dominated by expressions proportional to $\mathrm{Tr}(\mathbf{Y}_q \mathbf{Y}_q^\dagger)$ \cite{Azatov:2010pf,Goertz:2011hj,Carena:2012fk,Malm:2013jia}, and hence become insensitive to the magnitudes of the CP-violating phases in the Yukawa matrices. Indeed, according to (\ref{eqn:TrYY}) these traces are to good approximation given in terms of the single parameter $y_\ast$. So by assuming anarchic Yukawa couplings, the modification of the Higgs couplings will be largely dependent on just three parameters: $M_{\mathrm{KK}}$, $y_\ast$ and $\beta$. It is important in this context that we have defined the parameter $y_\ast$ in (\ref{ydimlesb}) and (\ref{eqn:YstarDef}) such that one can take the limit of large $\beta$ and still obtain the correct SM quark masses. In this way we ensure that for $\beta\gtrsim 10$ the predictions become approximately $\beta$ independent and match up with the results obtained in \cite{Azatov:2010pf,Frank:2013un,Malm:2013jia,Hahn:2013nza}. This can be seen in figures~\ref{fig:HGGvcbeta} and \ref{fig:HYYvcbeta}, where we show the new physics effects to the gluon-fusion cross section and the $H\to\gamma\gamma$ decay rate as a function of $\beta$. For the time being we do not include the constraints from electroweak precision tests, which would eliminate some points referring to small values of $M_{\rm KK}$ (see section~\ref{sec:EWpars}). For smaller values $\beta\lesssim 10$ the dependence on the Higgs scaling dimension can be significant. Indeed, for the two observables shown in the figure the new physics effects arising for a broad bulk Higgs are much smaller than those for a narrow one. In our analysis below, we will often consider the two representative values $\beta=1$ (broad Higgs profile) and $\beta=10$ (narrow Higgs profile). 

\subsection{\boldmath $H\to WW^*$ and $H\to ZZ^*$ Decays}
\label{subsec:WWZZ}

\begin{figure}
\begin{center}
\includegraphics[width=0.7\textwidth]{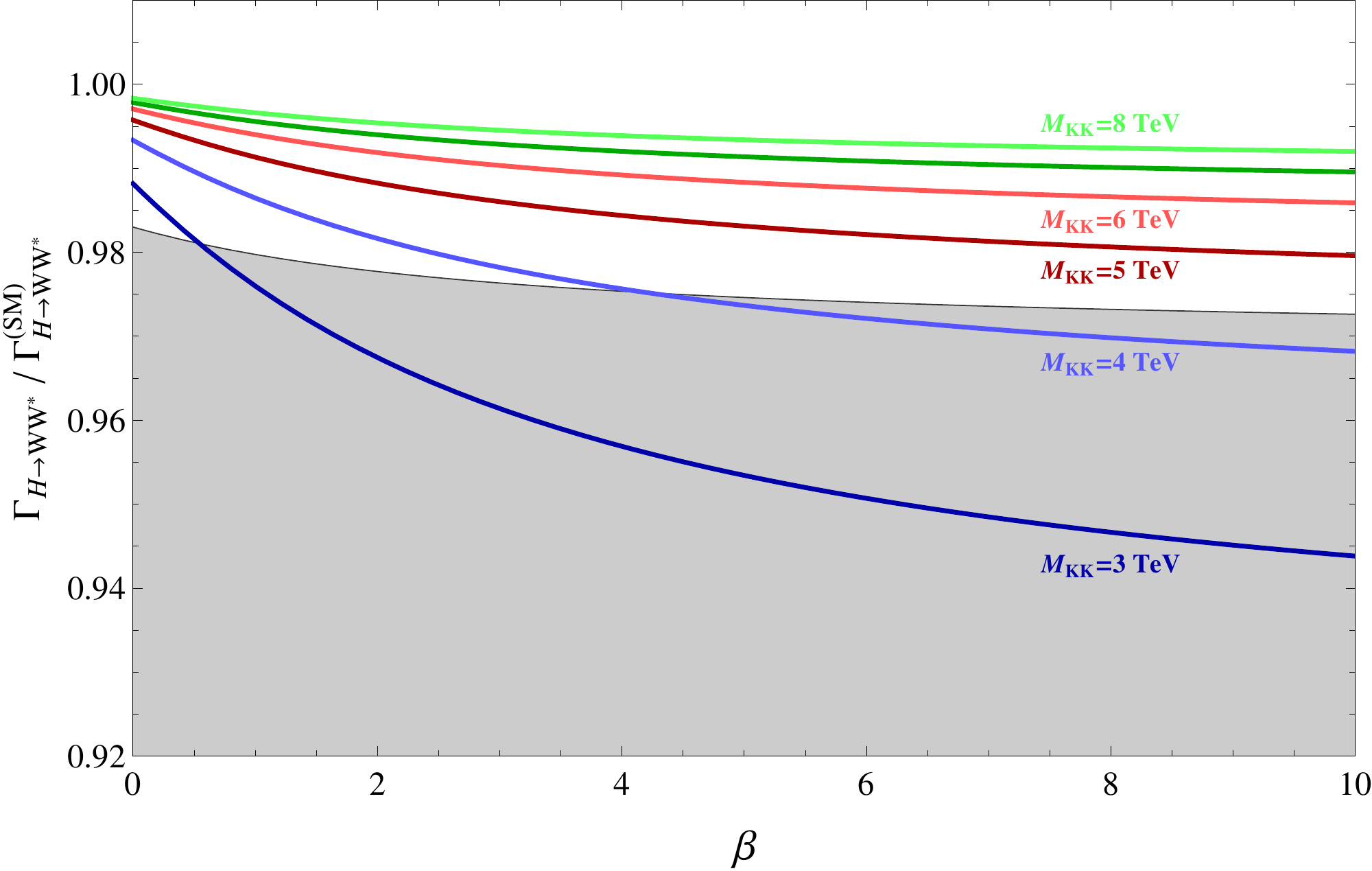}
\caption{\label{fig:HtoWW} 
Predictions for the ratio of the $H\to WW^*$ decay rate normalised to its SM value, for different values of the KK mass scale. The lightest KK gluon state has mass $M_{g^{(1)}}\approx 2.45\,M_{\rm KK}$. The analogous plot for $H\to ZZ^*$ decay looks very similar. The region in grey is excluded by the tree-level analysis of electroweak precision tests with $U$ not constrained to be zero.}  
\end{center}
\end{figure}

As was discussed in section \ref{sect:HtoWW}, in a warped extra dimension based on an AdS${}_5$ space the $H\to WW^*$ and $H\to ZZ^*$ decay rates are suppressed relative to the SM. Using the results of section~\ref{sect:AdS5Geo}, we can evaluate (\ref{HtoWWRw}) and (\ref{HtoWWPertApprox}) to find 
\begin{equation}\label{HtoWWDecayWidthApprox}
   \frac{\Gamma_{H\to WW^*}}{\Gamma_{H\to WW^*}^{\mathrm{(SM)}}}\approx 
    1 - \frac{m_W^2}{M_{\mathrm{KK}}^2} \left[ 
    \frac{3L(1+\beta)^2}{(2+\beta)(3+2\beta)} - 1 
    + \frac{1}{(2+\beta)^2} + \frac{1}{2L} \right] 
   \approx 1 - \frac{m_W^2}{\pi v^2} \left( 3 c_W^2 T - \frac{S}{2} \right) , 
\end{equation}  
which agrees with a corresponding result in \cite{Fichet:2013ola}. An analogous expression holds for $\Gamma_{H\to ZZ^*}$. This approximate result is in good numerical agreement with the exact result quoted in \cite{Malm:2014gha}. Note the close correlation between this result and the size of the corrections to the electroweak observables in (\ref{SandTeqn}). It is for this reason that the decay rate cannot deviate too far from its SM value.\footnote{We emphasise, however, that relation (\ref{HtoWWDecayWidthApprox}) no longer holds in RS models with custodial symmetry.} 
The size of the new physics correction is shown in figure~\ref{fig:HtoWW} as a function of $\beta$ and for different values of the KK mass scale. The grey-shaded region in the plot is excluded by the tree-level analysis of the $S$, $T$ and $U$ parameters at 95\% confidence level. When this constraint is taken into account, one finds that the $H\to WW^*$ and $H\to ZZ^*$ decay rates are reduced by at most 2--3\%.

\subsection{Higgs Decays to Fermions}
\label{subsec:Hff}

The previous discussion has shown that the Higgs couplings to $W$ and $Z$ bosons only receive moderate modifications in RS models. On the other hand, we find that for sizable values of $y_\ast$ significant corrections to the Higgs couplings to fermions can arise as a result of the mixing of the zero modes with the tower of KK resonances. For almost all choices of model parameters this results in a suppression of the effective Yukawa couplings and hence a reduction of the decay rates of the Higgs boson to two fermions relative to the rates in the SM. For the third-generation fermions, one finds the relation
\begin{equation}
   \frac{\Gamma_{H\to f\bar f}}{\Gamma_{H\to f\bar f}^{\rm (SM)}}
   = \frac{\big|\big(\tilde Y_f^{\rm mass}\big)_{33}\big|^2}{y_f^2} \,,
\end{equation}
where on the right-hand side $f=u,d,e$ in the numerator and $f=t,b,\tau$ in the denominator. Once again, we find that the size of this suppression is only mildly dependent on the degree of compositeness of the fermions. This is a consequence of assuming anarchic 5D Yukawa matrices of the same magnitude in the quark and lepton sectors. (We note in passing that the Higgs coupling to top quarks is reduced in a similar way, but the dependence on the degree of compositeness of the left-handed and right-handed top quarks plays a more relevant role in this case \cite{Malm:2014gha}.)

\begin{figure}
\begin{center} 
\includegraphics[height=4.7cm]{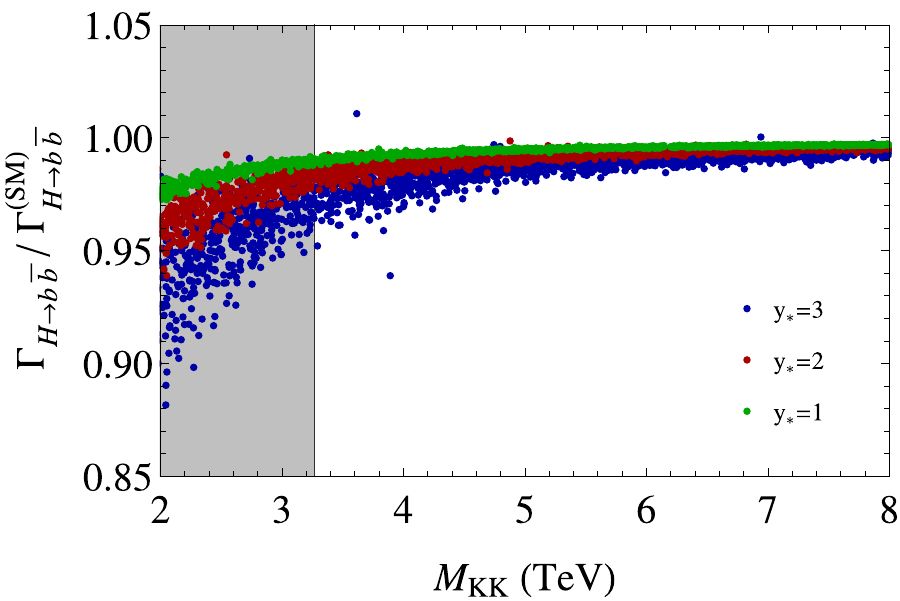} \hspace{1mm}
\includegraphics[height=4.7cm]{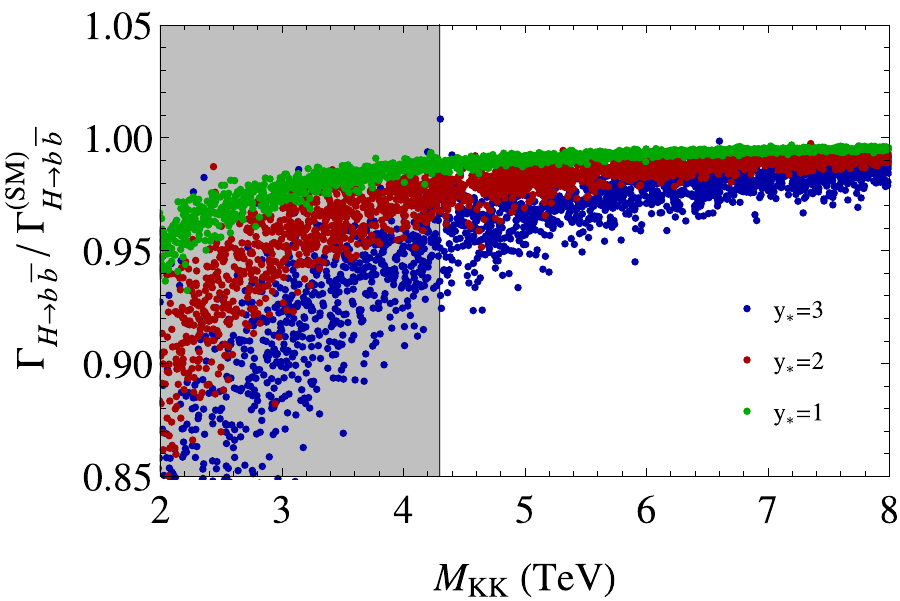}
\caption{\label{fig:HtoFermions}
Predictions for the $H\to b\bar b$ decay rate normalised to its SM value, as a function of $M_{\rm KK}$ and for different values of $y_\ast$. The left plot corresponds to a broad bulk Higgs with $\beta=1$, the right one to a narrow bulk Higgs with $\beta=10$. The grey-shaded are excluded by a leading-order analysis of electroweak precision observables.}
\end{center}
\end{figure}

In figure~\ref{fig:HtoFermions} we show the suppression of $H\to b\bar b$ decay rate as a function of the KK mass scale for different values of $y_\ast$ and $\beta$. The corresponding plots for the $H\to\tau^+\tau^-$ decay rate would look almost indistinguishable from the ones shown here. The left plot refers to $\beta=1$, the right one to $\beta=10$. The areas shaded in grey indicate the regions of parameter space excluded by the tree-level analysis of electroweak precision tests, as described in section~\ref{sec:EWpars}. The coloured sets of scatter points belong to three different values of $y_\ast$. To very good approximation the new physics effects scale like $y_\ast^2/M_{\rm KK}^2$. For a given value of this ratio, the corrections are significantly larger for the case of a very narrow Higgs profile ($\beta\gg 1$) than for a wide one ($\beta={\cal O}(1)$). However, one must also take into account that in the latter case significantly lower values of the KK mass scale are allowed by electroweak precision tests. When this is taken into account, we find that for $y_\ast=3$ the $H\to b\bar b$ decay rate can be reduced by up to about 5\% for $\beta=1$ and 8\% for $\beta=10$.

\subsection{\boldmath Higgs Production in Gluon Fusion and $H\to\gamma\gamma$ Decay}
\label{subsec:ggHgaga}

\begin{figure}
\begin{center} 
\subfigure[]{\label{fig:HGGvMKK}
 \includegraphics[height=4.7cm]{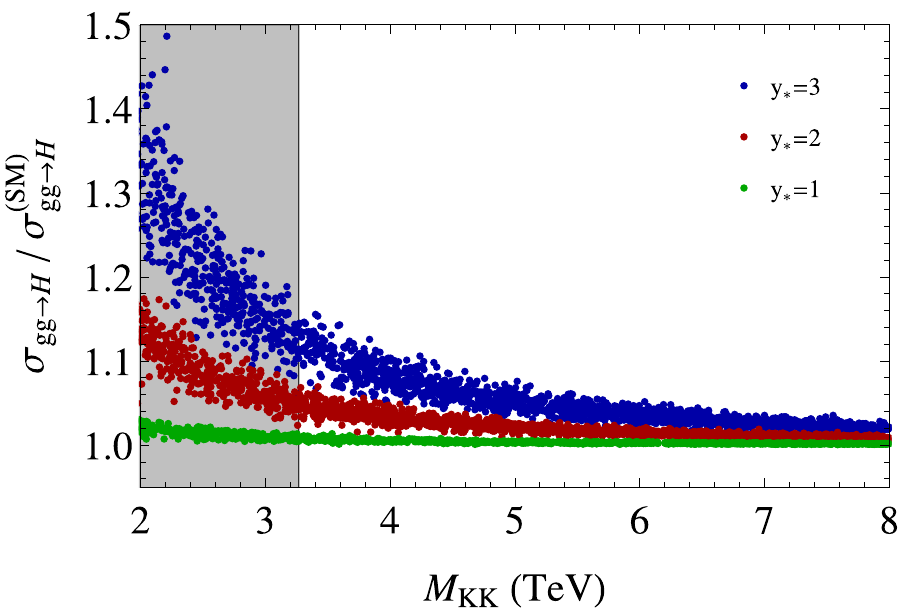} \hspace{3mm}
 \includegraphics[height=4.7cm]{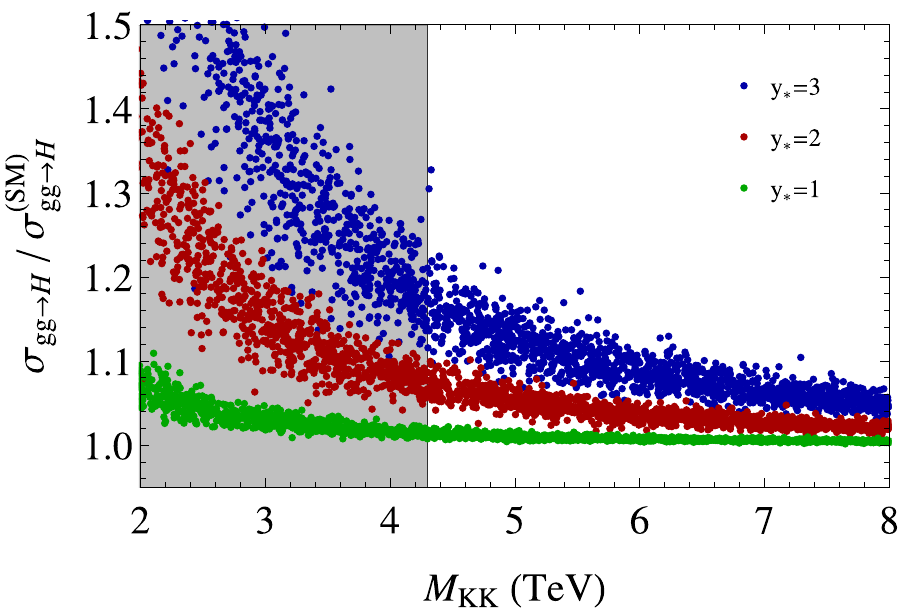}} \\
\subfigure[]{\label{fig:HGGvY}
 \includegraphics[height=4.7cm]{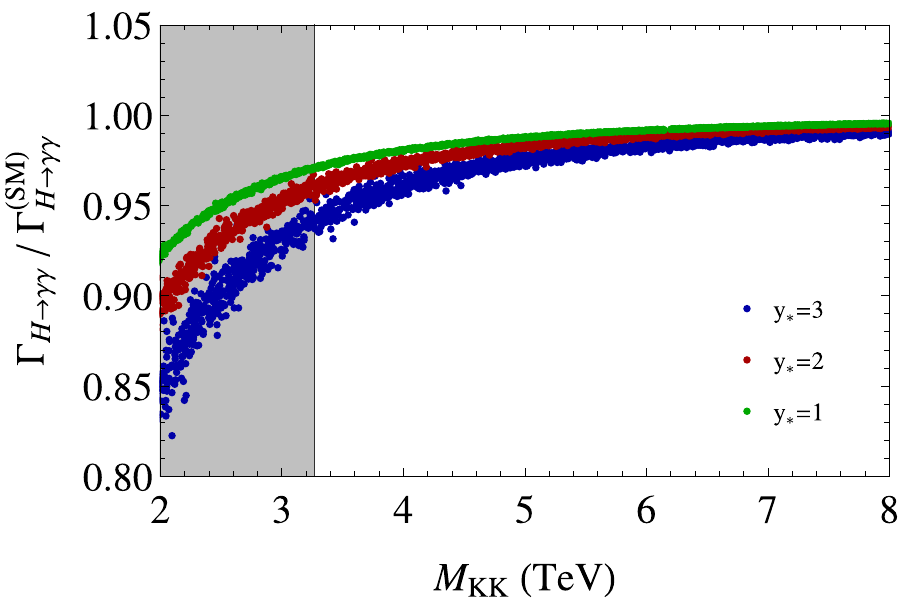} \hspace{0mm}
 \includegraphics[height=4.7cm]{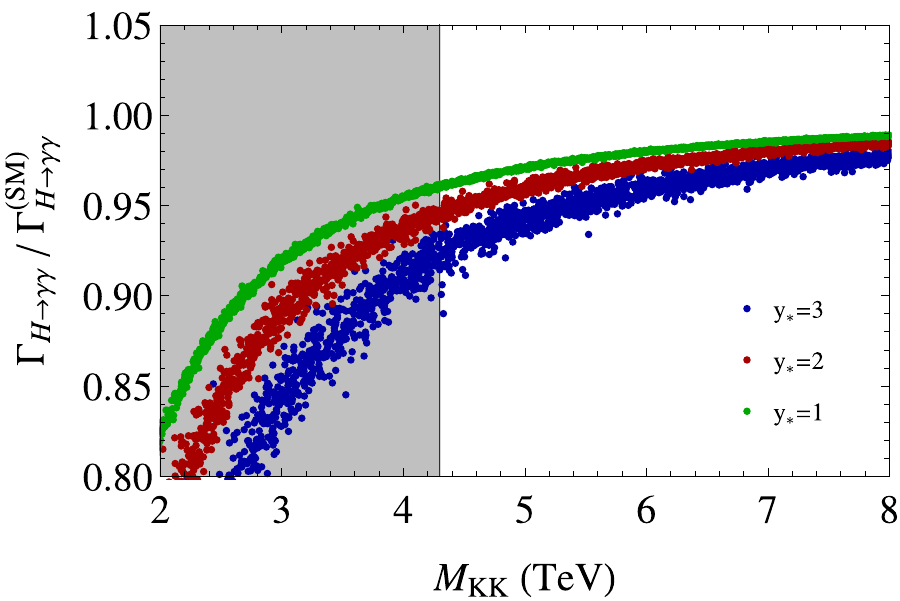}}
\caption{\label{fig:GGH_vs_Hyy}
Predictions for the gluon-fusion cross section (top) and the $H\to\gamma\gamma$ decay rate (bottom) normalised to their SM values, as functions of $M_{\rm KK}$ and for different values of $y_\ast$. The left plots correspond to a broad bulk Higgs with $\beta=1$, the right ones to a narrow bulk Higgs with $\beta=10$. The grey-shaded are excluded by a leading-order analysis of electroweak precision observables.}
\end{center}
\end{figure}  

The multitude of KK fermion resonances has a more profound impact on the loop amplitudes giving rise to Higgs production in gluon fusion and Higgs decay into two photons. Virtual KK resonances enter in these processes via the Feynman diagrams shown in figures~\ref{fig:GluonFusion} and \ref{fig:floop}. Their effects are two-fold in nature. Firstly, the top-quark zero mode mixes with the tower of KK resonances, and as mentioned earlier this results in a suppression in the effective top Yukawa coupling to the Higgs boson. Secondly, the full tower of KK fermion resonances also contributes in the loop and results in an enhancement of the amplitude. These two effects are of a similar order of magnitude ($\sim y_\ast^2\,\tilde v^2/M_{\rm KK}^2$), but the second one is generally more important. Indeed, according to (\ref{eqn:TrYY}) the contribution of the tower scales like $N_g^2$ (with $N_g=3$ the number of fermion generations), while the modification of the zero-mode contribution scales like $N_g$ \cite{Malm:2013jia,Hahn:2013nza}. As a result, we find an enhancement of the gluon-fusion amplitude. In the case of the $H\to\gamma\gamma$ decay amplitude the dominant contribution still comes from the $W$-boson loops (diagrams \ref{fig:Wloop} to \ref{fig:WGloop}), which interfere destructively with the fermion contribution. Hence an enhancement in the amplitude associated with diagram~\ref{fig:floop} results in a suppression of the $H\to\gamma\gamma$ decay rate. This suppression, as well as the correlated enhancement of the gluon-fusion cross section, is seen in figure~\ref{fig:GGH_vs_Hyy}, where we show the behaviour of these two observables (normalised to their SM values) as functions of the KK mass scale and for different values of $y_\ast$ and $\beta$. Once again, to good approximation the new physics effects scale like $y_\ast^2/M_{\rm KK}^2$, and for a given value of this ratio the corrections are significantly larger for the case of a very narrow Higgs profile than for a wide one. When the constraints on $M_{\rm KK}$ imposed by electroweak precision tests are taken into account, we find that for $y_\ast=3$ the gluon-fusion cross section can be enhanced by up to about 16\% for $\beta=1$ and 24\% for $\beta=10$. Likewise, the $H\to\gamma\gamma$ decay rate can be reduced by up to about 7\% for $\beta=1$ and up to 10\% for $\beta=10$. 

\subsection{Total Higgs Decay Width}
\label{subsec:Hwidth}

\begin{figure}
\begin{center} 
\includegraphics[height=4.7cm]{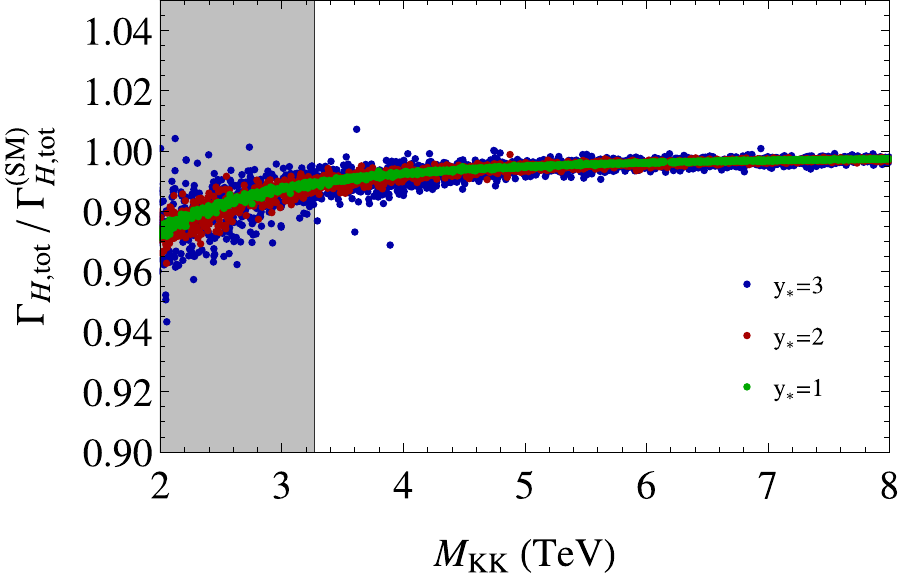} \hspace{1mm}
\includegraphics[height=4.7cm]{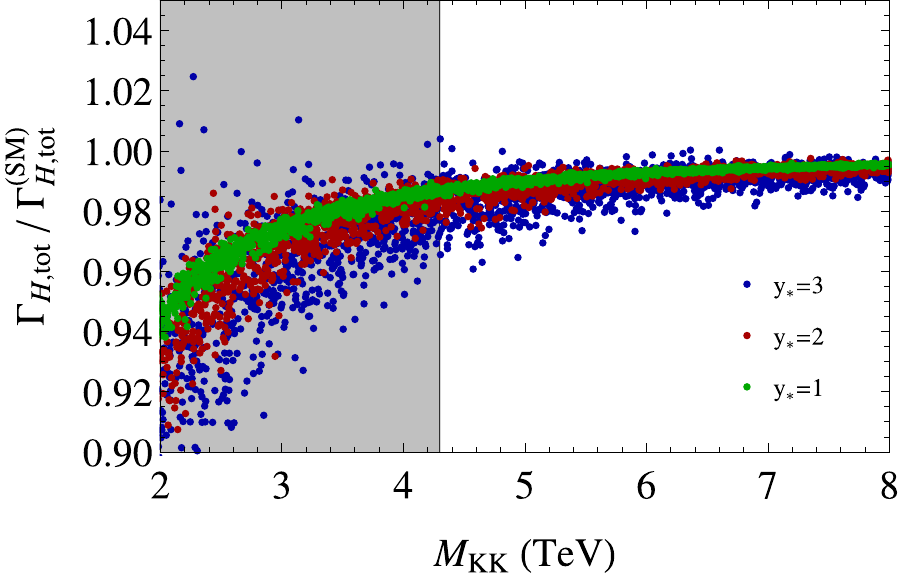}
\caption{\label{fig:RhovMKK} 
Predictions for the total Higgs width normalised to its SM value, as a function of $M_{\rm KK}$ and for different values of $y_\ast$. The left plot corresponds to a broad bulk Higgs with $\beta=1$, the right one to a narrow bulk Higgs with $\beta=10$. The grey-shaded are excluded by a leading-order analysis of electroweak precision observables.}
\end{center}
\end{figure}

Having studied the individual Higgs decay rates into pairs of gauge bosons and fermions, we now consider the total Higgs decay width given in (\ref{totrate}). With all the $H\to f\bar f$ and $H\to WW^*, ZZ^*$ decay rates suppressed and only the $H\to gg$ decay rate enhanced, we find that also the total decay width in our model is suppressed with regard to its value in the SM, but by a more modest amount than the dominant $H\to b\bar b$ decay rate. This is shown in figure~\ref{fig:RhovMKK}. Notice the peculiar behavior that increasing the value of $y_\ast$ only mildly enlarges the magnitude of the effects but mainly increases the spread of the scatter points. This is a consequence of the larger impact of the enhanced $H\to gg$ decay rate for larger values of $y_\ast$, which partially balances the effect of the suppressed $H\to b\bar b$ decay width.\footnote{In RS models with custodial symmetry, where the multiplicity of fermionic KK resonances is considerably larger, the enhancement of the $H\to gg$ rate dominates over the suppression of the $H\to b\bar b$ rate for large values of $y_\ast$, and this effect leads to an enhancement of the total Higgs width \cite{Malm:2013jia}.} 
The extent to which the total Higgs width can be suppressed is again limited by the lower bounds on the KK mass scale implied by electroweak precision tests, as indicated by the grey-shaded regions in the figure. When these bounds are taken into account, we find that the reduction can be at most about 3\% for $\beta=1$ and 4\% for $\beta=10$.
 
\subsection{LHC Phenomenology}

Having studied the individual Higgs decay rates, we now attempt to present our results in a form such that they can be compared with LHC measurements. In particular, it is instructive to study the correlated signal strengths, which are sensitive to the Higgs production and decay modes, as well as to the total decay width, see (\ref{Hrates}). The following analysis is based on two simplifying assumptions. Firstly, we assume that the corrections to the vector-boson fusion and associated production cross sections are equal and approximately given by the correction to the $HWW$ coupling. In practice, both processes receive additional (and different) corrections due to the tower of KK vector-boson resonances. As discussed in \cite{Malm:2014gha} for the large-$\beta$ limit, these additional corrections are subleading in $L$ and numerically insignificant. We expect this result to hold irrespectively of the value of $\beta$. Secondly, for the purpose of computing the total decay width, we assume that the impact of new physics effects on the strongly suppressed $H\to Z\gamma$ decay mode and on invisible Higgs decays is negligible. In particular, the absence of a signal in the current LHC data  supports the assumption of a small $H\to Z\gamma$ branching ratio, which in the SM is $1.54\times 10^{-3}$. 

\begin{figure}
\begin{center} 
\subfigure[$M_{\mathrm{KK}}=4$\,TeV (left) and $M_{\mathrm{KK}}=8$\,TeV (right) with $\beta=1$ (orange-brown) and $\beta=10$ (white-blue).]
{\label{fig:Corr_HYY_Yast}
 \includegraphics[width=0.42\textwidth]{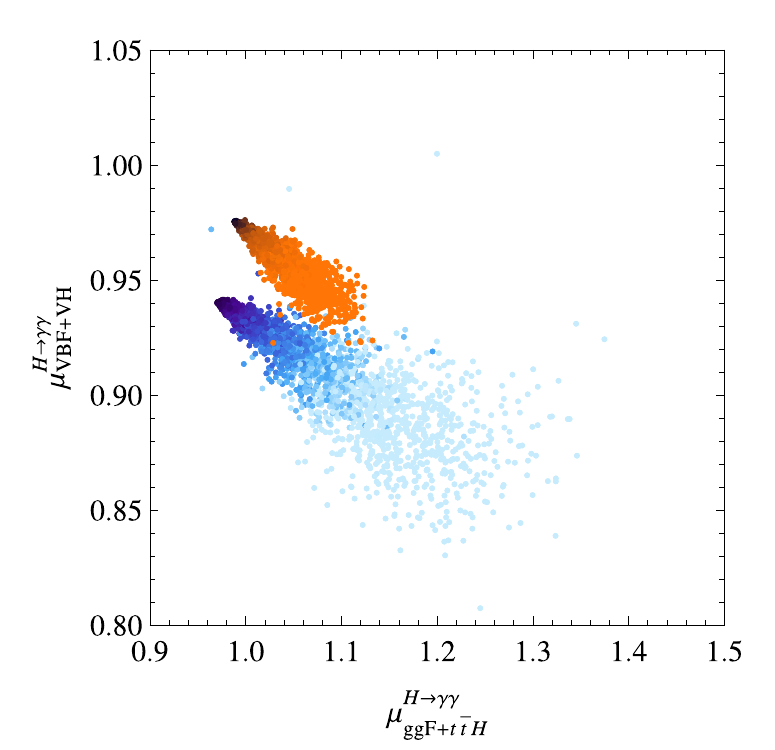}
 \includegraphics[width=0.42\textwidth]{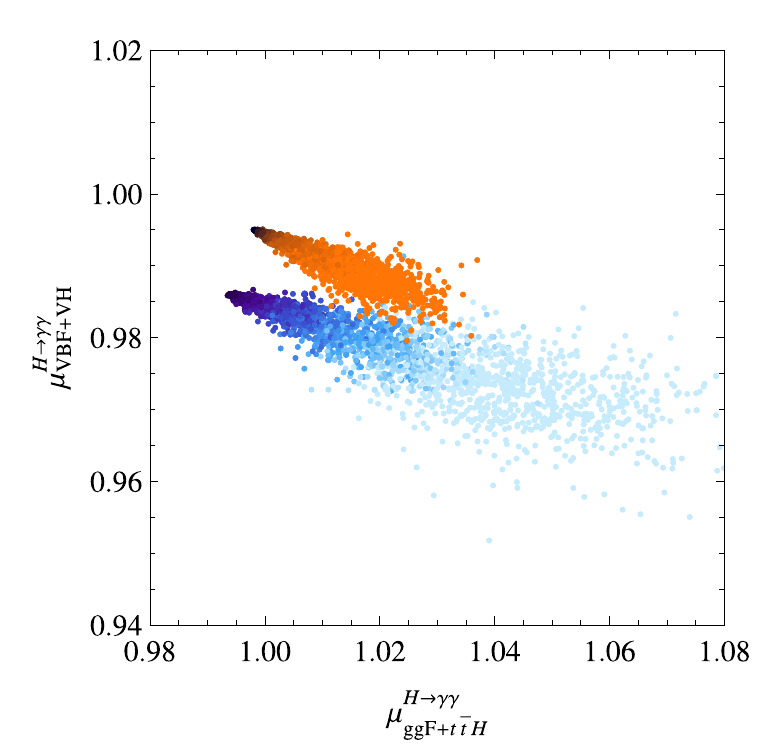}
 \includegraphics[width=0.0905\textwidth]{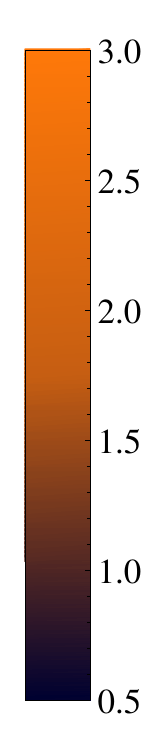}
 \includegraphics[width=0.132\textwidth]{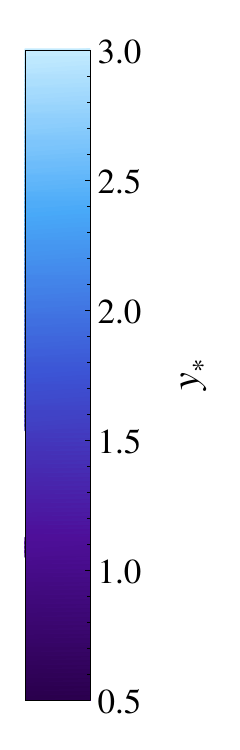}
}\\
\subfigure[$y_\ast=1$ (left) and $y_\ast=3$ (right) with $\beta=1$ (orange-brown) and $\beta=10$ (white-blue).]
{\label{fig:Corr_HYY_MKK}
 \includegraphics[width=0.42\textwidth]{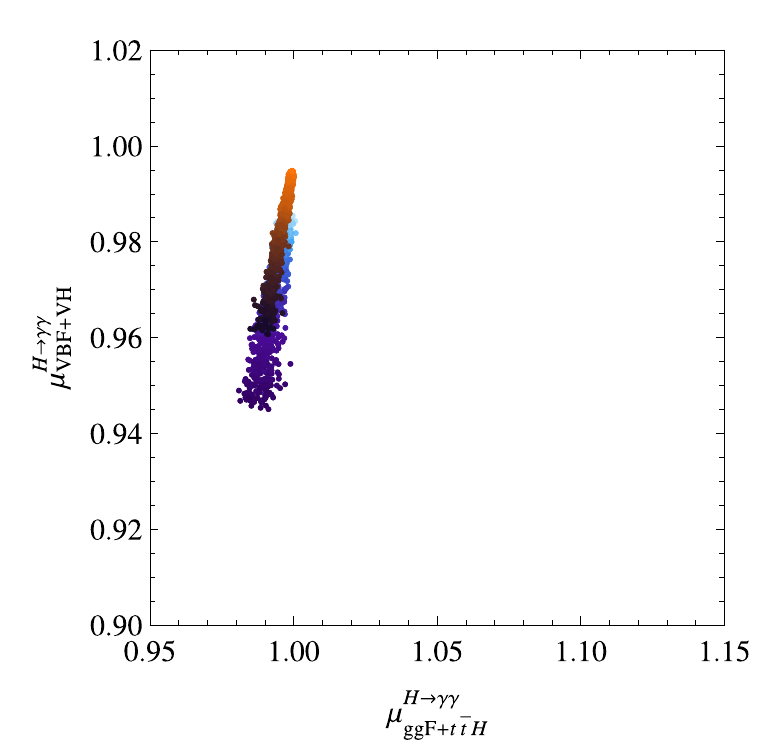}
 \includegraphics[width=0.42\textwidth]{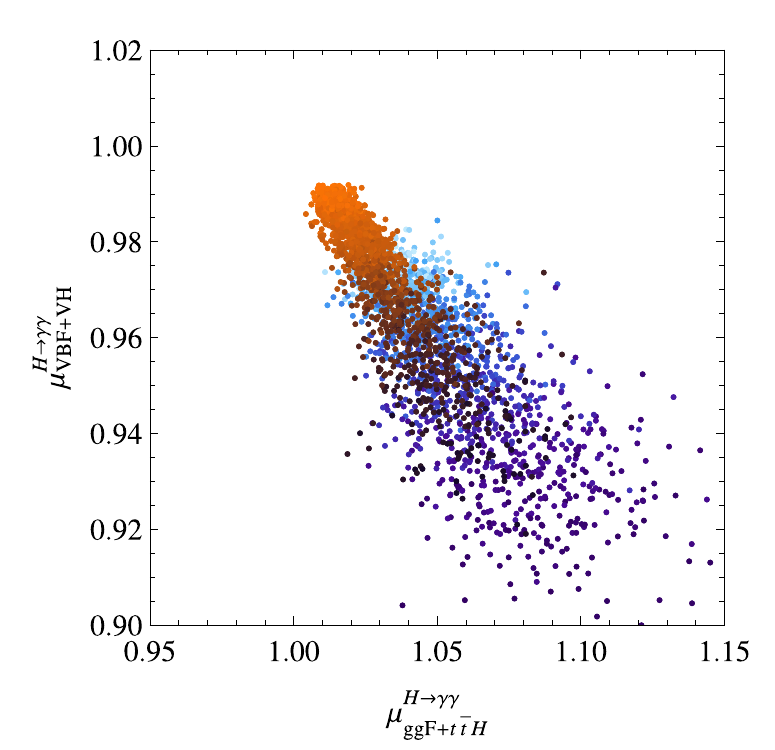}
 \includegraphics[width=0.0768\textwidth]{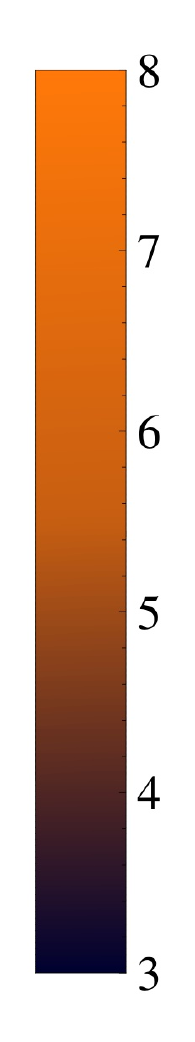}
 \includegraphics[width=0.118\textwidth]{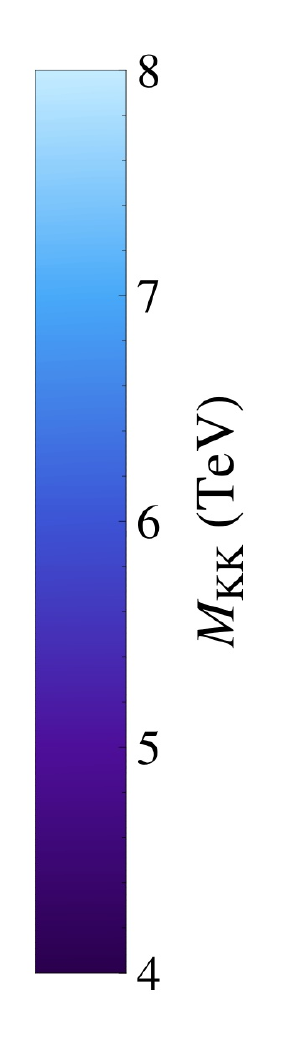}
}
\caption{\label{fig:YYCorrelationPlotsVsMKK} 
$H\to\gamma\gamma$ signal strength in the GF vs.\ VFB production channels for a broad bulk Higgs with $\beta=1$ (orange-brown colour shading) and a narrow bulk Higgs with $\beta=10$ (white-blue colour shading). Figure~\ref{fig:Corr_HYY_Yast} shows the signal strength for fixed values $M_{\rm KK}=4$\,TeV and 8\,TeV, with $y_\ast$ varied between~0 and~3. Figure~\ref{fig:Corr_HYY_MKK} shows results for fixed values $y_{\ast}=1$ and~3, with $M_{\rm KK}$ varied between 3.3\,TeV and 8\,TeV for $\beta=1$, and 4.3\,TeV and 8\,TeV for $\beta=10$. Here the lower bounds correspond to the tree-level analysis of electroweak precision observables in section~\ref{sec:EWpars}.}
\end{center}
\end{figure}

\begin{figure}
\begin{center} 
\subfigure[$M_{\mathrm{KK}}=4$ TeV (left plot) and $M_{\mathrm{KK}}=8$ TeV (right plot) with $\beta=1$ (orange-brown) and $\beta=10$ (white-blue).]
{\label{fig:Corr_HWW_Yast}
 \includegraphics[width=0.42\textwidth]{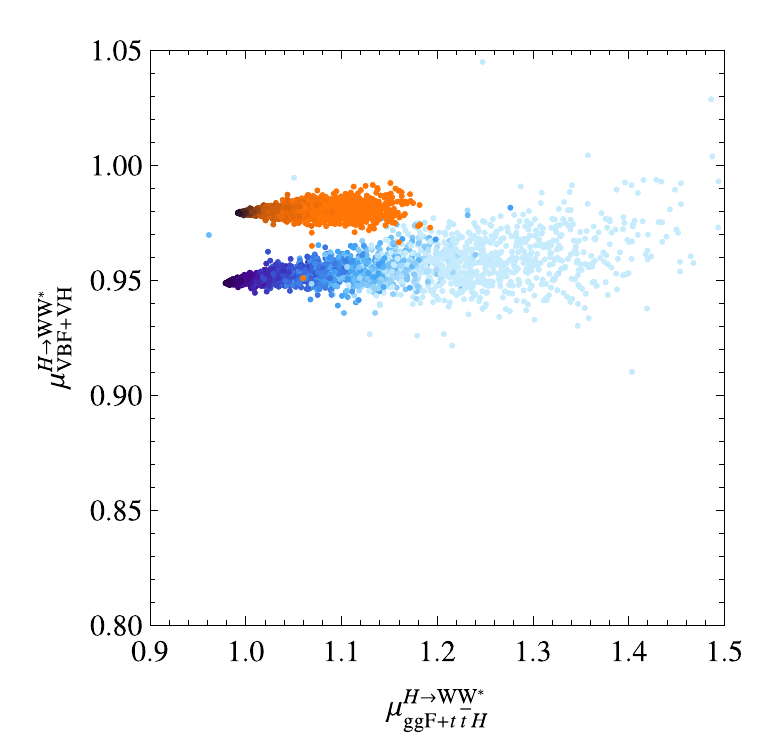}
 \includegraphics[width=0.42\textwidth]{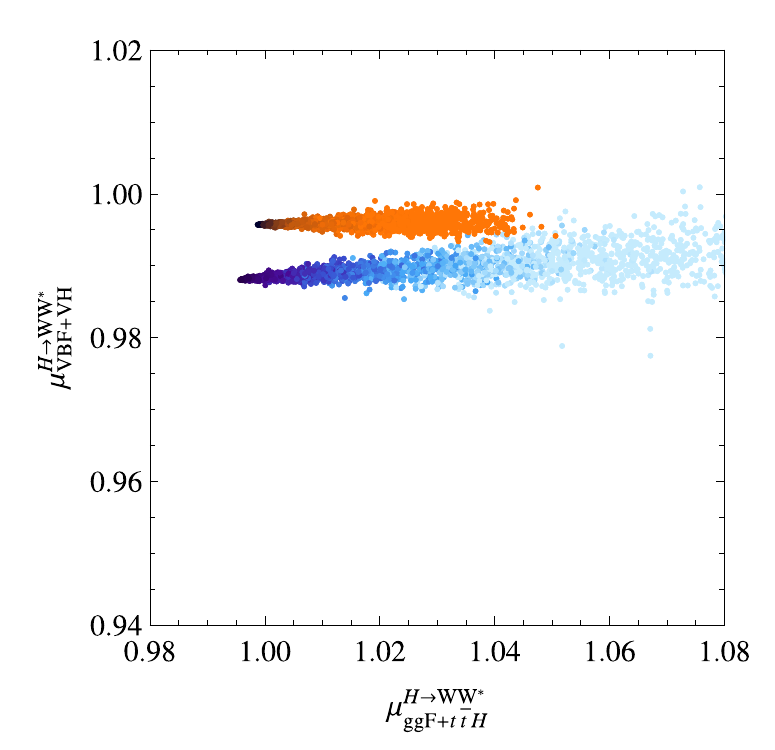}
 \includegraphics[width=0.0905\textwidth]{figures/HYY_Yastrun_colourBar1.pdf}
 \includegraphics[width=0.132\textwidth]{figures/HYY_Yastrun_colourBar2.pdf}
}\\
\subfigure[$y_\ast=1$ (left plot) and $y_\ast=3$ (right plot) with $\beta=1$ (orange-brown) and $\beta=10$ (white-blue).]
{\label{fig:Corr_HWW_MKK}
 \includegraphics[width=0.42\textwidth]{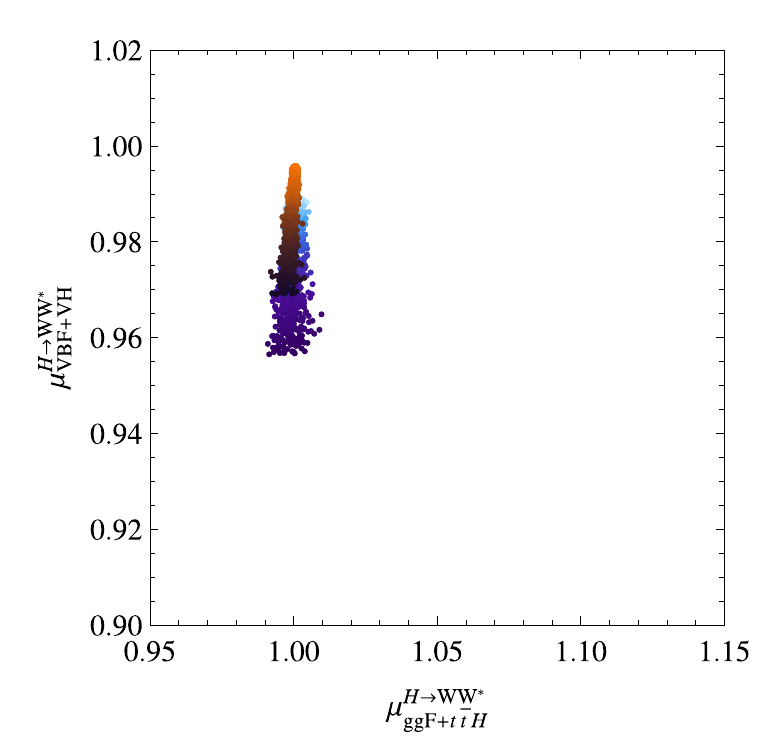}
 \includegraphics[width=0.42\textwidth]{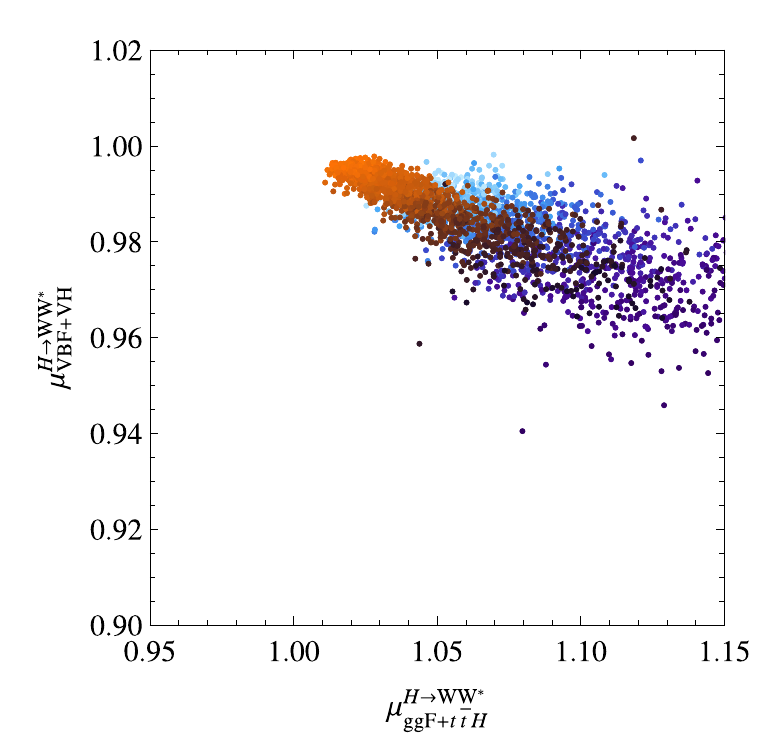}
 \includegraphics[width=0.0768\textwidth]{figures/MKKColourbar.pdf}
 \includegraphics[width=0.118\textwidth]{figures/MKKColourbar1.pdf}
 }
\caption{\label{fig:WWCorrelationPlotsVsMKK} 
Same as figure~\ref{fig:YYCorrelationPlotsVsMKK}, but for the $H\to WW^*$ decay mode.}
\end{center}
\end{figure}

\begin{figure}
\begin{center}  
\subfigure[$M_{\mathrm{KK}}=4$ TeV (left plot) and $M_{\mathrm{KK}}=8$ TeV (right plot) with $\beta=1$ (orange-brown) and $\beta=10$ (white-blue).]
{\label{fig:Corr_Htt_Yast}
 \includegraphics[width=0.42\textwidth]{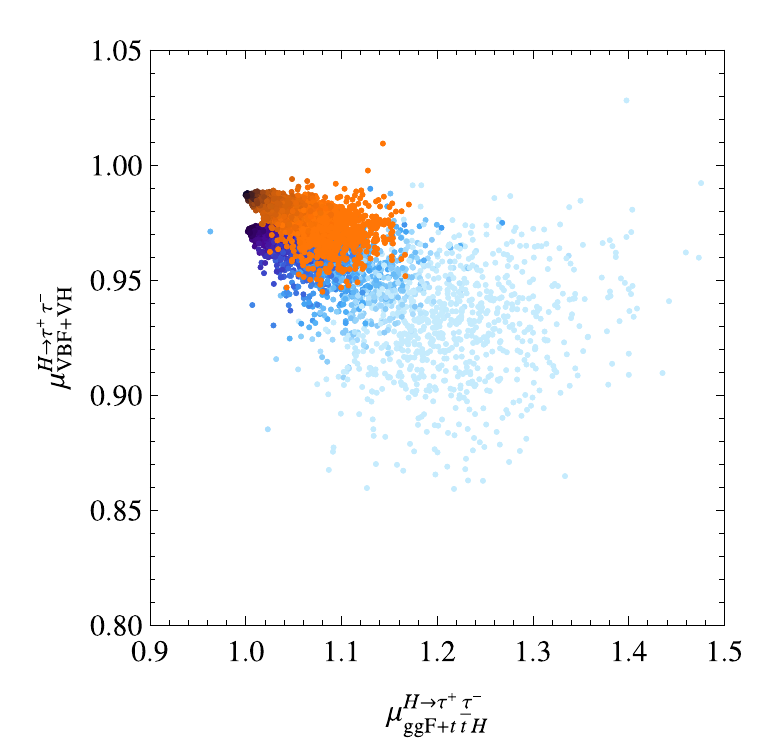}
 \includegraphics[width=0.42\textwidth]{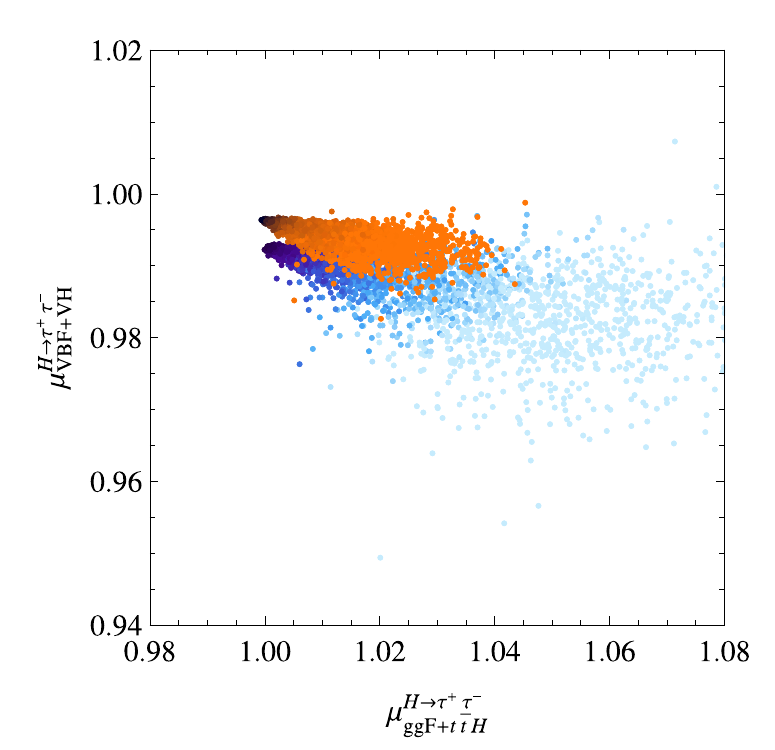}
 \includegraphics[width=0.0905\textwidth]{figures/HYY_Yastrun_colourBar1.pdf}
 \includegraphics[width=0.132\textwidth]{figures/HYY_Yastrun_colourBar2.pdf}
}\\
\subfigure[$y_\ast=1$ (left plot) and $y_\ast=3$ (right plot) with $\beta=1$ (orange-brown) and $\beta=10$ (white-blue).]
{\label{fig:Corr_Htt_MKK}
 \includegraphics[width=0.42\textwidth]{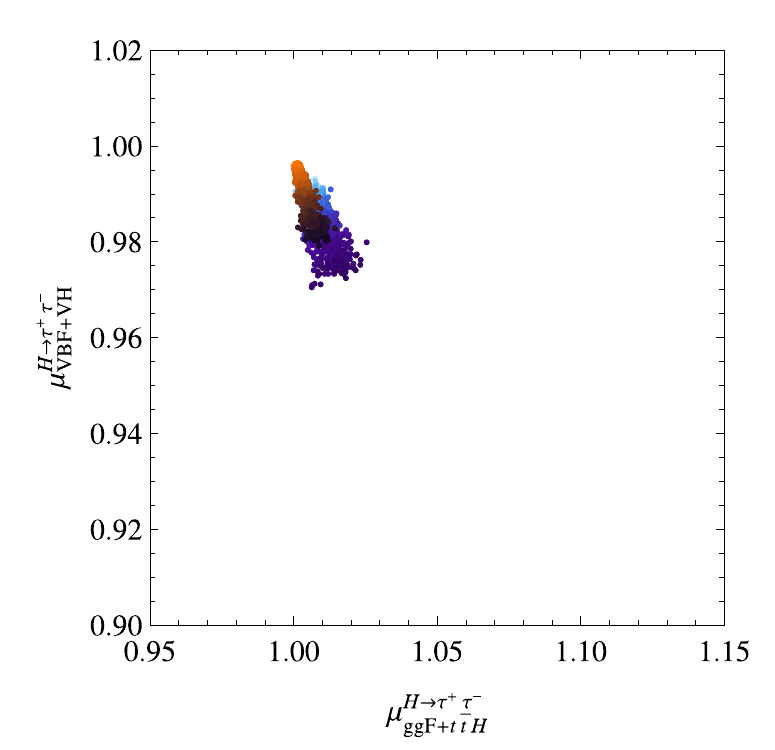}
 \includegraphics[width=0.42\textwidth]{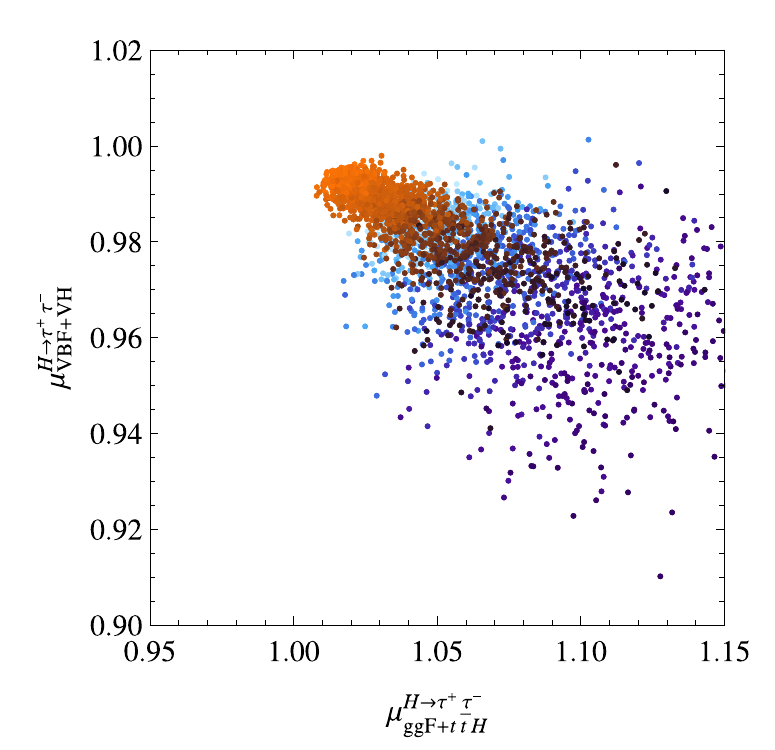}
 \includegraphics[width=0.0768\textwidth]{figures/MKKColourbar.pdf}
 \includegraphics[width=0.118\textwidth]{figures/MKKColourbar1.pdf}
}
\caption{\label{fig:TauTauCorrelationPlotsVsMKK}
Same as figure~\ref{fig:YYCorrelationPlotsVsMKK}, but for the $H\to \tau^+\tau^-$ decay mode.}
\end{center} 
\end{figure}

With these assumptions, the relevant signal strengths for the $H\to\gamma\gamma$, $H\to WW^*$ and $H\to\tau^+\tau^-$ decay modes, relative to the ones in the SM, are shown in figures~\ref{fig:YYCorrelationPlotsVsMKK}, \ref{fig:WWCorrelationPlotsVsMKK} and \ref{fig:TauTauCorrelationPlotsVsMKK}. In these plots we consider two representative values of $\beta$, which as discussed in section~\ref{subsec:numanal} correspond to a broad and a narrow bulk Higgs profile, respectively. We also distinguish two different sets of production mechanisms: Higgs production in gluon fusion together with $t\bar t H$ associated production (labeled ``$\mbox{ggF}+t\bar t H$'' in the plots and abbreviated as ``GF'' in our discussion), and Higgs production in weak vector-boson fusion together with $VH$ associated production (labeled ``$\mbox{VBF}+VH$'' and abbreviated as ``VBF''). In the SM the first mechanism mainly involves the $Ht\bar t$ coupling, while the second one is mainly sensitive to the $HVV$ couplings. In each figure, the upper two plots are shown for fixed $M_{\rm{KK}}$ values and varying $y_\ast$, while the lower two plots are shown for fixed $y_\ast$ values and varying $M_{\rm KK}$. In the latter case the lower bounds on the KK mass scale implied by electroweak precision tests are taken into account. From the discussion in section~\ref{subsec:WWZZ} it follows that the corresponding plots for the $H\to ZZ^*$ decay mode would look nearly identical to those for $H\to WW^*$ shown in figure~\ref{fig:WWCorrelationPlotsVsMKK}. Analogously, the discussion in section~\ref{subsec:Hff} implies that the plots for $H\to b\bar b$ or $H\to c\bar c$ would look very similar to those for $H\to\tau^+\tau^-$ shown in figure~\ref{fig:TauTauCorrelationPlotsVsMKK}. One can readily observe that, as $M_{\mathrm{KK}}$ is increased or $\beta$ is decreased, all production cross sections and decay widths become more SM like. On the other hand, as $y_\ast$ is decreased, the $H\to f\bar f$ decay rates, the gluon-fusion production cross section and the fermionic contribution to the $H\to\gamma\gamma$ decay rate become more SM like. The $H\to WW^*$ decay width and the bosonic contributions to the $H\to\gamma\gamma$ decay width, instead, remain unchanged under variations of $y_\ast$. In the following we will discuss these parametric dependences in more detail.

From figure~\ref{fig:Corr_HYY_Yast}, we observe an enhancement of the $H\to\gamma\gamma$ signal strength in the GF channel along with a suppression in the VBF channel for increasing values of $y_\ast$. This can be readily understood, since the heavy fermionic KK resonances yield a sizable enhancement of the gluon-fusion production cross section, which overcomes their opposite effect on the $H\to\gamma\gamma$ decay rate. In the VBF channel, the reduction of the signal strengths for increasing $y_\ast$ is due to the suppression of the $H\to\gamma\gamma$ decay rate. Note that for increasing $y_\ast$ the effect in the total decay width cannot compensate the reduction of the $H\to\gamma\gamma$ width. In figure~\ref{fig:Corr_HWW_Yast}, the same enhancement of the signal strength in the GF channel is observed for increasing values of $y_\ast$. The variation in the branching ratio of $H\to WW^*$ as a function of $y_\ast$ is due to the variation on the Higgs total width, arising from the combination of two competing effects: the suppression of the $H\to b\bar b$ decay rate and the enhancement of the $H\to gg$ decay rate due to the high multiplicity of KK resonances contributing in the loop. In both search modes, $H\to\gamma\gamma$ and $H\to WW^*$, we observe a clear separation of the predicted signal strengths in the VBF channel for the scenarios with a broad and a narrow Higgs profile. This effect can be understood from the fact that the $HWW$ coupling is very sensitive to the delocalisation of the Higgs profile, as can be seen from figure~\ref{fig:HtoWW}. Therefore, precise measurements of the $H\to\gamma\gamma$, $H\to WW^*$ and $H\to ZZ^*$ signal strengths in the GF and VBF production channels, together with knowledge of the KK mass scale $M_{\rm KK}$ from the discovery of a low-lying KK resonance, could shed light on the degree of partial compositeness of the Higgs boson. However, bounds from electroweak precision data imply that the discovery of such a resonance would most likely require a hadron collider beyond LHC. Figure~\ref{fig:Corr_Htt_Yast} shows that an enhancement of the signal strength in the GF channel along with a suppression in the VBF channel for increasing values of $y_\ast$ is also observed for the $H\to\tau^+\tau^-$ mode. However, in this case there is a significant overlap of the results corresponding to the two different values of $\beta$. This is due to the milder $\beta$ dependence together with the strong $y_\ast$ sensitivity of the $H\to\tau^+\tau^-$ decay rate. This makes it challenging to differentiate between the bulk and brane Higgs scenarios in this search mode. 

Considering figures~\ref{fig:Corr_HYY_MKK}, \ref{fig:Corr_HWW_MKK} and \ref{fig:Corr_Htt_MKK}, it can be noted that varying $\beta$ or $M_{\mathrm{KK}}$ for fixed values of $y_\ast$ results in an approximately correlated shift in all of the production cross sections and decay rates, yielding a large overlap in the signal strengths. Comparing the left and right plots in these figures we observe that, for sizable values of $y_{\ast}\sim 3$, the most important effect is an enhancement of the GF production channel due to the dominant contribution of heavy fermionic resonances in the loop. Additional effects are due to variations in the Higgs branching ratios. In the case of $H\to\gamma\gamma$ this translates in an enhancement of about 5\,--\,10\%, depending on the localization of the Higgs, while in the $H\to\tau^+\tau^-$ channel deviations of about 10\,--\,15\% may be allowed. In the $H\to WW^*$ channel one can observe an enhancement up to 20\% of the signal strength with respect the SM value. For small values $y_\ast\approx 1$ instead, the enhancement of the gluon-fusion cross section is no longer the dominant effect. It is counteracted by the reductions in the $H\to\gamma\gamma$, $H\to WW^*$ and $H\to\tau^+\tau^-$ branching ratios, which are largest (smallest) for the $\gamma\gamma$ ($\tau^+\tau^-$) channel. Quite generally, we observe that for all decay channels variations of at most a few percent are allowed in the GF production mode for small $y_\ast$. The VBF signal strengths are all suppressed with respect to the SM, independently of the value of $y_\ast$, by a few to several percent depending of the specific decay channel.
 
\begin{figure}
\begin{center}
 \includegraphics[width=0.37\textwidth]{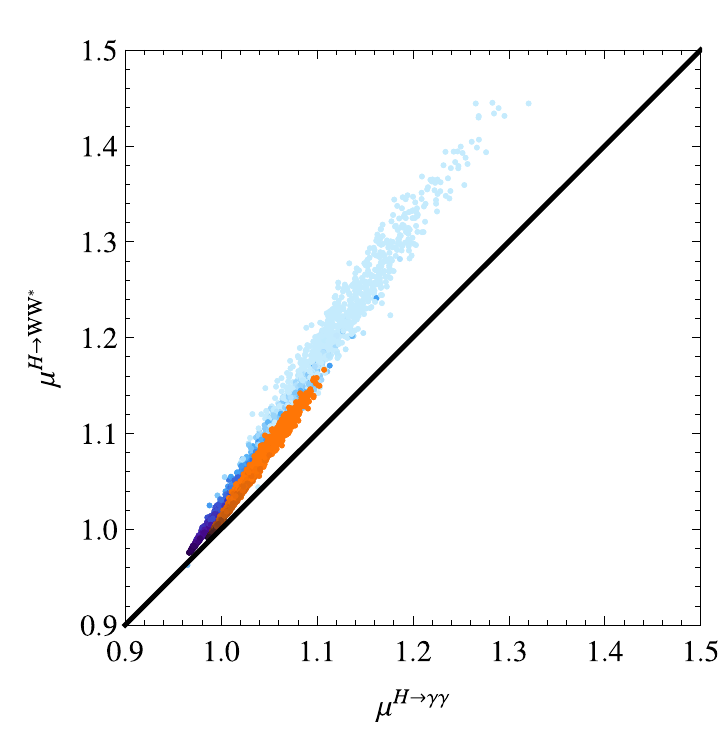}
 \includegraphics[width=0.37\textwidth]{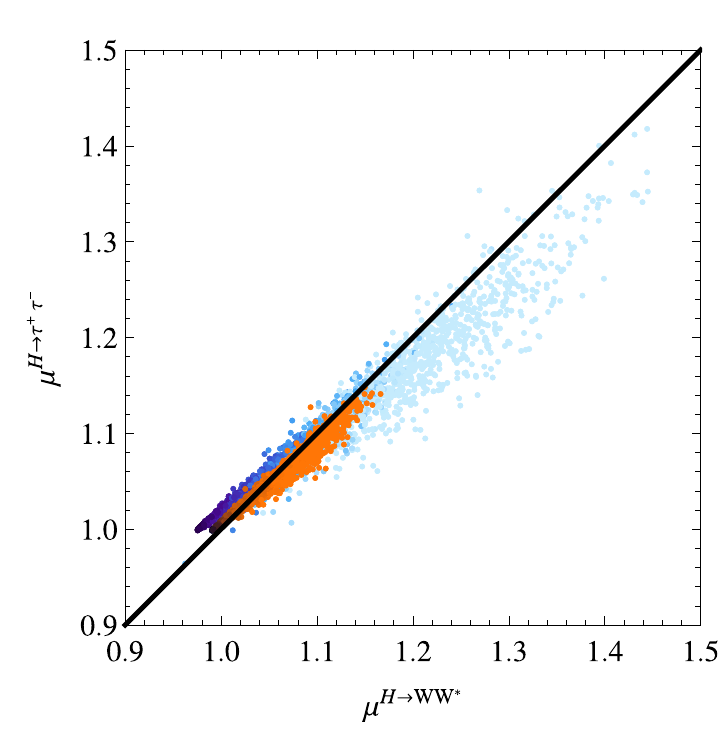}
 \includegraphics[width=0.0845\textwidth]{figures/HYY_Yastrun_colourBar1.pdf}
 \includegraphics[width=0.123\textwidth]{figures/HYY_Yastrun_colourBar2.pdf}
\caption{\label{fig:HYYvsOthersCorrelationPlotsVsMKK} 
Correlations between the signal strengths for $H\to\gamma\gamma$ vs.\ $H\to WW^*$ (left) and $H\to WW^*$ vs.\ $H\to\tau^+\tau^-$ (right) as a function of $y_\ast\in[0.5,3]$ for $\beta=1$ (orange-brown) and $\beta=10$ (white-blue). The signal strengths $\mu^{H\to X}$ are defined to include all production processes and are shown for a fixed value of $M_{\rm KK}=4$\,TeV.}
\end{center} 
\end{figure} 
 
Currently LHC data gives information on the cross sections times branching ratios for specific Higgs production mechanisms and decay modes. In order to study the correlations between ratios of partial decay widths normalized to their SM values, we consider in figure~\ref{fig:HYYvsOthersCorrelationPlotsVsMKK} the inclusive signal strengths $\mu^{H\to X}$ defined to include all production processes. We show results for a fixed value $M_{\rm KK}=4$\,TeV and varying $y_\ast\in[0.5,3]$ for the two representative values of $\beta$ considered throughout the paper. We observe strong correlations between the partial decay widths, such that $\frac{\Gamma_{H\to WW^*}}{\Gamma_{H\to WW^*}^{(\mathrm{SM})}}\gtrsim\frac{\Gamma_{H\to\gamma\gamma}}{\Gamma_{H\to\gamma\gamma}^{(\mathrm{SM})}}$ and $\frac{\Gamma_{H\to WW^*}}{\Gamma_{H\to WW^*}^{(\mathrm{SM})}}\sim \frac{\Gamma_{H\to \tau^+\tau^-}}{\Gamma_{H\to \tau^+\tau^-}^{(\mathrm{SM})}}$. Similar correlations would also be observed by varying $M_{\rm KK}$ for fixed values of $y_{\ast}$. Concerning the second ratio, we furthermore see that a small enhancement of $\mu^{H\to WW^*}$ over $\mu^{H\to\tau^+\tau^-}$ would favour large values of $y_\ast\sim 3$. Comparing the results of figure~\ref{fig:HYYvsOthersCorrelationPlotsVsMKK} with LHC data would enable us to test these correlations and obtain information about the model parameters.

\section{Discussion and Conclusions}
\label{sect:Conclusions}

The discovery of the Higgs boson, as well as the increasingly accurate information on its production and decay modes, provides invaluable input for testing BSM scenarios of electroweak symmetry breaking. In this paper we have studied the modification of Higgs couplings and the associated collider phenomenology in a class of scenarios, in which the Higgs propagates in a warped extra dimension. These models can be seen as the holographic description of a partially composite Higgs boson, providing thus a framework to quantitatively study  the implications of varying the scaling dimension of the Higgs operator. This is of particular interest since, as explicitly shown in section~\ref{sec:EWpars}, variations in the scaling dimension can significantly relax the constraints from electroweak precision data, allowing for smaller masses of the new heavy particles. On the other hand, considering scenarios of a Higgs boson propagating in the bulk of a 5D space allows us to explore the phenomenological differences between these scenarios and the more broadly studied narrow bulk-Higgs case of general RS models.

In this paper we have first focused on a 5D minimal realization of the SM with a generic warped background, specialising later to the particular case of an AdS$_5$ geometry for concreteness. We have first discussed several important features of these models that hold in general. In particular, such models include KK towers (without a zero mode) of additional charged and neutral scalar fields, whose masses depend on the scaling dimension of the Higgs operator and decouple in the limit $\beta\to\infty$ corresponding to a brane-localised Higgs sector. The explicit construction of the extended scalar sector and the derivation of the Feynman rules are some of the main technical accomplishments of this work. Moreover, we have observed that extending the gauge-boson/Goldstone equivalence theorem to five dimensions results in specific relations between the profiles of the gauge bosons and Goldstone bosons and their KK excitations. This fact plays an important role in ensuring gauge invariance. For the case of a general 5D metric, we find that the $HWW$ and $HZZ$ couplings, normalized to their SM values, are given by the ratio of the SM Higgs VEV over the corresponding parameter in the 5D model, $v/\tilde v$, times a factor that is always smaller than~1. In RS models based on an AdS$_5$ geometry, $v/\tilde v$ is also less than~1, and we believe this to be valid for a general 5D metric. In the case of the $Hf\bar f$ couplings, besides an overall suppression by $v/\tilde v$, there are three different contributions that should be evaluated for each specific 5D metric. Two of them -- denoted by $({\cal J}_f)_{ii}$ and $({\cal K}_f)_{ii}$ in the appendix -- always give rise to a suppression of the couplings. The third correction denoted by $({\cal I}_f)_{ii}$, which is due to the mixing with the heavy vector-like resonances, is found to be negative for most choices of model parameters. Therefore, we found quite generally that, for the AdS$_5$ case, the Higgs-boson decay rates for tree-level processes into SM gauge bosons and fermions are suppressed with respect to the SM.

We have also computed the loop-induced processes of Higgs production in gluon fusion and $H\to\gamma\gamma$ decay, explicitly showing that they are finite as expected from gauge invariance. For sizable values of the (properly defined) dimensionless 5D Yukawa couplings, we find that the dominant corrections are due to the effects of fermionic KK resonances. In the case of the $H\to\gamma\gamma$ amplitude they interfere destructively with the bosonic loop contributions. Using completeness relations, we have computed for the first time the fermionic contributions entering in both loop-induced processes for the general case of three fermion generations. The reduction of the $HWW$ coupling also has an impact on the $H\to\gamma\gamma$ decay amplitude, but this effects partially cancels against the contributions of bosonic KK resonances. It is thus only significant (compared with the fermionic effects) for relatively small values of the 5D Yukawa couplings.

In the last part of this work we have concentrated on the phenomenological implications of our results, focusing on the AdS$_5$ case. We have first derived analytical expressions for all relevant scalar and vector profiles. We have then performed a fit to a selected subset of the most precisely measured electroweak precision observables and derived the tree-level expressions for the parameters $S$, $T$ and $U$ in our model. Importantly, we have shown that the $T$ parameter can be reduced by up to a factor~3 compared with the brane-Higgs case. We have found that values of $M_{\rm KK}>2.5$~TeV ($M_{\rm KK}>4.3$~TeV) for $\beta=0$ ($\beta=10$) are allowed at 95\% confidence level. Regarding the fermion sector, we have discussed the most appropriate way of defining the dimensionless 5D Yukawa couplings, for different values of $\beta$, in terms of the original, dimensionful Yukawa couplings in the 5D Lagrangian. The dimensionless couplings are constrained from below by requiring to obtain the correct top mass, and from above by demanding that the theory remain perturbative up to energy scales around several times $M_{\mathrm{KK}}$. We have introduced the important parameter $y_\ast$, the maximum absolute value of each individual entry of the anarchic 5D Yukawa matrices. The larger the value of $y_\ast$, the more elementary the fermions are (i.e.\ the more the zero-mode profiles are shifted towards the UV brane) and the more flavour-changing neutral currents are suppressed. However, we have also found that the size of the corrections to the Higgs decay rates are quadratically sensitive to $y_\ast$. Hence, for a given KK scale, requiring SM-like Higgs couplings constrains $y_\ast$ in the opposite direction to the stringent constraints from flavour physics. Furthermore, in our model the corrections to the $H\to WW^*$ decay rate are directly correlated with the contributions to the $S$ and $T$ parameters. Hence, constraints from Higgs physics complement the constraints from flavour physics and electroweak precision tests in important ways.

In our analysis of Higgs physics we have focussed on the signal strengths, i.e.\ the cross section times branching ratio, for different production and decay channels. For sizable values $y_\ast\gtrsim 1.5$, the dominant new physics effects arise from fermionic KK-resonance contributions to the gluon-fusion cross section, the $H\to\gamma\gamma$ decay amplitude, and the individual Higgs decay rates into SM fermions (in this order). The latter effect also has an important impact on the total Higgs width, which affects all signal strengths equally. Even though the individual 5D Yukawa couplings are random complex parameters in our model, we find that the dominant effects involve expressions such as $\mathrm{Tr}(\mathbf{Y}_q\mathbf{Y}_q^\dagger)$, which can be estimated with good accuracy in terms of $y_\ast$. Consequently, we have obtained a reasonably accurate description of our phenomenological results in terms of just three parameters: $M_{\mathrm{KK}}$, $\beta$ and $y_\ast$. In the opposite case of small $y_\ast\lesssim 1$, the modifications of the observable signal strengths are predominantly due to the corrections to the Higgs couplings to electroweak gauge bosons and the value of the Higgs VEV. These effects are governed by $M_{\mathrm{KK}}$ and $\beta$. We have shown that they are reduced as the Higgs field is moved further into the bulk, i.e.\ as $\beta$ is lowered. Overall our analysis has thus resulted in rather definite predictions for the observable signal strengths, which in the near future will be measured with increased precision at the LHC. Our findings can be straightforwardly confronted with those of other approaches. For example, assuming that some definite deviations from the SM predictions will show up in future analyses, it will be possible to experimentally distinguish the partially-composite Higgs models discussed here from models in which the Higgs is a pseudo Nambu-Goldstone boson. Further still, we have shown that with prior knowledge of the KK mass scale from the direct observation of a KK resonance, it would also be possible to distinguish between the bulk-Higgs scenario explored here and models in which the scalar sector is localised on the IR brane, thus potentially shedding some light on the important question of the degree of partial compositeness of the Higgs particle.
\vfil

\acknowledgments
We are grateful to Florian Goertz, Raoul Malm, Eduardo Ponton and Christoph Schmell for useful discussions. A.C.~and M.C.~would like to express special thanks to the Mainz Institute for Theoretical Physics (MITP) for its hospitality and support during the final stages of this project. P.A.~thanks ITP at ETH Zurich for hospitality during the completion of this work. P.A., M.C.~and M.N.~gratefully acknowledge KITP Santa Barbara for hospitality and support during the early phase of this project. The research of P.A.\ and M.N.\ has been supported by the Advanced Grant EFT4LHC of the European Research Council (ERC), the Cluster of Excellence {\em Precision Physics, Fundamental Interactions and Structure of Matter\/} (PRISMA -- EXC 1098), and grant 05H12UME of the German Federal Ministry for Education and Research (BMBF). M.C.~gratefully acknowledges the generous support of the Alexander von Humboldt Foundation, which has helped to initiate this project. The work of A.C.~was supported by the Swiss National Science Foundation under contract 200021-143781. Fermilab is operated by the Fermi Research Alliance, LLC, under contract DE-AC02-07CH11359 with the United States Department of Energy. 
\vfill

\clearpage
\appendix

\section{Using Completeness Relations to Evaluate the Fermion Loop}
\label{sec:Completness}

To compute the fermion contribution to the one-loop processes $gg\to H$ and $H\to \gamma \gamma$,  we need to evaluate the different sums appearing in  $\kappa_{t,b}$ and $\kappa_{\rm KK}^q$, with $q=u,d$. These can be written as follows \cite{delAguila:2000rc}
	\begin{eqnarray}
		 \kappa_{t}&=&1-\left(\mathcal{J}_u\right)_{33}-\left(\mathcal{K}_u\right)_{33}+\left(\mathcal{I}_u\right)_{33} \,, \label{eq:klonet}\label{Appen_kappat}\\
		 \kappa_{b}&=&1-\left(\mathcal{J}_d\right)_{33}-\left(\mathcal{K}_d\right)_{33}+\left(\mathcal{I}_d\right)_{33} \,, \label{eq:kloneb}\label{Appen_kappab}\\
  \kappa_{\rm KK}^q&=&\mathrm{Tr}\,\mathcal{J}_{q}+\mathrm{Tr}\,\mathcal{K}_q-\mathcal{H}_q \,; \quad  q=u,d,\label{eq:kkk}
	\end{eqnarray}
	where we have defined
\begin{eqnarray}
	 \mathcal{H}_q&=& \sum_{k,l}\sum_{n,m=1}^{\infty}\frac{Y_{Q_L^k q_R^l}^{(n,m)}Y_{Q_R^k q_L^l}^{(n,m)\ast}}{m_n^{(Q^k)}m_m^{(q^l)}} \,,\qquad	\left(\mathcal{I}_q\right)_{ij}=\sum_{k,l}\sum_{n,m=1}^{\infty}\frac{Y_{Q_L^i q_R^k}^{(0,m)}Y_{Q_R^l q_L^k}^{(n,m)\ast}Y_{Q_L^l q_R^j}^{(n,0)}}{\lambda_j^q m_n^{(Q^l)}m_m^{(q^k)}} \,,\quad \label{eqn:CompletenessSums1}\\
 \left(\mathcal{J}_q\right)_{ij}&=&\frac{1}{2}\sum_{k}\sum_{n=1}^{\infty}\frac{Y_{Q_L^iq_R^k}^{(0,n)}Y_{Q_L^jq_R^k}^{(0,n)\ast}}{m_n^{(q^k)2}} \,, \qquad  
 \hspace{1mm}
 \left(\mathcal{K}_q\right)_{ij}=\frac{1}{2}\sum_{k}\sum_{n=1}^{\infty} \frac{Y_{Q_L^kq_R^i}^{(n,0)\ast}Y_{Q_L^k q_R^j}^{(n,0)}}{m_n^{(Q^k)2}} \,. \label{eqn:CompletenessSums2}
\end{eqnarray}
In order to simplify the explicit form of the expressions appearing above, we have rotated the zero-mode profiles in such a way that $Y_{Q_L^i q_R^j}^{(0,0)}=\lambda^q_i\delta_{ij}$ for $q=u,d$.\footnote{This corresponds to rotate (\ref{UpMass}) and its analogue in the down sector to a basis where both upper-left blocks are diagonal.} 

Due to the well-behaved integrands present in the bulk Higgs case, we can permute the integrals and the infinite sums in the expressions above obtaining,  for $q=u,d$,
\begin{eqnarray}
\mathcal{H}_q &=&  \tilde v^2 \sum_{k,l}(Y^{5D}_{q})_{kl} (Y^{5D\dagger}_{q})_{lk}\int dr dr^\prime b(r)b(r^{\prime})\left[ h(r)\sum_{m=1}^{\infty}\frac{ f_m^{(q_R^l)}(r)f_m^{(q_L^l)\ast}(r^{\prime})}{m_m^{(q^l)}}\times\right.\nonumber\\
																							 &&\qquad \left.\sum_{n=1}^{\infty}\frac{f_n^{(Q_R^k)}(r^{\prime})f_n^{(Q_L^k)\ast}(r)}{m_n^{(Q^k)}} h(r^{\prime})\right],\label{eq:h}\\ 
	\left(\mathcal{I}_q\right)_{ij}&=& \tilde v^2\sum_{k,l}(Y^{5D}_q)_{ik} (Y_q^{5D\dagger})_{kl}(Y_q^{5D})_{lj}\int dr dr^{\prime}dr^{\prime\prime}b(r)b(r^{\prime})b(r^{\prime\prime})\left[f_0^{(Q_L^i)\ast}(r)h(r)\times \right.\nonumber\\
																																														   &&\left. \sum_{m=1}^{\infty}\frac{f_m^{(q_R^k)}(r)f_m^{(q_L^k)\ast}(r^{\prime})}{m_m^{(q^k)}}h(r^{\prime})\sum_{n=1}^{\infty}\frac{f_n^{(Q_R^l)}(r^{\prime})f_n^{(Q_L^l)\ast}(r^{\prime\prime})}{m_n^{(Q^l)}}h(r^{\prime\prime})f_0^{(q_R^j)}(r^{\prime\prime})\right](\lambda^{q}_j)^{-1},\quad\label{eq:i}\\
 \left(\mathcal{J}_q\right)_{ij}&=& \tilde v^2 \frac{1}{2}\sum_{k} (Y^{5D}_q)_{ik} (Y_q^{5D\dagger})_{ kj}\int dr dr^{\prime}b(r)b(r^{\prime})\left[f_0^{(Q_L^i)\ast}(r)h(r)\sum_{n=1}^{\infty}\frac{f_n^{(q_R^k)}(r)f_m^{(q_R^k)\ast}(r^{\prime})}{m_n^{(q^k)2}}\times \right.\nonumber\\
											 &&\qquad \left.h(r^{\prime})f_0^{(Q_L^j)}(r^{\prime})\right]\label{eq:j},\\
	\left(\mathcal{K}_q\right)_{ij}&=&  \tilde v^2 \frac{1}{2}\sum_{k} (Y_{q}^{5D\dagger})_{ik} (Y^{5D}_{q})_{kj}\int dr dr^{\prime}b(r)b(r^{\prime})\left[ f_0^{(q_R^i)\ast}(r)h(r)\sum_{n=1}^{\infty}\frac{f_n^{(Q_L^k)}(r^{\prime})f_n^{(Q_L^k)\ast}(r)}{m_n^{(Q^k)2}}\times\right.\nonumber\\
																										  &&\left.\qquad h(r^{\prime})f_0^{(q_R^j)}(r^{\prime})\right].\label{eq:k}
\end{eqnarray}

Before we proceed, let us make the following redefinition of the fermion profiles,  
\begin{eqnarray}
	f_n^{(\Psi_{L,R})}(r)=b_{M_{\Psi}}^{\pm 1/2}(r) \hat{f}_n^{(\Psi_{L,R})}(r) \,,
\end{eqnarray}
where
\begin{eqnarray}
	b_{M}(r)=\mathrm{exp}\left[-2\int_{r_{\rm UV}}^r dr^{\prime} b(r^{\prime})M\right]=b_{-M}^{-1}(r) \,,
\end{eqnarray}
and the new functions $\hat{f}_n^{(\Psi_{L,R})}$ satisfy
\begin{eqnarray}
	b_{M_{\Psi}}^{\pm 1} \partial_r \hat{f}_{n}^{(\Psi_{L,R})}(r)=\pm \frac{b}{a} m_n^{(\Psi)} \hat{f}_n^{(\Psi_{R,L})}(r) \,.
\end{eqnarray}

Up to some possible prefactors due to the chosen normalization for the fermion profiles, the sums in (\ref{eq:h}-\ref{eq:k}) are just 5D propagators evaluated at zero-momentum once we have subtracted the possible contribution from the zero modes. In particular, if we define
\begin{eqnarray}
\hat{P}^{LL[+,+]}_{p;M_{Q}}(r_1,r_2)&=&\sum_{n=1}^{\infty}\frac{\hat{f}_n^{(Q_L)}(r_1)\hat{f}_n^{(Q_L)\ast}(r_2)}{p^2-m_n^{(Q)2}} \,,\\
\hat{P}^{RR[-,-]}_{p;M_{q}}(r_1,r_2)&=&\sum_{n=1}^{\infty}\frac{\hat{f}_n^{(q_R)}(r_1)\hat{f}_n^{(q_R)\ast}(r_2)}{p^2-m_n^{(q)2}} \,; \quad q=u,d,\\
								  \hat{P}_{p;M_{Q}}^{RL[+,+]}(r_1,r_2)&=&\sum_{n=1}^{\infty}m_n^{(Q)}\frac{\hat{f}_n^{(Q_R)}(r_1)\hat{f}_n^{(Q_L)\ast}(r_2)}{p^2-m_n^{(Q)2}} \,,\\
									  \hat{P}_{p;M_{q}}^{RL[-,-]}(r_1,r_2)&=&\sum_{n=1}^{\infty}m_n^{(q)}\frac{\hat{f}_n^{(q_R)}(r_1)\hat{f}_n^{(q_L)\ast}(r_2)}{p^2-m_n^{(q)2}} \,;\quad  q=u,d,
\end{eqnarray}
the previous expressions read 
\begin{eqnarray}
	\mathcal{H}_q&=&  \tilde v^2 \sum_{k,l}(Y_{q}^{5D})_{kl} (Y^{5D\dagger}_q)_{lk}\int dr dr^{\prime}b(r)b(r^{\prime})\left[h(r)b_{ M_{q}^{l}-M_{Q}^{k}}^{-1/2}(r)\hat{P}_{0;M_q^l}^{RL[--]}(r,r^{\prime})\right.\nonumber\\
													 &&\left.b_{ M_{q}^{l}-M_{Q}^{k}}^{1/2}(r^{\prime})\hat{P}_{0;M_Q^k}^{RL[++]}(r^{\prime},r)h(r^{\prime})\right],\quad \\ 
	\left(\mathcal{I}_q\right)_{ij}&=&  \tilde v^2 \sum_{k,l}(Y^{5D}_q)_{ik}(Y_{q}^{5D\dagger})_{kl}(Y^{5D}_{q})_{lj}\int dr dr^{\prime} dr^{\prime\prime}b(r)b(r^{\prime})b(r^{\prime\prime})\left[f_0^{(Q_L^i)\ast}(r)h(r)b_{M_{q}^{k}}^{-1/2}(r)\right.\nonumber\\
																																																 &&\left.\hat{P}_{0;M_q^k}^{RL[--]}(r,r^{\prime}) b_{ M_{q}^{k}-M_{Q}^{l}}^{1/2}(r^{\prime})h(r^{\prime})\hat{P}_{0;M_Q^l}^{RL[++]}(r^{\prime},r^{\prime\prime})b_{M_{Q}^l}^{1/2}(r^{\prime\prime})h(r^{\prime\prime})f_0^{(q_R^j)}(r^{\prime\prime})\right] /(\lambda_{j}^q),\qquad\quad \\
	\left(\mathcal{J}_q\right)_{ij}&=&-  \tilde v^2 \frac{1}{2}\sum_{k}(Y^{5D}_q)_{ik}(Y^{5D\dagger}_q)_{kj}\int dr dr^{\prime}b(r)b(r^{\prime})\left[f_0^{(Q_L^i)\ast}(r)h(r)b_{ M_{q}^{k}}^{-1/2}(r)\hat{P}_{0;M_q^k}^{RR[--]}(r,r^{\prime})\right.\nonumber\\
																											 &&\left.b_{M_{q}^{k}}^{-1/2}(r^{\prime})h(r^{\prime})f_0^{(Q_L^j)}(r^{\prime})\right]\label{eq:j2},\\
	\left(\mathcal{K}_q\right)_{ij}&=&- \tilde v^2 \frac{1}{2}\sum_{k}(Y^{5D\dagger}_q)_{ik} (Y^{5D}_q)_{kj}\int dr dr^{\prime}b(r)b(r^{\prime})\left[ f_0^{(q_R^i)\ast}(r)h(r)b_{M_{Q}^k}^{1/2}(r)\hat{P}_{0;M_Q^k}^{LL[++]}(r,r^{\prime})\right.\nonumber\\
																											 &&\left.b_{M_{Q}^k}^{1/2}(r^{\prime})h(r^{\prime})  f_0^{(q_R^j)}(r^{\prime})\right],
\end{eqnarray}
where  again $q=u,d$.  Note that the different propagators exactly match those defined in appendix~A of \cite{Carmona:2011rd} for the special case $a(r)=b(r)$.\footnote{ Besides a missing $L_1\leftrightarrow r_{\rm IR}$ factor for the LL and RR ones. Moreover, in the notation of \cite{Carmona:2011rd}, $\left(\mathcal{J}_x\right)_{ij}=\frac{1}{2}v^2\beta^{X ;R}_{q_L^{i}q_L^{i}}, \left(\mathcal{K}_x\right)_{ij}=\frac{1}{2}v^2\beta^{Q;L}_{x_R^{i}x_R^{j}}$ and $\left(\mathcal{I}_x\right)_{ij}=-v^2\gamma_{q_L^{i}x_R^{j}}^{X Q;RL}$, with $x=u,d$.} In general, when $a(r)\neq b(r)$,   using the completeness conditions
\begin{eqnarray}
	\delta(r-r^{\prime})=\frac{b}{a} b_{M_{\Psi}}\sum_{n}\hat{f}_n^{(\Psi_L)}(r)\hat{f}_n^{(\Psi_L)\ast}(r^{\prime})=\frac{b}{a} b_{M_{\Psi}}^{-1}\sum_{n}\hat{f}_n^{(\Psi_R)}(r)\hat{f}_n^{(\Psi_R)\ast}(r^{\prime})
\end{eqnarray}
for the hated profiles, where  $ba^{-1} b_{M}^{\pm}$ can be evaluated at either $r$ or $r^{\prime}$, we obtain
\begin{eqnarray}
	\hat{P}_{0; M_{Q}}^{RL[++]}(r_1,r_2)&=&\theta(r_1-r_2)-\frac{1}{L_{M_{Q}}}\int_{r_{\rm UV}}^{r_1} d\hat{r}\frac{b}{a}b_{M_{Q}} \,,\\
 \hat{P}_{0; M_{q}}^{RL[--]}(r_1,r_2)&=&-\theta(r_2-r_1)+\frac{1}{L_{-M_{q}}}\int_{r_{\rm UV}}^{r_2} d\hat{r}\frac{b}{a}b_{M_{q}}^{-1} \,,
\end{eqnarray}
and
\begin{eqnarray}
	\hat{P}_{0; M_{Q}}^{LL[++]}(r_1,r_2)&=&-\int_{r_{\rm UV}}^{\mathrm{min}(r_1,r_2)} dr^{\prime}\frac{b}{a}b_{ M_{Q}}^{-1}+\frac{1}{L_{ M_{Q}}}\int_{r_{\rm UV}}^{r_1} dr^{\prime}\frac{b}{a}b_{ M_{Q}}^{-1}\int_{r^{\prime}}^{r_{\rm IR}} dr^{\prime\prime}\frac{b}{a}b_{ M_{Q}}\nonumber\\
																																																			   &&\mbox{}+ \frac{1}{L_{ M_{Q}}}\int_{r_{\rm UV}}^{r_{\rm IR}} d\hat{r}\frac{b}{a}b_{ M_{Q}}\int_{r_{\rm UV}}^{\mathrm{min}(\hat{r},r_2)} dr^{\prime}\frac{b}{a}b_{M_{Q}}^{-1}\nonumber\\
																																												 &&\mbox{}- \frac{1}{L_{M_{Q}}^2}\int_{r_{\rm UV}}^{r_{\rm IR}} d\hat{r}\frac{b}{a}b_{M_{Q}}\int_{r_{\rm UV}}^{\hat{r}} dr^{\prime}\frac{b}{a}b_{M_{Q}}^{-1}\int_{r^{\prime}}^{r_{\rm IR}} dr^{\prime\prime}\frac{b}{a}b_{M_{Q}} \,,\label{eq:pll}\\
	\hat{P}_{0; M_{q}}^{RR[--]}(r_1,r_2)&=&-\int_{r_{\rm UV}}^{\mathrm{min}(r_1,r_2)} dr^{\prime}\frac{b}{a}b_{M_{q}}+\frac{1}{L_{-M_{q}}}\int_{r_{\rm UV}}^{r_1} dr^{\prime}\frac{b}{a}b_{M_{q}}\int_{r^{\prime}}^{r_{\rm IR}} dr^{\prime\prime}\frac{b}{a}b_{M_{q}}^{-1}\nonumber\\
																																																			&&\mbox{}+ \frac{1}{L_{-M_{q}}}\int_{r_{\rm UV}}^{r_{\rm IR}} d\hat{r}\frac{b}{a}b_{M_{q}}^{-1}\int_{r_{\rm UV}}^{\mathrm{min}(\hat{r},r_2)} dr^{\prime}\frac{b}{a}b_{M_{q}}\nonumber\\
																																																&&\mbox{}- \frac{1}{L_{-M_{q}}^2}\int_{r_{\rm UV}}^{r_{\rm IR}} d\hat{r}\frac{b}{a}b_{M_{q}}^{-1}\int_{r_{\rm UV}}^{\hat{r}} dr^{\prime}\frac{b}{a}b_{M_{q}}\int_{r^{\prime}}^{r_{\rm IR}} dr^{\prime\prime}\frac{b}{a}b_{M_{q}}^{-1} \,,\label{eq:prr}
\end{eqnarray}
where  we have defined 
\begin{eqnarray}
	L_{\pm M}=\int_{r_{\rm UV}}^{r_{\rm IR}} dr \frac{b}{a}b_{\pm M} \,.
\end{eqnarray}
To obtain (\ref{eq:pll}) and (\ref{eq:prr}) we have used that, applying the corresponding boundary conditions and after some algebra, we can write
\begin{eqnarray}
	\hat{f}_{n}^{(Q_L})(r)&=&-\frac{m_{n}^{(Q)2}}{L_{ M_{Q}}}\int_{r_{\rm UV}}^{r_{\rm IR}} d\hat{r}\frac{b}{a}b_{ M_{Q}}\int_{r_{\rm UV}}^{\hat{r}} dr^{\prime}\frac{b}{a}b_{M_{Q}}^{-1}\int_{r^{\prime}}^{r_{\rm IR}} dr^{\prime\prime}\frac{b}{a}b_{ M_{Q}} \hat{f}_{n}^{( Q_L)}(r^{\prime\prime})\nonumber\\
																																						   &+&m_n^{( Q)2}\int_{r_{\rm UV}}^r dr^{\prime}\frac{b}{a}b_{ M_{Q}}^{- 1}\int_{r^{\prime}}^{r_{\rm IR}} dr^{\prime\prime}\frac{b}{a}b_{ M_{Q}} \hat{f}_{n}^{(Q_L)}(r^{\prime\prime}) \,,\\
 \hat{f}_{n}^{( q_R)}(r)&=&-\frac{m_n^{(q)2}}{L_{-M_{q}}}\int_{r_{\rm UV}}^{r_{\rm IR}} d\hat{r}\frac{b}{a}b_{M_{q}}^{-1}\int_{r_{\rm UV}}^{\hat{r}} dr^{\prime}\frac{b}{a}b_{ M_{q}}\int_{r^{\prime}}^{r_{\rm IR}} dr^{\prime\prime}\frac{b}{a}b_{M_{q}}^{-1} \hat{f}_{n}^{(q_R)}(r^{\prime\prime})\nonumber\\
																																						&+&m_n^{(q)2}\int_{r_{\rm UV}}^r dr^{\prime}\frac{b}{a}b_{M_{q}}\int_{r^{\prime}}^{r_{\rm IR}} dr^{\prime\prime}\frac{b}{a}b_{M_{q}}^{-1} \hat{f}_{n}^{(q_R)}(r^{\prime\prime}) \,.
\end{eqnarray}

In the AdS$_5$ case (\ref{ADS5def}), we can get analytical formulas for the expressions defined in eqs.~(\ref{eq:klonet}), (\ref{eq:kloneb}) and (\ref{eq:kkk}) above. They are
{\small  \begin{eqnarray}
	\mathcal{H}_{q}&\approx&\sum_{k,l}\frac{\tilde{v}^2}{M_{\rm KK}^2}\frac{Y_{q}^{kl}Y_{q}^{\dagger lk} (1+\beta)^2}{\mathfrak{p}_2(-c_Q^l+c_{q}^k)\left(1-\Omega^{1-2c_Q^l}\right)\left(1-\Omega^{1+2c_{q}^k}\right)}\left[\left(\frac{1}{\mathfrak{q}_4(0)}-\frac{1}{\mathfrak{p}_2(c_Q^l-c_{q}^k)}\right)\right.\nonumber\\
																   &+&\left.\left(\frac{1}{\mathfrak{q}_5(2c_{q}^k)}+\frac{1}{\mathfrak{q}_5(-2c_Q^l)}-\frac{1}{\mathfrak{p}_4(-c_Q^l+c_{q}^k)}-\frac{1}{\mathfrak{q}_4(0)}\right)\Omega^{2-2c_Q^l+2c_{q}^k}\right.\nonumber\\
																   &+&\left.\left(\frac{1}{\mathfrak{p}_3(c_{Q}^l+c_{q}^k)}-\frac{1}{\mathfrak{q}_5(2c_{q}^k)}\right)\Omega^{1+2c_{q}^k}+\left(\frac{1}{\mathfrak{p}_3(-c_{Q}^l-c_{q}^k)}-\frac{1}{\mathfrak{q}_5(-2c_{Q}^l)}\right)\Omega^{1-2c_{Q}^l}\right] , \\
\left(\mathcal{I}_{q}\right)_{ij}&\approx&\sum_{k,l}\frac{\tilde{v}^2}{M_{\rm KK}^2}\frac{Y_{q}^{ik}Y_{q}^{\dagger kl}Y_{q}^{lj} (1+\beta)^3\sqrt{1-2c_Q^{i\phantom{j}}}\sqrt{1+2c_{q}^j}\mathfrak{p}_2(-c_Q^i+c_{q}^k)^{-1}\mathfrak{p}_2(-c_Q^l+c_{q}^j)^{-1}}{\lambda_{i}^{q}\left(1-\Omega^{1-2c_Q^l}\right)\left(1-\Omega^{1+2c_{q}^k}\right)\sqrt{-1+\Omega^{1-2c_Q^i}}\sqrt{-1+\Omega^{1+2c_{q\phantom{Q}}^j}}} \nonumber\\
																																	   &\times&\left[\left(\frac{1}{\mathfrak{q}_4(c_Q^l-c_Q^i)}+\frac{1}{\mathfrak{q}_4(-c_{q}^k+c_{q}^j)}-\frac{1}{\mathfrak{p}_2(c_Q^l-c_{q}^k)}	-\frac{1}{\mathfrak{r}_6(-c_Q^i+c_{q}^j)}\right)\Omega^{1-c_Q^i+c_{q}^j}\right.\nonumber\\
																													 &+&\left.\left(\frac{1}{\mathfrak{q}_5(c_{q}^k+c_{q}^j)}+\frac{1}{\mathfrak{q}_5(-c_{Q}^l-c_{Q}^i)}-\frac{1}{\mathfrak{p}_4(-c_Q^l+c_{q}^k)}-\frac{1}{\mathfrak{r}_6(-c_Q^i+c_{q}^j)}\right)\Omega^{3-2(c_{Q}^l- c_{q}^k)-c_{Q}^i+c_{q}^j}\right. \nonumber\\
																																	   &+&\left.\left(\frac{1}{\mathfrak{p}_3(c_{Q}^l+c_{q}^k)}-\frac{1}{\mathfrak{q}_4(c_Q^l-c_Q^i)}-\frac{1}{\mathfrak{q}_5(c_{q}^k+c_{q}^j)}+\frac{1}{\mathfrak{r}_6(-c_Q^i+c_{q}^j)}\right)\Omega^{2+2c_{q}^k+c_{q}^j-c_{Q}^i}\right.\\
																																	   &+&\left.\left(\frac{1}{\mathfrak{p}_3(-c_{Q}^l-c_{q}^k)}-\frac{1}{\mathfrak{q}_4(-c_{q}^k+c_{q}^j)}-\frac{1}{\mathfrak{q}_5(-c_{Q}^l-c_{Q}^i)}+\frac{1}{\mathfrak{r}_6(-c_Q^i+c_{q}^j)}\right)\Omega^{2-2c_{Q}^l+c_{q}^j-c_{Q}^i}\right] , \nonumber\\
\left(\mathcal{J}_{q}\right)_{ij}&\approx&\sum_k\frac{1}{2}\frac{\tilde{v}^2}{M_{\rm KK}^2}\frac{Y_{q}^{ik}Y_{q}^{\dagger kj}\left(1+\beta\right)^2\sqrt{1-2c_{Q}^i}\sqrt{1- 2c_{Q}^j} \mathfrak{p}_2(-c_Q^i+c_{q}^k)^{-1} \mathfrak{p}_2(-c_Q^j+c_{q}^k)^{-1}}{\sqrt{-1+\Omega^{1-2c_{Q}^i}}\sqrt{-1+\Omega^{1-2c_Q^j}}\left(1-\Omega^{1+2c_{q}^k}\right)^2} \nonumber\\
																									&\times &\left[\frac{(1+2c_{q}^k)^2\Omega^{2c_{q}^k-c_Q^i-c_Q^j}}{(-1+2c_{q}^k)(3+2c_{q}^k)}-\frac{ \mathfrak{q}_6(-c_{Q}^i-c_{Q}^j-2 c_{q}^k)\mathfrak{p}_2(-c_Q^i+c_{q}^k)\mathfrak{p}_2(-c_Q^j+c_{q}^k)  \Omega^{1-c_{Q}^i-c_{Q}^j}}{(-1+2 c_{q}^k) \mathfrak{q}_5(-c_{Q}^i-c_{Q}^j)\mathfrak{p}_3(-c_{Q}^i-c_{q}^k) \mathfrak{p}_3(-c_{Q}^j-c_{q}^k)   }\right.\nonumber\\
   &-&\left. 
   \frac{(4 c_{q}^k (c_{q}^k+1)-3)  \mathfrak{q}_7(-c_{Q}^i-c_{Q}^j)\Omega^{2+2
   c_{q}^k-c_{Q}^i-c_{Q}^j}\mathfrak{S}(c_Q^i,c_Q^j,c_{q}^k)}{(-1+2 c_{q}^k) (3+2 c_{q}^k) \mathfrak{q}_5(-c_{Q}^i-c_{Q}^j) \mathfrak{p}_3(-c_Q^i-c_{q}^k) \mathfrak{p}_3(-c_Q^j-c_{q}^k)\mathfrak{p}_4(-c_Q^i+c_{q}^k) \mathfrak{p}_4(-c_Q^j+c_{q}^k) }\right.\nonumber\\
   &+&\left.\frac{\mathfrak{p}_1(-c_{Q}^i-c_{q}^k) \mathfrak{p}_1(-c_{Q}^j-c_{q}^k)\mathfrak{q}_8(-c_Q^i-c_Q^j+2c_{q}^k)\Omega^{3+4c_{q}^k-c_Q^i-c_Q^j}}{(3+2c_{q}^k)\mathfrak{p}_4(-c_Q^i+c_{q}^k) \mathfrak{p}_4(-c_Q^j+c_{q}^k)\mathfrak{q}_5(-c_Q^i-c_Q^j) } \right],\\
   \left(\mathcal{K}_{q}\right)_{ij}&=& \left(\mathcal{J}_{q}\right)_{ij}\left(c_Q^{i,j}\to -c_{q}^{i,j}, c_{q}^k\to -c_{Q}^{k}\right).
   \end{eqnarray}}
In the above expressions $Y_q$ are the dimensionless Yukawa matrices defined in (\ref{ydimlesb}), and for the sake of notational simplicity we have defined the functions
\begin{eqnarray}
	\mathfrak{S}(x,y,z)&=&x^2 y^2+(-2 \beta -5) x y (x+y)+(x+y)^2 \left(\beta ^2+5 \beta -z^2-z+5\right)\nonumber\\
								   &&\mbox{}+ x y \left(4 \beta ^2+20 \beta +2 \left(-\beta ^2-5 \beta
   +z^2+z-5\right)+25\right)\nonumber\\
   &&\mbox{}+ (x+y) \left(-2 \beta ^3-15 \beta ^2-35 \beta +2 \beta  z^2+5 z^2+2 \beta  z+5 z-25\right)\nonumber\\
   &&\mbox{}+ z^4+2 z^3-2 \beta ^2 z^2-10 \beta  z^2-13 z^2-2 \beta ^2 z-10 \beta  z-14 z+24\nonumber\\
   &&\mbox{}+ \beta ^4+10 \beta ^3+35 \beta ^2+50 \beta \,,
\end{eqnarray}
and
\begin{equation}
	\mathfrak{p}_k(c)=(k+c+\beta) \,, \qquad 
	\mathfrak{q}_k(c)=(k+c+2\beta) \,, \qquad 
	\mathfrak{r}_k(c)=(k+c+3\beta) \,.
\end{equation}

\bibliographystyle{JHEP}
\bibliography{bibliography}   

\end{document}